\documentclass{aastex63} 

\usepackage{multirow}

\received{xx 2021}
\revised{xxx}
\accepted{xxx}
\submitjournal{ApJ}

\shorttitle{Prompt emission of short gamma ray bursts}
\shortauthors{Iyyani et al.}

\begin{document}

\title{Study of prompt emission of short gamma ray bursts using multi-color blackbody: a clue to the viewing angle}

\correspondingauthor{Shabnam Iyyani, Vidushi Sharma}
\email{shabnam@iucaa.in, vidushi@iucaa.in}

\author[0000-0002-0786-7307]{Shabnam Iyyani}
\affil{Inter-University center for Astronomy and Astrophysics, Pune, Maharashtra, 411007, India}

\author[0000-0002-0786-7307]{Vidushi Sharma}
\affiliation{Inter-University center for Astronomy and Astrophysics, Pune, Maharashtra, 411007, India}







\begin{abstract}
The prompt emission of short gamma ray bursts (sGRBs) with known
redshifts are analyzed using the model of multi-color blackbody which 
is interpreted as the emission from a non-dissipative photosphere 
taking into account a power law jet structure and the viewing geometry of the jet.
We find nearly $69\%$ and $26\%$ of the sample are consistent with 
multi-color blackbody and a pure blackbody model, respectively. Using this interpretation, we infer
that nearly $57\% \, (18\%)$ of the sGRBs in our sample are observed 
within (or along the edge of) the jet core.
The sGRB jets are deduced to possess a narrow core with a median $\theta_c \sim 3^{\circ}$. This suggests the rate of sGRBs that would be viewed within the jet core to be $1.8 - 26 \, \rm 
Gpc^{-3} \, yr^{-1}$. The power law index of the decreasing Lorentz factor profile of the jet structure is deduced to be $1.3 - 2.2$.
The intrinsic luminosity is found to range between $10^{48} - 10^{53}\, \rm erg/s$. The average values of Lorentz factor and nozzle radius of the sGRB jets are inferred to be $210\, (85)$ and $10^{7.7} \, (10^{9.6}) \, \rm cm$ for the cases when the photosphere forms in the coasting (accelerating) phase respectively. 
The viability of the inferred values of the different parameters of the GRB outflow and viewing geometry within this physical interpretation, enhances the prospect of the photospheric emission model to explain the observed GRB spectrum.

 
\end{abstract}



\section{Introduction} \label{sec:intro}

A short Gamma ray burst (sGRB) is an extremely intense gamma ray flash that is observed for a duration
of less than $2$ seconds \citep{Kouveliotou_etal_1993}. They are hypothesized to be produced by the 
merger of binary compact objects such as neutron star -neutron star and neutron star - black hole. 
The detection of gravitational waves along with GRB 170817A \citep{Abbott_etal_2017_170817A}, has 
confirmed that at least a fraction of the observed sGRBs are produced by the merger of binary neutron
stars. The extensive follow-up observations of the GRB 170817A have confirmed the following key aspects: (i) the presence of a relativistic jet 
\citep{Mooley_etal_2018_170817A_jetsign,Mooley_2018_Nat_superluminal,Duffell_etal_2018_successfuljet}
; (ii) the jet has some structure beyond the jet opening angle 
\citep{Alexander_etal_2018_170817A_structuredjet} and (iii) it is viewed off-axis 
\citep{Granot_etal_2018_170817A_offaxis}. This affirms that the observed prompt emission spectrum has
a strong dependence on the viewing geometry and the structure of the jet. There are different possible jet structures such as Gaussian \citep{Resmi_etal_2018,Troja_etal_2018,Cunningham_etal_2020}, power law \citep{Lundman_etal_2013,Gottlieb_etal_2021}, hollow-cone \citep{Takahashi_etal_2020a,Takahashi_etal_2020b} etc. Recent 3D simulation studies of the hydrodynamic evolution of the GRB jet as it penetrates through the stellar core, have found that it is more likely to form jet structures characterised by simple power law distributions \citep{Gottlieb_etal_2020,Gottlieb_etal_2021}.

The radiation process during the prompt emission of GRBs is still ambiguous. Because of the non-thermal
nature of the spectrum, they are generally tried to be explained by the synchrotron emission process 
\citep{Tavani1996,Bosnjak_etal_2009,Yu_etal_2015,Zhang_etal_2020}. Recently, there have been several studies where a 
model of synchrotron emission is directly tested with the GRB data 
\citep{burgess_etal_2014a,burgess_etal_2020} as well as studies where doubly smoothly broken 
power laws are used to model the GRB spectra which later are interpreted as synchrotron emission 
\citep{Oganesyan_etal_2017,Ravasio_etal_2018}. Even though these works show that the GRB spectra can 
be fitted with synchrotron emission, it is worth noting that the microphysical parameters required to
achieve these fits are either unrealistic or require some fine tuning. 
Thus, there are some major drawbacks in interpreting the GRB spectra as synchrotron radiation. 

An alternative emission model that has been widely studied is the photosphere model. Photospheric
emission is inherent in the classical fireball model where the thermalized emission from the 
photosphere forms a subdominant part of the entire prompt gamma ray emission \citep{meszaros2006gamma}. The emission from the 
photosphere is expected to be completely thermalized, closely consistent with a blackbody spectrum, 
when no significant dissipation of the kinetic/ Poynting flux happens at moderate optical depths close to 
the photosphere. This is generally referred to as the non-dissipative photospheric emission \citep{ryde2004cooling,ryde2009quasi,guiriec2011detection,axelsson2012grb110721a,iyyani2013variable,Acuner_etal_2019,Acuner_etal_2020}.
However, several studies have shown that if continuous or localized dissipation of the kinetic or
Poynting flux energy happens at optical depths close to the photosphere can result in a 
non-thermalized emission which would look much broader than a Planck function (blackbody). These 
models are generally referred to as the subphotospheric dissipation models \citep{Pe'er_Waxman2004,Pe'er_Waxman2005,Giannios2006,beloborodov2013regulation,iyyani2015extremely,Bjorn_etal_2015,Bjorn_etal_2019}. 

\cite{Lundman_etal_2013} studied the non-dissipative photospheric emission from relativistic jets
that possess angle dependent Lorentz factor profile structure characterized by a power law distribution. They found that the observed emission 
from the photosphere can significantly alter from a blackbody spectrum depending on the viewing 
geometry and jet profile structure. The obtained spectrum is a superposition of several blackbodies 
with varying temperatures which brings about a significant broadening as well as softening of the low
energy part of the spectrum below the peak. Their work, thus, predicted the low energy power law 
indices of the spectrum, expected from different power law distribution jet profiles when viewed outside the jet core. 

\begin{figure}
    \centering
    \includegraphics[scale=0.5]{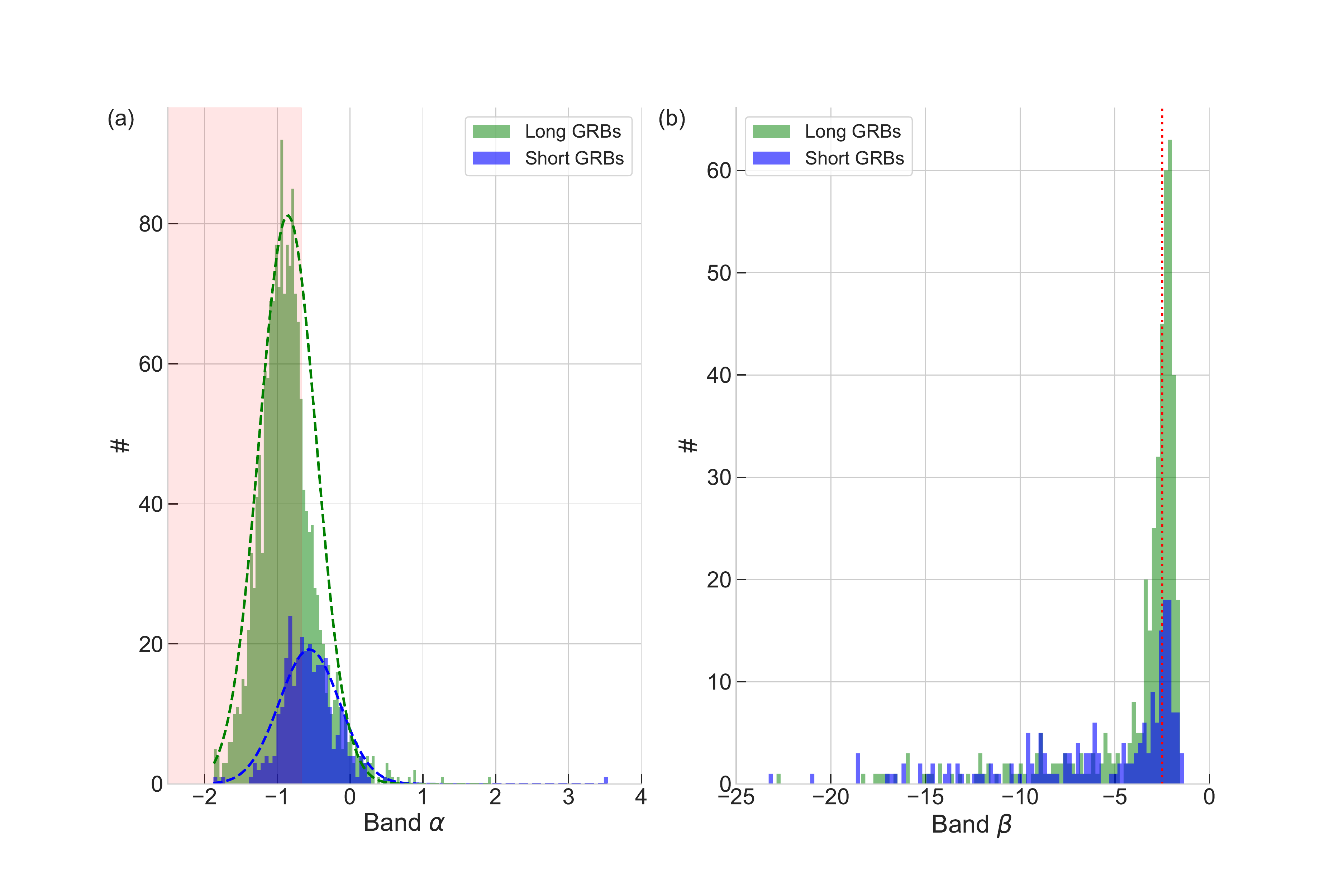}
    \caption{(a) The distribution of spectral low energy index, $\alpha$ of the Band function fits
    done to time integrated emission ($\rm T_{90}$ region) of the long (green) and short (blue) GRBs detected 
    by {\it Fermi} are shown. The $\langle \alpha 
    \rangle$ for short (long) GRBs is found to be $-0.56 \pm 0.42 \, (-0.86 \pm 0.39)$. The region shaded in red represents the values of $\alpha<-0.67$, the 'line of death' of synchrotron emission.  \\
    (b) The distributions of the high energy index, $\beta$, of the Band function fits of the long and short GRB sample
    whose $\alpha > -0.67$ are shown. $\beta = -2.5$ is marked by the vertical dotted line.  
    }
    \label{fermi_alpha_catalog}
\end{figure}

The Band function is widely used to empirically model the prompt gamma ray emission of GRBs \citep{band1993batse}.
The low energy power law index $\alpha$ of the Band function  fits done to the
short GRB spectra are largely found to be harder than the line of death ($\alpha = -0.67$) of 
synchrotron emission \citep{preece1998synchrotron}. This is demonstrated in the Figure 
\ref{fermi_alpha_catalog}a where we have plotted the distribution of 
$\alpha$ of the Band function fits done to the time integrated emission (T90 region) of the short 
GRBs (blue) detected and reported by {\it Fermi} till 15th July 2018 in their spectral catalog. The 
plotted distribution includes only those cases where the $\alpha$ is well constrained (i.e error 
measured on $\alpha$ is less than $0.5$) and the peak energy, $E_{peak} \ge 20\, \rm keV$. This 
ensures that the hard $\alpha$ values obtained in the spectral fits are not an artifact of having the
$E_{peak}$ close to the edge of the energy limits of the observation window. We find the mean of the 
$\alpha$ of short (long) GRBs to be $-0.56 \pm 0.42$ ($-0.86 \pm 0.39$). We also find that $61\%$ 
($25\%$) of the short (long) GRBs possess an $\alpha > -0.67$. Interestingly, we also note that 
$69\%$ ($53\%$) of these short (long) GRBs with hard $\alpha$ values also possess steep high energy 
power law indices, $\beta < -2.5$. A similar assessment of $\alpha$ distribution is
also found in \cite{Husne_etal_2020} where they find that a significantly large 
fraction of the upper limit of $\alpha$ in their sample violates the line of death 
of synchrotron emission. This kind of spectral features suggest a narrow spectrum
and thereby strongly point towards the emission coming from the jet photosphere. In addition, the cluster analysis conducted on all {\it Fermi} detected GRBs by \cite{Acuner_etal_2018}, found that the cluster that contains predominantly short bursts is found to be consistent with photospheric emission origin. 

Testing the different physical models directly with spectral data would be the proper course of action, however, this process is computationally very intensive. In the traditional method of spectral analysis, the spectrum is analysed using different models and the best fit model is determined by the best fit statistics. However, we note that the spectral analyses of most of the GRB data show that many of the different empirical and physical models are statistically equally consistent with the data \citep{Iyyani_etal_2016,Meng_etal_2018,Acuner_etal_2020} and as a result the best fit model remains ambiguous. 
In such scenarios of degeneracy between different models, the way forward will be to examine if the particular model of choice is consistent with the data and then assess the implications of the results within the physical interpretations of that model.


With these above motivations, in contrast to the traditional methodology, in this work, we first investigate if the time integrated as well as the peak count
spectra of sGRBs with known redshifts, detected by {\it Fermi} gamma ray space telescope and {\it 
Niel Gehrels Swift} is consistent with the multi-color blackbody model or not. Later, this model is interpreted within the physical scenario presented in \cite{Lundman_etal_2013}. The analysis results are presented in section \ref{spectral_analyzes}. The physical model and its correlation with the multi-color blackbody is described in section \ref{structuredjet}. Within the physical interpretation, the assessment of the different parameters such as the viewing geometry, power law index of the Lorentz factor jet profile, the jet opening angle and the core angle of the jet are presented in section \ref{view_angle} and section \ref{Thetaj}. Finally, discussions and conclusions are presented in section \ref{discussion} and section \ref{conclusion} respectively. 

\section{Sample selection and spectral analyzes}
\label{spectral_analyzes}
\begin{figure}
\includegraphics[width=\columnwidth]{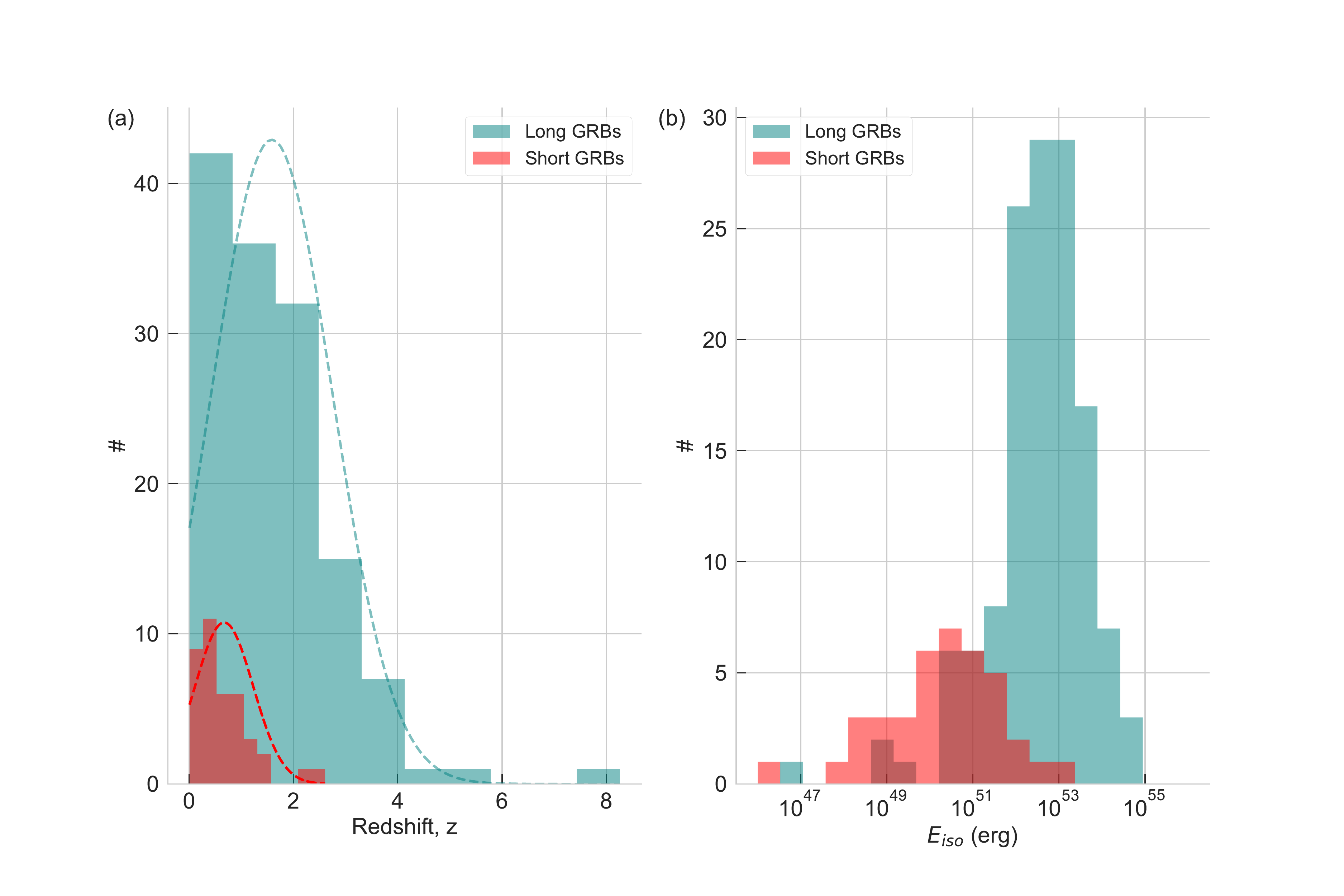}
    \caption{The distribution of (a) redshift, $z$ with a mean of $0.67 \pm 0.56$, and (b) isotropic
    prompt gamma ray emission, $E_{iso}$ with a mean of $(5 \pm 22) \times 10^{51} \, \rm erg $, 
    measured for short GRBs in the sample, are shown (red). For the purpose of comparison the redshift
    ($\langle z \rangle = 1.58 \pm 1.16$) and $E_{iso}$ ($\langle E_{iso} \rangle = (2.5 \pm 5.6)
    \times 10^{53} \, \rm erg$) of long GRBs are also shown in teal color.}
    \label{z_Eiso}
\end{figure}

All $39$ short GRBs with known redshifts detected by {\it Fermi} and {\it Niel Gehrels Swift} until
December 2018 are included in our sample (see Table \ref{sample_def} for more details). This enables us to have strong constraints on the energetics of the bursts. However, this makes the sample a mixture 
of bursts with varying energy fluences ranging between orders of $10^{-8} \, \rm erg/cm^2 - 10^{-5} 
\, \rm erg/cm^2$. The distribution of the redshift is shown in Figure \ref{z_Eiso}a and we find the 
average redshift of the sGRBs to be $\langle z \rangle = 0.67 \pm 0.56$ \citep{Berger2014_sGRB_review}. 

{\it Niel Gehrels Swift} BAT (Burst Alert Telescope) data is available for all sGRBs in the sample except for GRB 170817A.
There are 12 sGRBs which are observed by both {\it Fermi} and {\it Niel Gehrels Swift} BAT and in 
these cases, joint spectral analyzes are done using the prompt emission data from both the missions 
and composite light curves are made (see Figures \ref{lcs_1},\ref{lcs_2},\ref{lcs_3},\ref{lcs_4} in appendix).
For BAT data analyzes, the light curves and spectra are generated using the standard
procedures\footnote{https://swift.gsfc.nasa.gov/analysis/threads/bat threads.html} and data within 
the energy range $15 -150$ keV are used. For {\it Fermi} data analyzes, the three bright sodium 
iodide ({\tt NaI}) detectors with source angles $< 50^{\circ}$ and the brightest bismuth germanate ({\tt BGO}) detector are chosen and data within energy ranges, NaI: $8 - 800$ keV; BGO: $250$ 
keV - $30$ MeV are used\footnote{For GRB 090510, {\it Fermi} Large Area Telescope low energy (LLE) 
data was available and the data in the energy range $30 - 100 \, \rm MeV$ were used in the spectral 
analyzes.}. The {\it Fermi} spectral files are generated using Burst Analysis GUI v 02-01-00p1 
(gtburst3)\footnote{ https://fermi.gsfc.nasa.gov/ssc/data/analysis/scitools/gtburst.html}. 

The time interval for integrated spectral analysis is chosen subjectively by looking at each burst
lightcurve such that it includes the entire prompt emission of the burst. 
The start and stop times of the intervals as well as the results of the spectral analyzes are
mentioned in Table \ref{integ_analyzes}. The interval used for time integrated analysis of each 
burst is marked in shaded red color in the burst lightcurves shown in Figures \ref{lcs_1}, 
\ref{lcs_2}, \ref{lcs_3}, \ref{lcs_4} in appendix. The time integrated analyzes enables us to model the average 
shape of the overall burst emission and thereby estimate the total isotropic energy of the burst, 
$E_{iso}$ (Table \ref{integ_analyzes}). We assume a flat universe with cosmology parameters: 
$H_0$ = 67.4, $\Omega_m$ = 0.315 and $\Omega_{\lambda} =0.685$ for estimating the luminosity 
distances \citep{2018_planck}. The distribution of the estimated $E_{iso}$ is shown in Figure \ref{z_Eiso}b and we find 
the average isotropic total energy of sGRBs to be $\langle E_{iso} \rangle = (5  \pm 22) \times 
10^{51} \, \rm erg$. 

The burst emission can undergo significant spectral evolution within the total duration of the burst
and therefore, the underlying actual spectrum can be identified by doing time resolved spectral
analysis. However, the brevity of sGRBs makes the time resolved spectral analysis highly challenging
due to photon scarcity in each time resolved interval. Therefore, in this work, we conduct the
spectral analysis of the peak count interval of each burst where the photon counts are the highest.
Since the sample consists of sGRBs of varying intensity, we employ signal-to-noise ratio
(SNR)\footnote{The Bayesian block binning method, on the other hand, primarily follows the changes in counts of the light curve to decide the time intervals. In case of a sGRB with low intensity, this 
method may not give us a peak count spectral time interval with enough signal-to-noise ratio to
constrain the spectral model parameters.} binning method on the BAT light curves which are available
for all sGRBs except GRB 170817A where the method is employed on the brightest NaI detector. Later 
from the time resolved bins, the interval with highest counts is identified for spectral analysis. 
The start and stop times of the intervals as well as the results of the spectral analyzes are
mentioned in the Table \ref{peak_analysis}. The interval used for peak count spectral analysis of
each burst is marked in shaded black color in the burst light curves shown in Figures \ref{lcs_1},
\ref{lcs_2}, \ref{lcs_3}, \ref{lcs_4} in Appendix.
We point out that a strict limit of SNR could not be exercised on the entire sample because of the
varying intensity of the bursts. Thus, we choose the best possible SNR values subjectively for each 
burst such that the overall variability of the light curve is retained and enough signal is obtained 
in the peak interval to constrain the spectral parameters of most of the models that are fitted. The 
SNR values ranging between $3 \, \sigma - 25 \, \sigma$ are used for different bursts to obtain their
respective peak count spectrum (see the table \ref{sample_SNR} in appendix).  

The spectral analyzes are conducted on both the time integrated and time resolved peak count spectra
of each burst using the Multi-Mission Maximum Likelihood ({\tt 3ML}) software 
\citep{Vianello_etal_2015}. We use the maximum likelihood estimation technique to estimate the best 
fit spectral parameters and follow Akaike Information Criteria (AIC) for model selection 
\citep{akaike1974}.

\subsection{Results of spectral analyzes}
We analyze the time integrated and the peak count spectra of each burst in our sample using the 
empirical models such as power law ({\tt PL}), power law with exponential cutoff ({\tt CPL}), Band 
function ({\tt Band}) and physical models like blackbody ({\tt BB}) and multicolor blackbody 
({\tt mBB}, the model named as {\tt diskpbb} in {\tt Xspec}. See the section \ref{structuredjet} for 
more details). The functions of the different models are given in the appendix 
\ref{spec_models}. 
We note that the aim of this spectral analyzes is to inspect if physical models like multi-color
blackbody or simple blackbody can fit the data consistently or not, and does not aim to rule out
other possible interpretations to the same data.

\begin{figure}
    \centering
    \includegraphics[scale=0.5]{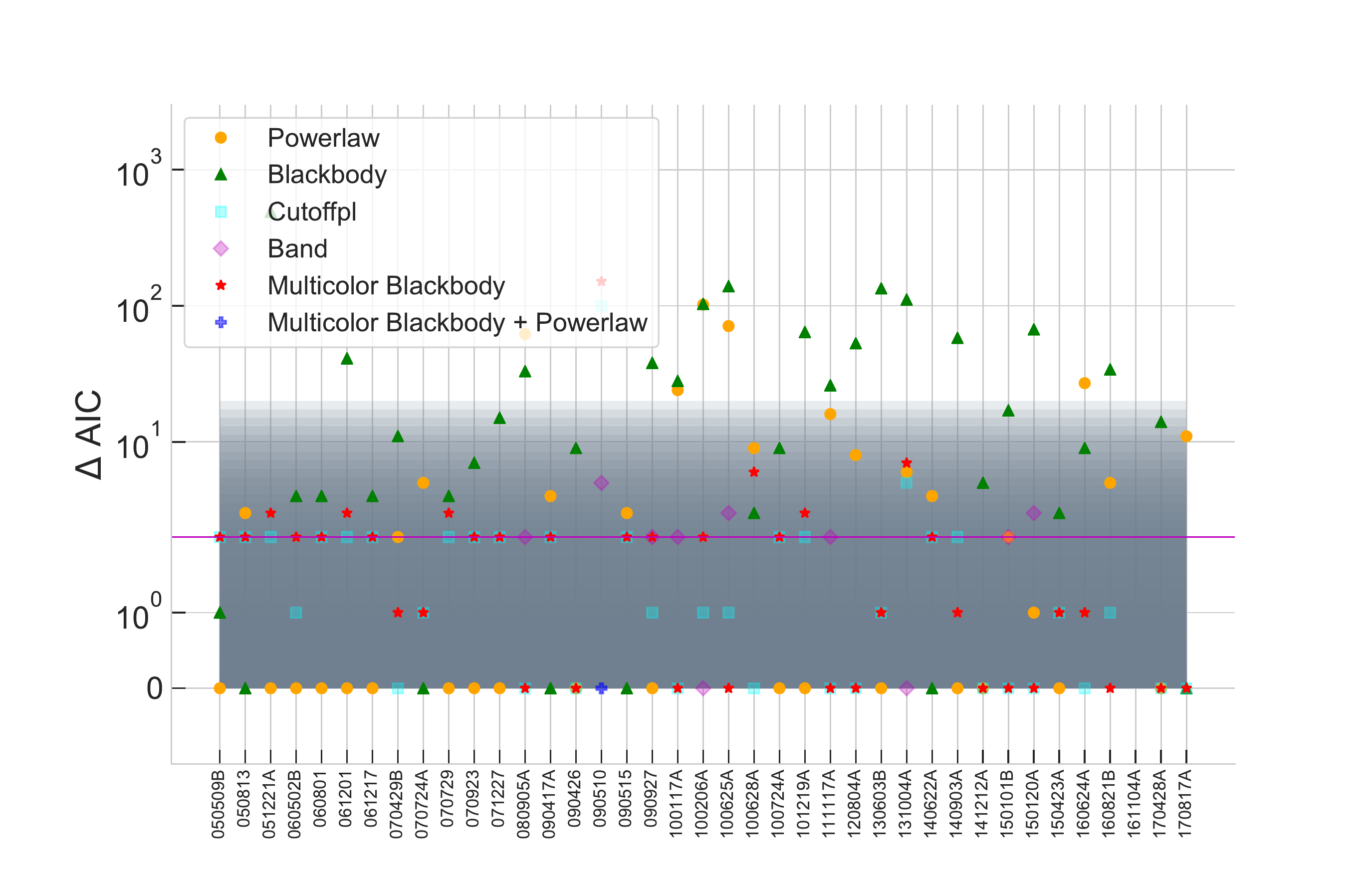}
    \includegraphics[scale=0.5]{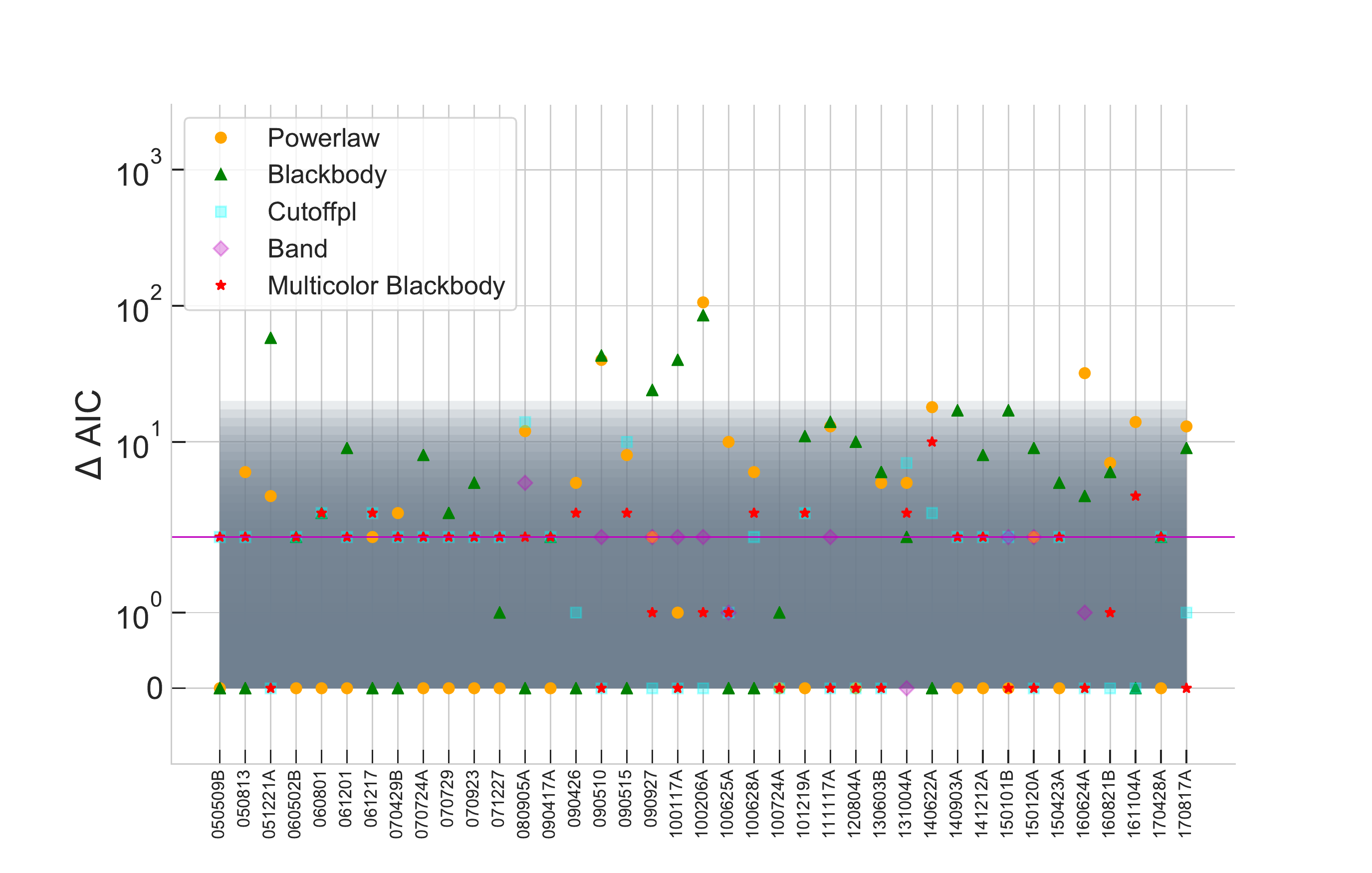}
    \caption{The $\Delta$AIC estimated for different models in case of different bursts for (a) time integrated spectra (top panel) and (b) peak count spectra (bottom panel) are shown.}
    \label{AIC}
\end{figure}

The spectral fits of the different models are compared by estimating the $\Delta \, \rm
AIC_{model} = AIC_{model} - AIC_{min}$, where $\rm AIC_{model}$ is the AIC obtained for each 
spectral model (see Table \ref{AIC_integ} and Table \ref{AIC_peak} in appendix for time integrated and peak spectral analyzes respectively) and $\rm AIC_{min}$ is the 
minimum of the AIC values obtained for all the spectral models that are in comparison. The 
$\Delta \, \rm AIC_{model}$ is estimated for each burst and is plotted in Figure \ref{AIC}. 

Generally, the plausible models are those whose $\Delta \rm AIC < 4$ (region shown in dark grey 
area), those models with $\Delta \rm AIC > 4 $ in the light grey area have inconclusive evidence and
those models with $\Delta \rm AIC $ in the white area are considered implausible 
\citep{Burnham_etal_2011}. We note that most of the tried models are largely consistent with the data, which leads to the scenario of degeneracy between models (see Figure \ref{residuals} in Appendix). Since the best fit model is ambiguous in this work, we consider the limit of $\Delta \rm AIC =2$ to determine 
if the model of our choice i.e {\tt mBB} or {\tt BB} is consistent with the data or not. If we find the $\Delta \rm AIC_{mBB}$
or $\Delta \rm AIC_{BB} \le 2$, then we choose ${\tt mBB}$ or ${\tt BB}$ as the best fit model of 
the respective spectral analysis. If $\Delta \rm AIC_{mBB \, or\,  BB} > 2$, then the model with the least
$\Delta \rm AIC$ is chosen as the best fit model for that analyzes. The results for the time
integrated and peak count spectra are shown in Table \ref{integ_analyzes} and  Table
\ref{peak_analysis} respectively. 
With this methodology, we find that the peak count spectra of GRB 101219A and GRB
060801 are best modeled using {\tt PL} model, whereas the
remaining sGRBs in the sample are found to be consistent with either {\tt mBB} or {\tt BB}
model. We find that $69\%$ of the sample is consistent with multi-color blackbody spectrum,
whereas $26\%$ of the sample is consistent with a simple blackbody spectrum. Thus, only $37$ GRBs in the sample whose peak count spectra are consistent with either {\tt mBB} or {\tt BB} are studied further within the structured jet model. Throughout the paper, until otherwise mentioned, the spectrum used for study is only the peak count spectrum. 

\subsection{Results of Multi-color Blackbody ({\tt mBB}) fits}
The appropriate method would have been to directly fit the output of the Monte Carlo simulations of the \cite{Lundman_etal_2013} model with the data, however, this process is computationally very intensive and also, time consuming when it has to be employed on a large sample of bursts. In addition, it has also been shown in several studies \citep{Iyyani_etal_2016,Meng_etal_2018,Acuner_etal_2020} that different physical models can be equally fitted to the data, thereby being unable to resolve the ambiguity between different spectral models that are analysed. In this scenario, we find it judicious to use the readily available multi-color blackbody, {\tt diskpbb}, model adopted from {\tt Xspec} whose spectral shape is produced by the superposition of many blackbodies analogous to that expected in \cite{Lundman_etal_2013}. This {\tt diskpbb} model was originally devised to model the spectrum coming from an accretion disc. Therefore, we note that the interpretations that we do to this model in the current work are totally different and have no connection other than the similar spectral shapes.

\begin{figure}
    \centering
    \includegraphics[scale=0.5]{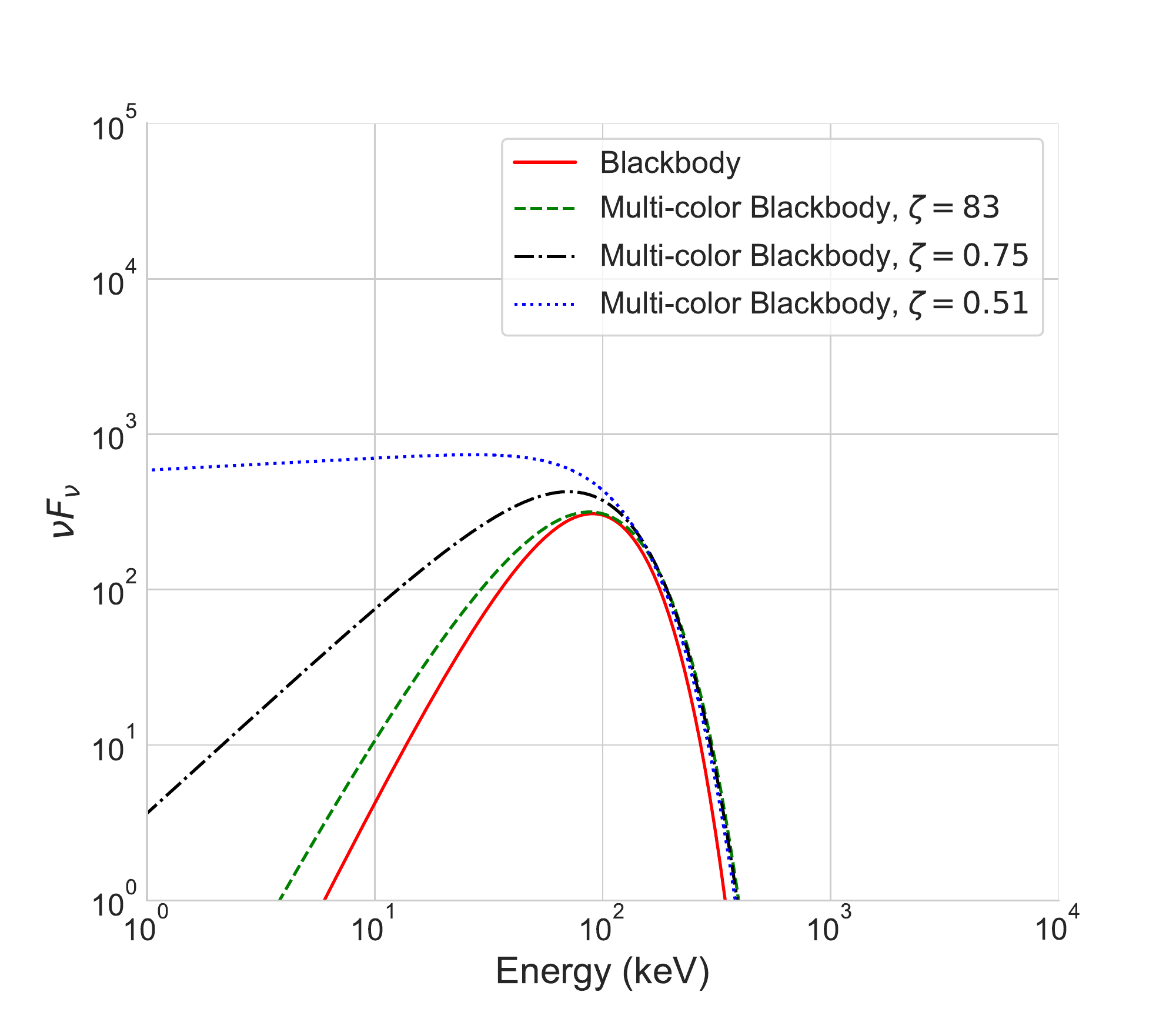}
    \caption{The $\nu F_{\nu}$ plot of  multicolor-blackbody ({\tt mBB}) for different values of $\zeta$ are
    shown. For comparison, blackbody spectrum (red/solid line) is also plotted. Spectral width of 
    blackbody and ${\tt mBB}$ with $\zeta =(83,0.75,0.51)$ spectra are $3.5$ and $(3.9,7.1,27)$ 
    respectively. Shallow decay of temperature would produce a multicolor-blackbody spectrum with 
    broader $\nu F_{\nu}$ peak (blue dot-dot line).}
    \label{mbb}
\end{figure}

The multi-color blackbody has three fit parameters: $T_p$ is the peak temperature (keV), $\zeta$
is the power law index of the radial dependence of the temperature\footnote{For a different formulation, refer \cite{Hou_etal_2018}, wherein instead of probability distribution of temperatures, they consider a multi-color blackbody whose distribution of luminosity is a function of temperature with a power-law form. In contrast to {\tt diskpbb}, in their model, they have the minimum temperature, $T_{min}$ also as a free parameter, thereby allowing the inclusion of the Rayleigh-Jean component of the spectrum below $E < kT_{min}$, in the modelling of the observed spectral data.} 
\begin{equation}
    T \propto r^{-\zeta}
\end{equation}
and $K$ is the normalization (the detailed model expression is given in the Appendix A). 
The parameter space of $\zeta$ is configured to $0-100$ (the default range is $0.5 - 1$ for the accretion disk, \citealt{mBB_reference}) in order to incorporate the wide range of spectral shapes that are possible in case of the GRB 
spectra. The minimum and maximum $\zeta$ values obtained in this sample study are $0.51$ and $83$ respectively. When $\zeta$ tends to infinity, theoretically the $\tt mBB$ tends to a $\tt BB$ 
spectral shape. The {\tt mBB} spectral shapes obtained for different $\zeta =[0.51, 0.75, 83]$ and $T_p = 30 \, \rm keV$ are plotted in the Figure \ref{mbb}.  

We simulate a large number of {\tt mBB} spectral model assuming normal distributions for the fit
parameters with the best fit values as the mean and the errors as its standard deviation. For 
each instance of the simulated {\tt mBB} spectral model, a power law model is fitted to the low 
energy part of the spectrum below $0.3 \, T_p$ and thus, the power law spectral index ({\tt mBB} 
$\alpha$) corresponding to a given {\tt mBB} spectral shape is measured. This is repeated for 
all the simulated {\tt mBB} spectra and thereby obtaining a distribution of {\tt mBB} $\alpha$. 
We then fit a Gaussian function to this distribution to obtain the mean as the {\tt mBB} 
$\alpha$ value and the standard deviation as its error.     

\begin{figure}
    \centering
    \includegraphics[scale=0.4]{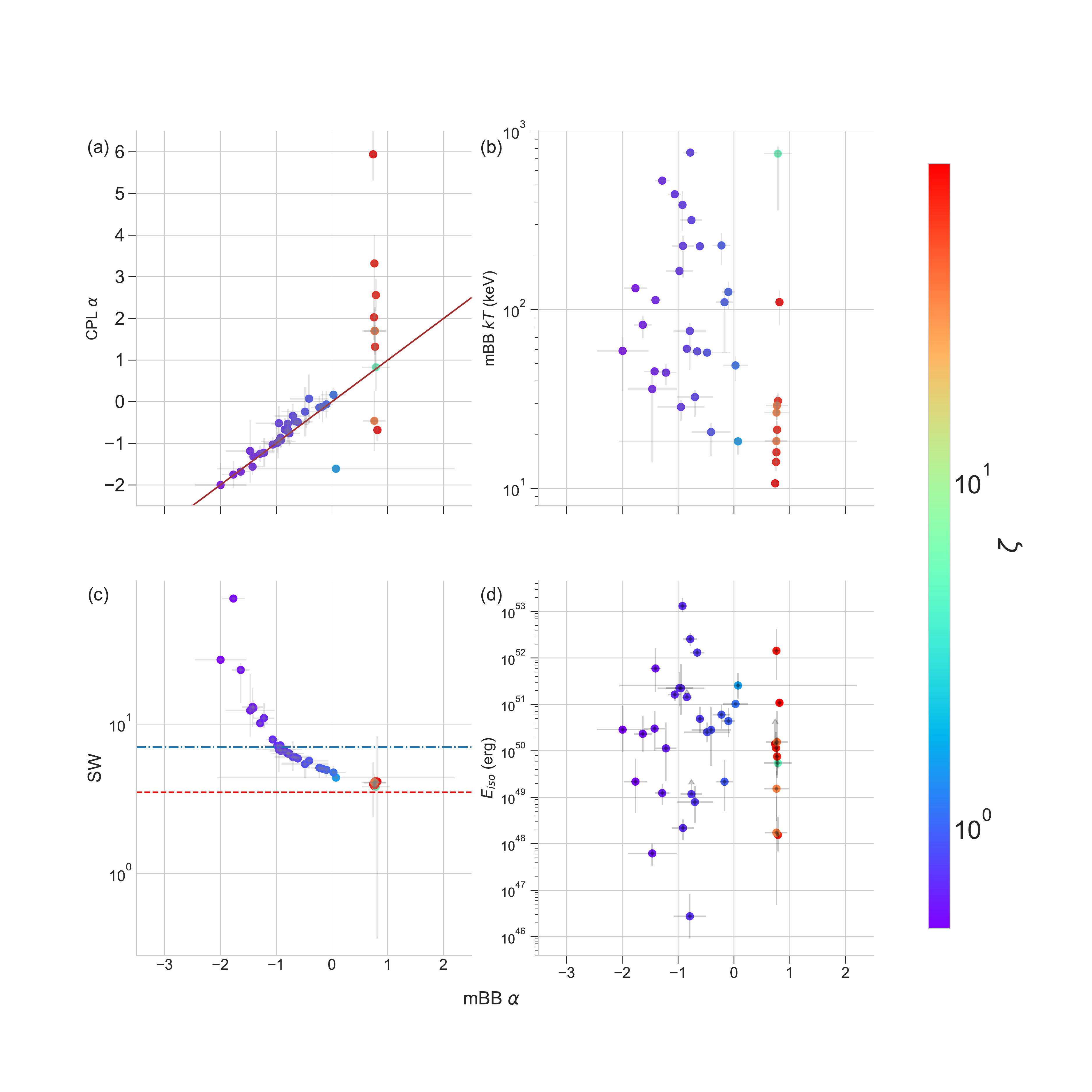}
    \caption{(a) The correlation between the low energy power law index determined from a cutoffpl fit
    ({\tt CPL} $\alpha$) and that from a multicolor blackbody fit ({\tt mBB} $\, \alpha$) is shown. The 
    hard low energy power indices in case of spectral shapes that are similar to blackbody are not properly recovered by the
    {\tt CPL} fits. The brown solid line marks when {\tt CPL} $\alpha$ $=$ {\tt mBB} 
    $\alpha$. (b) The variation of the peak temperature of the {\tt mBB} fits with respect to the {\tt mBB} \, 
    $\alpha$ is shown. We find that there is no distinct correlation between these two parameters. (c) 
    The spectral width of the $\nu\, F_{\nu}$ peak of the {\tt mBB} spectral fits are shown. We can see
    that softer $\alpha$ leads to broader spectrum whereas harder $\alpha$ tends to narrower spectral 
    shapes closer to that of a blackbody. The blue dash dot line represents $SW=7$ which is the spectral width of the photospheric emission from a spherically symmetric coasitng wind. The red dashed line is the spectral width of blackbody, $SW=3.5$. (d) The isotropic energy, $E_{iso}$ of sGRBs versus $\alpha$ is 
    plotted. We find that both soft and hard $\alpha$ are found across the range of observed values of $E_{iso}$. All the plots are color coded with respect to the power law index $\zeta$ of the 
    variation of the temperature in the multicolor blackbody.}
    \label{analyzes_results}
\end{figure}

The {\tt mBB} model is flexible enough to capture the spectral features, similar to the 
empirical model {\tt CPL}. This comparison is shown in the Figure \ref{analyzes_results}a where 
the low energy power law indices, {\tt CPL} $\alpha$, of cutoff power law function are plotted 
against the low energy power law indices, {\tt mBB} $\alpha$, obtained for the multi-color 
blackbody fits. We find that the spectra with soft low energy spectral slopes ($\alpha <0$) 
are consistently reproduced by both the models 
within errors. However, we find that the spectra which tend to have hard spectral slopes 
($\alpha >0$) are not well captured by {\tt CPL} mostly in cases where the peak of the spectrum 
is also close to the edge of the energy window, which results in unphysical hard {\tt CPL} 
$\alpha$ values\footnote{The narrowest spectrum that can be obtained is that of a 
blackbody (Planck function) whose low energy spectral slope is $\alpha = +1$} $\gg +1$. On the other
hand, these hard spectra when modeled using {\tt mBB} produce good fits and has $\alpha$ values
$\le +1$ which are more realistic. This shows that {\tt CPL} is not always a good enough
empirical approximation for blackbody model particularly when the model finds it difficult to 
constrain the cutoff peak energy at low energies close to the edge of the energy window.  

We note that there is no correlation found between the peak temperature, $T_p$ of {\tt mBB} fits
with the spectral slope, {\tt mBB} $\alpha$ (Figure \ref{analyzes_results}b) indicating that there is no energy window bias. We also studied
the spectral width\footnote{Spectral width is defined as the ratio of the limiting energies 
bounding the full width half maximum of the $\nu F_{\nu}$ spectral peak
\citep{Axelsson_Borgonovo2015,iyyani2015extremely}} of the {\tt mBB} model fits. We find that as
$\alpha$ gets harder, the width of the spectrum becomes narrower. A significant fraction of the 
GRBs have peak count spectrum that is as narrow as the non-dissipative photospheric emission 
from a spherically symmetric wind ($SW = 7$) and even narrower as blackbody ($SW= 3.5$) \citep{iyyani2015extremely}. The broadest spectrum is found to be of $SW = 69 $ (the errors on the 
spectral width are not well constrained in this case) with the softest $\alpha \sim -2$ in the 
sample. We note that we do not find any spectral width as large as a few hundreds as expected in
the fast cooling synchrotron emission (power law electron distribution; $SW \sim 370$) and slow
cooling synchrotron emission (Maxwellian electron distribution; $SW \sim 25$) \citep{Axelsson_Borgonovo2015,iyyani2015extremely}. 
We also do not find any correlation between the isotropic burst energy, $E_{iso}$ and {\tt mBB}
$\alpha$ (Figure \ref{analyzes_results}d). It is interesting that a given $\alpha$ has a large
range of $E_{iso}$ of at least $2-3$ order of magnitude. This suggests that the intrinsic total
burst energy of the sGRB events vary case to case.


\section{Structured Jet}
\label{structuredjet}
\begin{figure}
    \centering
    \includegraphics[scale=0.4]{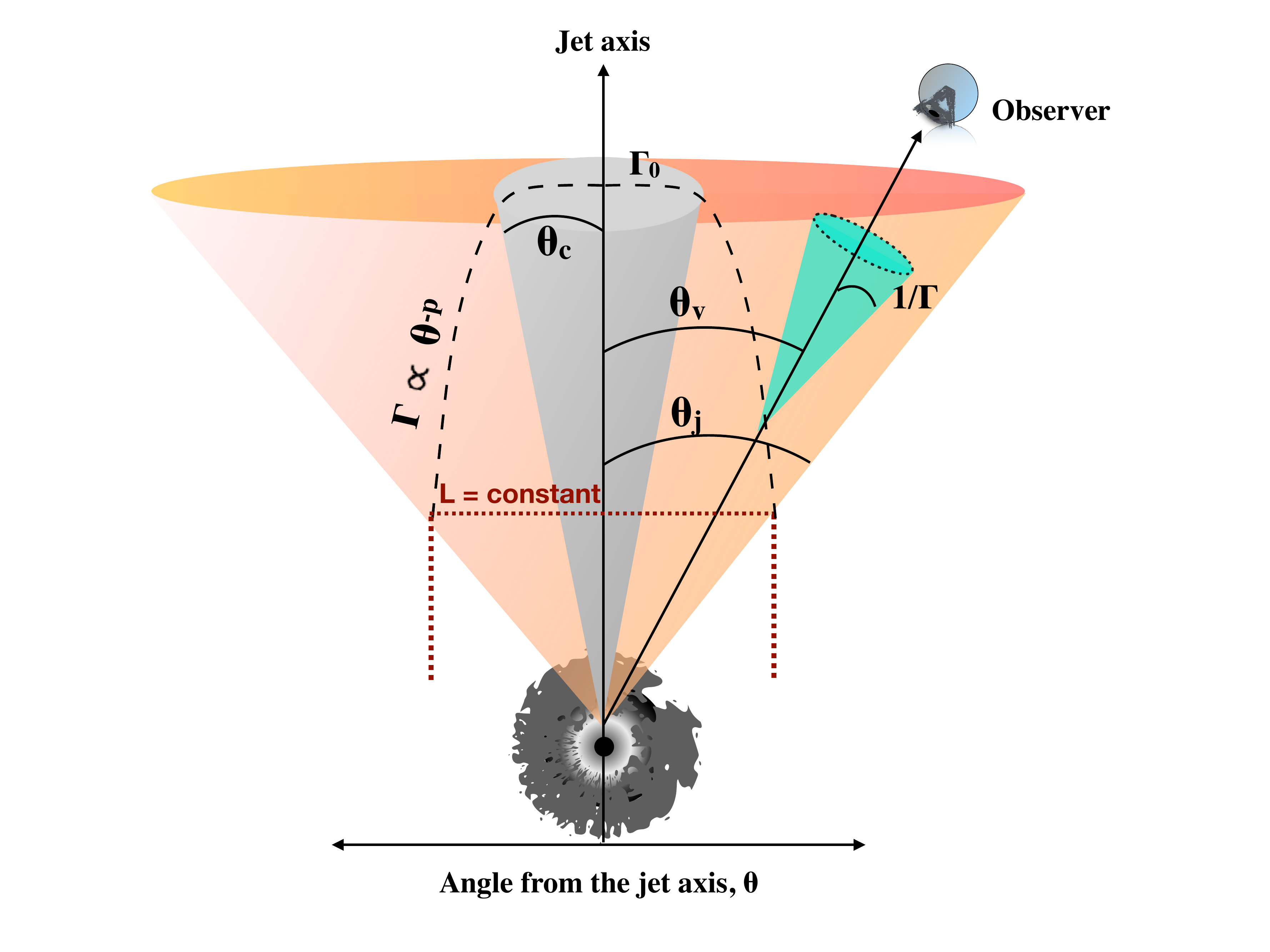}
    \caption{A schematic diagram of the structured jet model based on \cite{Lundman_etal_2013} is shown above. The dark shaded grey region represents the core of the jet ($\theta_c$) with Lorentz factor, $\Gamma_0$ and the surrounding orange shaded regions represent the extending jet structure such that $\Gamma \propto \theta^{-p}$ (shown in long dashed black line) until the jet opening angle ($\theta_j$). The burst luminosity ($L$), on other hand, remains angle independent with $\theta_j$. The viewing angle is represented by $\theta_v$ and the observer sees emission within the cone of $1/\Gamma$ along the line of sight of the observer. This model is used to interpret the multi-color blackbody spectral fits done in this work.}
    \label{schematic_fig}
\end{figure}

The observations of GRB 170817A have affirmed that a successful relativistic jet surrounded by a
mildly relativistic structure was produced in the sGRB event
\citep{Mooley_etal_2018_170817A_jetsign,170817A_offaxis}.
Previously, several hydrodynamic simulations have shown that as the jet emerges out, by undergoing
significant interactions with its surrounding ejecta (or stellar core in case of long GRBs), the jet
develops some angular profile structure with varying Lorentz factor, mass density and total energy
\citep{Zhang_Woosley_MacFadyen_2003, Morsony_Lazzati_Begelman_2007,Mizuta_Nagataki_Aoi_2011}.     
Inspired by these simulations, \cite{Lundman_etal_2013} did a Monte Carlo simulation of the
non-dissipative photospheric emission from a structured jet with an angular Lorentz factor profile
such that within the jet core, the Lorentz factor remains constant and beyond the
core, the Lorentz factor decreases as a power law until the jet opening angle\footnote{When the core region extends till $\theta_j$, the jet structure becomes equivalent to a top-hat jet.} ($\theta_j$, see Figure \ref{schematic_fig}). The Lorentz factor profile structure
is analytically defined as the following 
\begin{equation}
    (\Gamma - \Gamma_m)^2 = \frac{(\Gamma_0 - \Gamma_m)^2}{(\theta/\theta_c)^{2p} +1}
    \label{jet_profile}
\end{equation}
where $\Gamma_0$ is the Lorentz factor of the outflow within the jet core, $\theta_c$, $p$ is the power law index of the decreasing Lorentz factor profile beyond the
$\theta_c$ and $\Gamma_m = 1.2$ is the lowest Lorentz factor (slightly differing from unity) considered in the above definition of the jet profile. In this structured jet model, the outflow luminosity per solid angle of the central engine is considered to be angle independent\footnote{The observed jet luminosity, however, would decrease with increasing viewing angle \citep{Lundman_etal_2014}.} within $\theta_j$ and thereby, the outflow temperature at the nozzle radius, $r_0$, also remains angle independent. Beyond $\theta_j$ there is no injection of burst energy (Figure \ref{schematic_fig}) and therefore, any observer at a viewing angle, $\theta_v > \theta_j + 1/\Gamma(\theta_j)$, would not see any prompt emission from the GRB. 
Such a luminosity profile is inspired by the hydrodynamical simulations presented in \cite{Zhang_Woosley_MacFadyen_2003}.

They computed the observed spectrum from the photosphere when it forms in the coasting phase (that is
the photosphere lies above the saturation radius), taking into account the jet angular profile and
the geometry of how the jet is viewed by the observer. They found that when a jet is viewed off-axis
along the edge or beyond the jet core, such that the emission from the wings of the jet significantly
contribute to the resultant spectrum, the observed spectrum from the photosphere is much
broader than a pure blackbody ($\alpha = +1$) with the low energy part of the spectrum below the peak
energy as soft as $-1 \le \alpha \le -0.5$. The broadening of the observed spectrum from a pure blackbody
is due to the fact that the resultant spectrum is a superposition of multiple blackbodies of
decreasing temperatures coming from the emitting regions of lower Lorentz factors. However, when viewed within the jet core, the spectrum is found to be
relatively narrower with a low energy power law index $\alpha =+0.4$, similar to the spectrum
expected from a spherical wind \citep{Beloborodov2011}. 

\subsection{Correlation between {\tt mBB} and the structured jet}
\cite{Lundman_etal_2013} model found that for values of $ -1<\alpha<-0.5$, the low energy power
law index of the observed spectrum, $\alpha$ is related to the jet's Lorentz factor profile index,
$p$ by the relation 
\begin{equation}
    \alpha = \frac{-1}{4}\left(1+\frac{3}{p}\right),
    \label{alpha_1}
\end{equation}
in case of a narrow jet (where $\theta_c \le 3/\Gamma_0$) such that the observed emission is dominated by the emission coming from the region outside the jet core and thereby depends on the profile of the structure of the jet. In case of a wide jet ( $\theta_c \gg 3/\Gamma_0$), when viewed within the jet core ($\theta_v \ll \theta_c$), the observed emission has minimum superposition of blackbodies of varying temperatures, and the spectrum would be close to a thermal shape ($\alpha = +0.4$). Even in case of wide jets, if viewed outside or close to the jet core ($\theta_v \ge \theta_c$), the spectrum would be dominated by the emission coming from the region of decreasing Lorentz factor and this would produce significant softening in the low energy part of the spectrum. 

\begin{figure}
    \centering
    \includegraphics[scale=0.4]{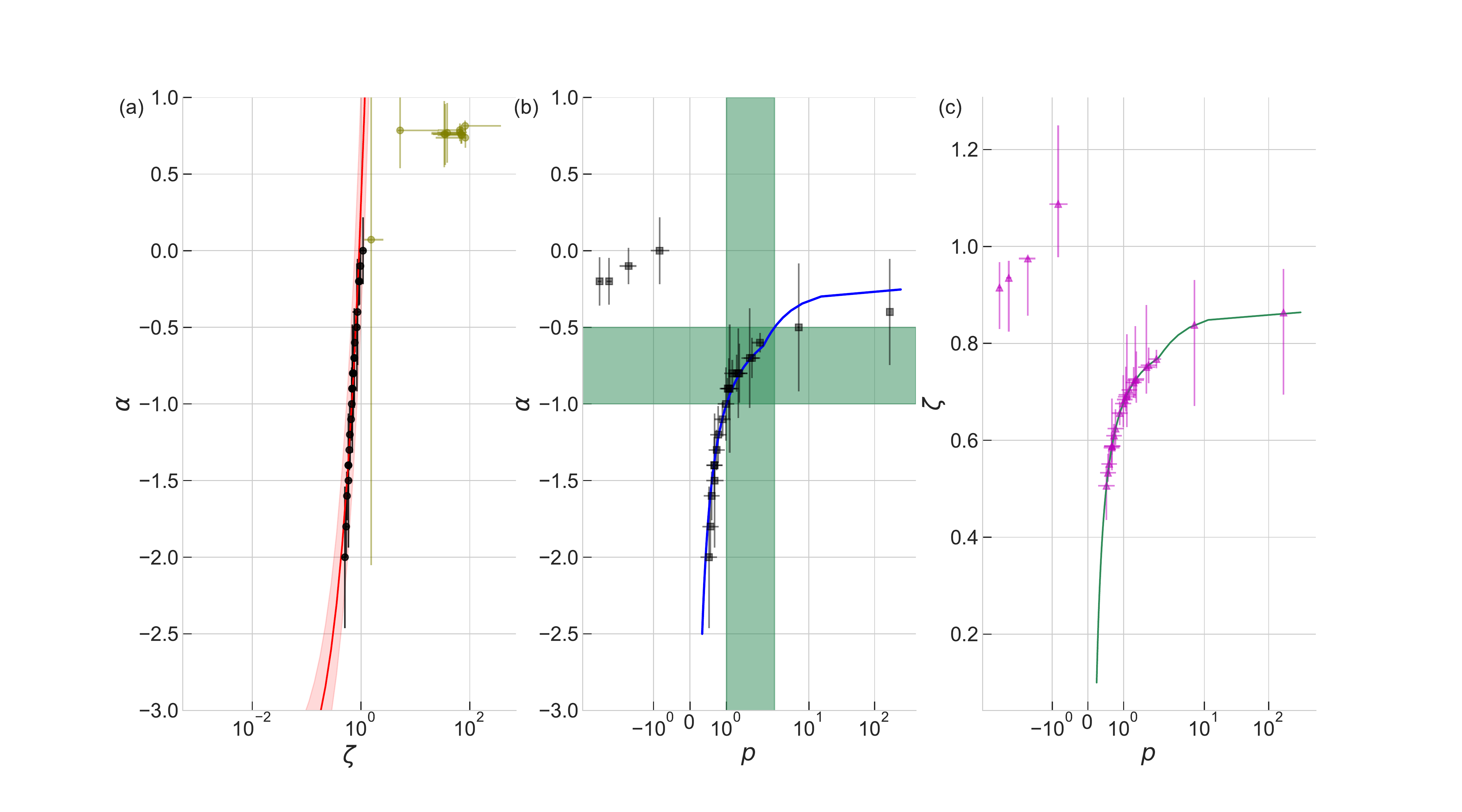}
    \caption{(a) The values of $\alpha$ versus $\zeta$ are plotted. The obtained linear fit (equation \ref{alpha_2}) and its
    $68\%$ confidence region are plotted in solid red line and shaded red respectively. The
    values of $\alpha> 0$ that are found to be inconsistent with the linear fit are plotted in olive 
    green color. (b) The estimates of $p$ obtained using the equation \ref{p_zeta} versus {\tt mBB} 
    $\alpha$ are shown in black squares. The curve according to the equation \ref{alpha_1} as 
    suggested in \cite{Lundman_etal_2013} is shown in blue solid line. The green shaded region marks 
    the physically reasonable parameter space of $\alpha$ and $p$ expected for the spectra that are 
    viewed off-axis outside the jet core. (c) The values of $\zeta$ versus $p$ are shown. The curve obtained by the equation \ref{p_zeta} is shown in solid green line.}
    \label{alpha_p_zeta}
\end{figure}

From the {\tt mBB} spectral fits, we find that $-2 <\alpha <0$ is linearly correlated to 
$\zeta$. 
Using a linear fit, we find the correlation to be
\begin{equation}
    \alpha = 4.025 \zeta - 3.748
    \label{alpha_2}
\end{equation}
and is shown in the Figure \ref{alpha_p_zeta}a. 
We note that $\alpha >0$ deviates from this linear relationship. This deviation arises because
when $\zeta$ equals large values, it means that the temperature varies steeply, thereby, producing less
softening of the spectrum below the peak. As a result, the spectrum tends
towards a blackbody (see Figure \ref{mbb}). Due to the limiting value of $\alpha =
+1$, we do not find a corresponding increase in $\alpha$ for large values of $\zeta$. 
By equating equations \ref{alpha_1} and \ref{alpha_2}, we find the relation between the fit parameter, $\zeta$ and the Lorentz factor profile index, $p$ to be 
\begin{equation}
    p = \frac{-3}{4} \frac{1}{(4.025\, \zeta -3.748 +0.25)}
    \label{p_zeta}
\end{equation}
We estimate the $p$ values by using this relationship and the $\zeta$ values obtained from the
multi-color blackbody fits. The estimates are shown in the Figure \ref{alpha_p_zeta}b as a function of
$\alpha$.
We note that $\alpha > -0.2$, gives negative values of $p$ which suggest an increasing
Lorentz factor profile structure beyond the jet core. However, even in such a jet profile structure, if viewed outside the jet core, the spectrum would be expected to have softer $\alpha$ values. On the other hand, the hard values of $\alpha$ suggest that there is less emission from outside the jet core contributing to the observed spectrum. We, thus, note that the negative $p$ values obtained in these cases are not physically feasible and is obtained as the result of the break down of the approximate relation between $\alpha$ and $p$ found in \cite{Lundman_etal_2013}.  Also, for values of $-0.5 < \alpha \le -0.2$ gives rise to large values of
$p$ which suggests a top hat jet and in which case, the jet would be visible only when it is viewed within the 
jet cone. Thus, these hard $\alpha$ values and the corresponding $p$ values clearly indicate that 
these spectra are obtained when viewed within the jet core.  
For $\alpha <-1$ gives values of $p <1$ which suggest that the decrease of  
the Lorentz factor beyond the jet core is very weak, in other words, no significant decrease in $\Gamma$ exists beyond the jet core.
One possibility is that such spectrum is obtained by viewing a narrow jet ($\theta_c <1/\Gamma_0$) on-axis such that within the viewing cone a significant part of the emission is from the region outside the jet core. \cite{Lundman_etal_2013} have shown that such scenarios can produce spectrum with very soft $\alpha$ value. However, considering $\Gamma_0 = 210$ (see section \ref{outflow}), we find that in such cases, $\theta_c \ll 1^{\circ}$ which is unlikely. 
Another possibility is that the peak count spectra that are studied in these cases could be the integration of evolving spectra in the time bin, leading to soft $\alpha$ values. It is, thus, likely that in these cases the underlying instantaneous spectrum may possess an $\alpha \ge -1$. 

The jet profile structure can be estimated only if the observed spectrum is shaped dominantly by the
emission coming from the regions outside the jet core. The physically reasonable range of $p$ values that can be derived, thus, lie between $1-3$ which corresponds to $\alpha$ values 
ranging between $-1$ to $-0.5$ as suggested in \cite{Lundman_etal_2013}. These regions are marked in shaded 
green color and the curve (blue solid line) corresponding to the equation \ref{alpha_1} is shown in 
the Figure \ref{alpha_p_zeta}b. The values of $\zeta$ versus $p$ are plotted in the Figure
\ref{alpha_p_zeta}c. 

\section{Determination of the viewing angle, $\theta_v$}
\label{view_angle}
Following the above discussion, the viewing geometry, whether the GRB is viewed within or outside the jet core, is assessed using the following criteria:\\
(i) GRBs with observed $\alpha \le -0.5$ determined from the {\tt mBB} spectral fits\footnote{We note that $\alpha \approx -1$ can also be obtained when wider jets are viewed near or outside the edge of the jet core (\cite{Lundman_etal_2013,Meng_etal_2019}, also see section \ref{within_jc}).}, 
and the non-observance of jet break in the afterglow emission are classified to possess a narrow jet 
with the line of 
sight of the observer lying outside the jet core angle, 
$\theta_c$.\\ 
(ii) Those GRBs with spectral index, $\alpha>-0.5$ or has a jet break in the afterglow
emission are classified as GRBs that are viewed within the core of the jet.

We find that $16/37$ GRBs ($\sim 43\%$ of the sample) are viewed from outside the
$\theta_c$ and the remaining $21$ GRBs in the sample are viewed within the jet core. 

\subsection{GRBs viewed outside the jet core and distribution of $p$}
With this understanding, we then estimate the viewing angle, $\theta_v$ of the GRBs 
that are viewed from outside the jet core, using the X-ray afterglow emission observed 
by $\it Swift$ X-ray Telescope (Figure \ref{xrt_afterglow} in appendix). 
When the afterglow emission produced by a structured jet is viewed off-axis, the peak 
of the afterglow light curve is observed when $1/\Gamma = \theta_v$. Let $t_0$ and $t$ 
be the timescales of afterglow emission as observed by the observer who is placed 
on-axis and off-axis from the direction of the jet respectively. The ratio of these 
timescales are given as $t_0/t \sim 1/(1+\Gamma^2 \theta_v^2)$ \citep{Granot_etal_2002_offaxis_jet}. When the peak of the 
emission is observed (i.e $1/\Gamma = \theta_v$), we find that $t_{peak} = 2 \times
t_0$ and when $1/\Gamma \gg \theta_v$, we find $t \sim t_0$ where $t_0 = R/(2\Gamma^2
c)$ where $R$ is the radius from where the afterglow emission is observed and it is
measured from the center of the GRB explosion. 

In most of these cases except GRB 170817A and GRB 150101B, the afterglow emission
consists of decreasing flux which suggests that the peak of the emission has already
gone past the start time ($t_{start}$) of the XRT afterglow observations and what we are observing is the decay phase. This means $t_{peak} < t_{start}$ and by equating 
\begin{equation}
   t_{start} = \frac{R}{2 \Gamma^2 c}, 
\end{equation}
we estimate the $\Gamma$ at the $t_{start}$ which we denote as $\Gamma_{min}$ as the Lorentz factor at the time of peak emission would have been greater than this value. 
Since the $t_{start}$ of XRT observations in these GRBs are only a few hundred seconds,
we can consider that the $R$ has not significantly evolved to larger values from the 
radius of deceleration. We, thus, assume $R \sim 10^{15-16} \rm cm$ and estimate the $\Gamma_{min}$. Since during the decay phase, $1/\Gamma > 
\theta_v$, thus, with the above estimate of the Lorentz factor, we find the probable range 
of viewing angle to be $\theta_c < \theta_v < 1/\Gamma_{min}$ and the values obtained are reported in the Table \ref{geometry_outside}. 

In GRB 170817A and GRB 150101B, the peak of the afterglow emission is observed at $150\, \rm days$ and $8.79 \, \rm days$ \footnote{$https://www.swift.ac.uk/xrt$\_$live$\_$cat/00020464$} (we consider this peak even though it is not statistically very significant in the data). 
Such long time after the onset of afterglow emission suggests that $\Gamma$ and $R$ should have significantly decreased and increased with time respectively. In case of GRB 170817A (GRB 150101B), by assuming that the initial Lorentz factor $\sim 100$ and following the standard afterglow emission properties in \cite{Huang_etal_2000_afterglow}, we expect $R$ to have increased from a deceleration radius of $10^{15} - 10^{17} \, \rm cm$  by an order of $1.5 (1.2)$ and Lorentz factor would have decreased to $\Gamma  \ge 1 (\le 10)$ by the time, when the peak of the afterglow emission has been observed by the observer. The Lorentz factor at $t_{peak}$ has to $\ge  1$, this brings in a constraint that $10^{17.6}< R <10^{18.5} \, \rm cm$ ($10^{16.4} < R < 10^{18.3}$) which gives a $1< \Gamma(150 \, \rm days) < 3$ ($1< \Gamma(8.79 \, \rm days) < 10$) which in turn constraints the viewing angle to be $20^{\circ} < \theta_v < \theta_j$ 
($6^{\circ} < \theta_v < \theta_j$) 
for GRB 170817A (GRB150101B), where the upper limit of $\theta_j = 45^{\circ}$ \citep{Nagakura_etal_2014}. 
We note that the robust way to determine the viewing angle would be by modeling the afterglow data directly using the physical models. The above mentioned method gives only a rough estimate of the probable viewing angles. We also note that there were not enough XRT data points to allow us to make the estimations of $\theta_v$, $\theta_c$ and $\Gamma_0$ for GRB 100206A, GRB 090417A and GRB 070923A. 

\begin{figure}
    \centering
    \includegraphics[scale=0.5]{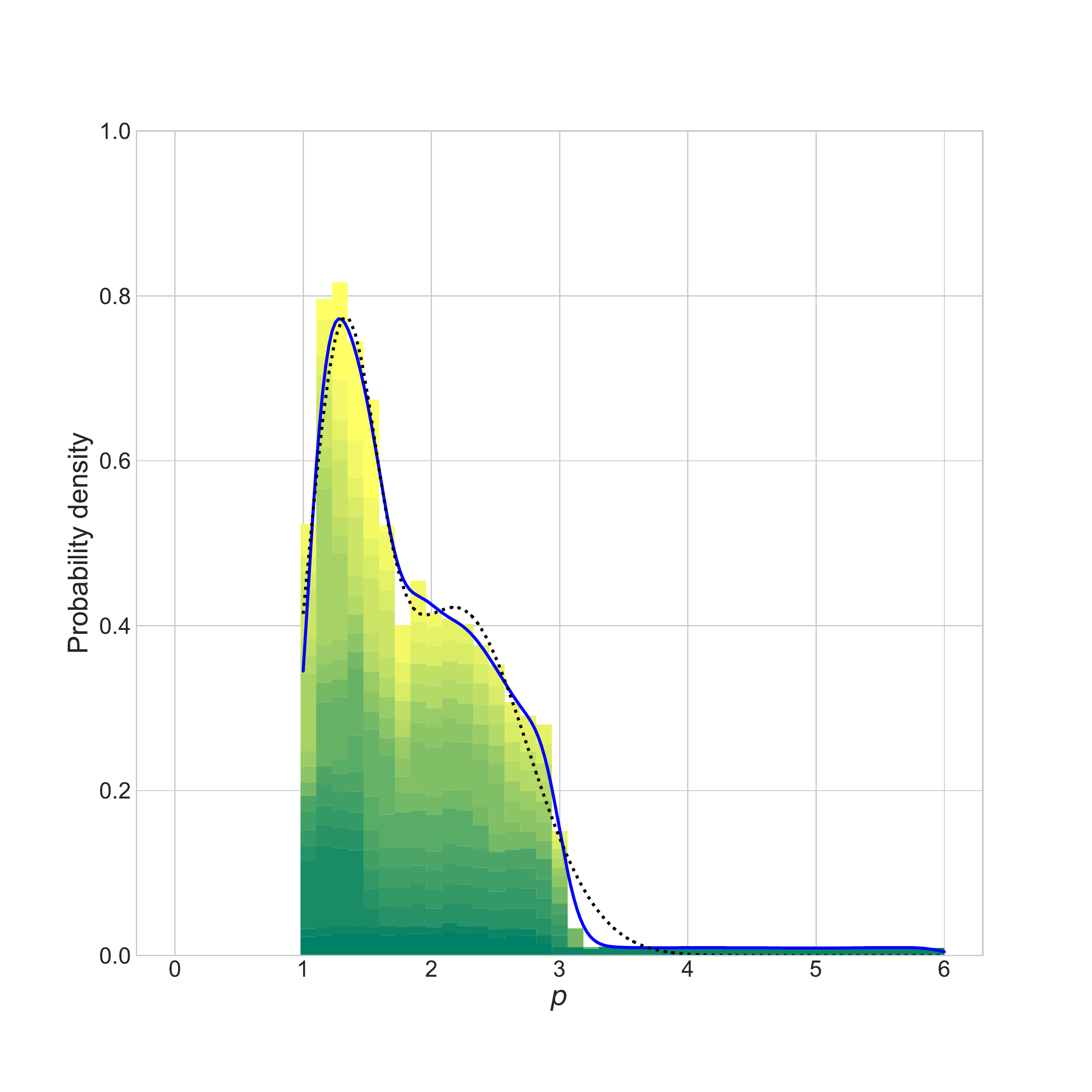}
    \caption{The cumulative distribution of the probable values of the power law index of the Lorentz factor profile,$p$ (given in equation \ref{jet_profile}), determined in this sample is plotted above. In solid blue line, we have plotted the kernel density curve corresponding to the cumulative histogram. The best fit bimodal Gaussian distribution is shown in dotted black line with peaks at $p = 1.30 \pm  0.28 $ and $2.23 \pm  0.53$.}
    \label{p_dist}
\end{figure}

For these GRBs that are viewed outside the jet core (also the cases where the jet is viewed along the edge of the jet core, see section \ref{within_jc}), the Lorentz factor jet profile ($p$) is found from the
relation given in equation \ref{p_zeta} and the possible range is determined by taking 
into account the errors of $\zeta$ and $\alpha$. The values are reported in the Table \ref{geometry_outside} and Table \ref{geometry_inside}. In order to get a generic perspective of what the jet profile index would be, we made a cumulative distribution of the possible values of $p$ by assuming a uniform distribution between the minimum and maximum values of the $p$ estimated for each GRB. The obtained cumulative distribution is shown in Figure \ref{p_dist}. We find the most probable peak at $p =  1.30 \pm  0.28$ and a secondary peak at $p = 2.23 \pm  0.53$. We further note that from Figure \ref{analyzes_results}a, the $\alpha$ measured from CPL (the empirical model) and the mBB fits are similar and consistent with each other when the low energy spectrum is mostly soft. Since, the $p$ is estimated for $\alpha < -0.5$, we note that the $\alpha-p$ estimates done here are, therefore, largely spectral model-independent.    

\subsection{GRBs viewed within the jet core}
\label{within_jc}
Depending on the burst dynamics, the optical depth of the outflow may fall to unity either below or above the saturation radius ($r_s$), thereby leading to the formation of the photosphere in either the accelerating or coasting phase respectively. 
When viewed within the jet core, the spectrum expected from the photosphere formed in the coasting phase would possess an $\alpha = +0.4$. On the other hand, the spectrum expected from the photosphere formed in the accelerating phase would possess a more harder spectrum which would be closely similar to a blackbody \citep{Beloborodov2011,Ryde_etal_2017}. 

Out of the $21$ GRBs that are viewed within the jet core, we find 10 GRBs in which {\tt mBB} fits produce extremely narrow spectra with $+0.4 < \alpha \le +1$ (Figure \ref{analyzes_results}(a)) and they are found to be equally consistent with {\tt BB}. We find that in some of these cases {\tt BB} gives a much lower AIC than {\tt mBB}, however, we find that the $\Delta \rm AIC$ values of both models are within $2$. This is mainly because {\tt BB} has lower number of free parameters in comparison to {\tt mBB}. A pure blackbody is not expected from a GRB outflow \citep{Ryde_etal_2017} and therefore, we find the {\tt mBB} fits to be more physically reasonable. Such extremely narrow and hard spectrum is expected from a photosphere formed in the accelerating phase. 

Based on the observance of jet breaks, the following GRB 160821B, GRB 140903A, GRB 090510, GRB 061201 and GRB 051221A are included as cases where the jet is viewed within the jet core even though they possess $\alpha <-0.5$. These soft $\alpha$ can be understood as the artifact of two possibilities: (i) the jet may be viewed close to the core edge such that the emission from higher latitudes with lower Lorentz factors become significant, thereby softening the low energy part of the spectrum; (ii) due to poor signal-to-noise ratio, it is likely that there is significant integration of time dependent variation of the spectrum within the analyzed peak time interval which thereby leads to softer $\alpha$.

\begin{figure}
    \centering
    \includegraphics[scale=0.5]{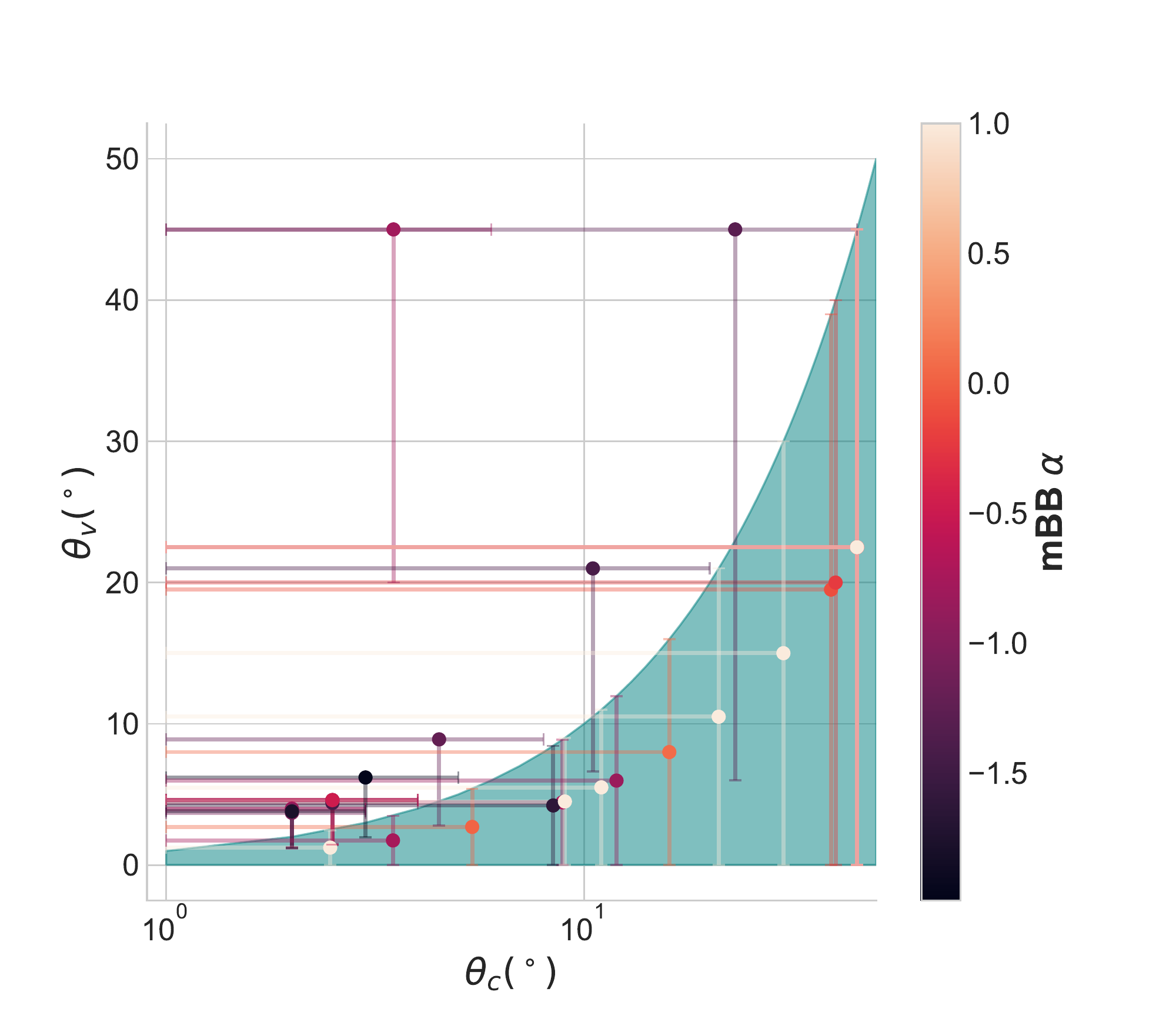}
    \caption{The range of $\theta_v$ and $\theta_c$ values estimated for the sample, color coded according to the respective $\alpha$ values is shown above. 
    The teal color shaded region marks the parameter space where the jet is viewed within the jet core. The average value of the possible range of $\theta_v$ and the maximum value of the possible range of $\theta_c$ are marked by the solid circles in the plot. The error bars mark the possible range of $\theta_v$ and $\theta_c$. There is a certain number of overlapping of points in the above plot.}
    \label{inter_1}
\end{figure}

\section{Determination of the jet opening angle, $\theta_j$ and jet core, $\theta_c$}
\label{Thetaj}
\subsection{GRBs viewed outside the jet core}
In case of GRBs that are viewed outside the jet core, the 
possible range of $\theta_c$ is considered as those values such that $\theta_c < \theta_v$. In case of GRB 170817A, we have 
considered the possible range of $\theta_c$ values to be 
between $1^{\circ} - 6^{\circ}$ based on the values that were 
earlier reported in the literature by modeling the afterglow 
emission \citep{DAvanzo_etal_2018}. The obtained values are listed in Table \ref{geometry_outside}. Figure \ref{thetaj_thetav}(a) shows the cumulative probability density distribution of the possible range of estimates of $\theta_c$ made from the GRBs that are viewed outside the jet core.  
We find that the median of the $\theta_c$ distribution is $\sim 3^{\circ}$ and the most probable value of $\theta_c$ is $2^{\circ} \pm 1^{\circ}$.  

In GRBs that are viewed outside the jet core, we cannot make a clear estimate about the jet opening angle, $\theta_j$, but the possible range would be between $\theta_c \le \theta_j < 45^{\circ}$, which is the hard upper limit of the jet opening angle \citep{Nagakura_etal_2014}.

\subsection{GRBs viewed within the jet core}
In our sample, there are $7$ GRBs (GRB 051221A, GRB 061201, GRB 090426, GRB 090510, GRB 
130603B, GRB 140903A, GRB 160821B) with jet breaks ($t_{break} = 5, 0.027,  0.016, 0.4,0.47,1, 0.7 \, \rm days$ respectively) that have been previously reported and studied as well. Among them apart from GRB 090426 and GRB 130603B, all others have $\alpha<-0.5$. 

The afterglow light curves produced by the structured jet with profiles similar to the one discussed in this work, when viewed outside the jet core are expected to exhibit an increasing flux which later on at a point in time turns over and start decreasing. This jet break is associated to the point when the Lorentz factor along the line of sight of the observer becomes equal to $1/\theta_v$.  
In contrast to this behaviour, in the above 7 GRBs in our sample, a decreasing light curve and a sharp jet break is observed later in time. This suggests that these GRBs are viewed within the jet core, $\theta_c$. In such cases, the observed jet break refers to the scenario when $1/\Gamma_0 \sim \theta_j$ \citep{Peng_etal_2005} \footnote{The jet break corresponding to $\theta_c$ is likely to be smeared out and less distinguishable in the afterglow light curve \citep{Peng_etal_2005}}. 
Thus, wherever the jet break is observed in the afterglow emission in this sample, the opening angle of the jet is estimated using the standard formula given in \cite{sari1999_jets,wang2015_break} which is based on the temporal evolution of the $\Gamma_0$ during the afterglow, 
\begin{equation}
    \theta_j = 0.057\, \left(\frac{3.5}{1+z)}\right)^{3/8} \,   \left(\frac{\kappa}{0.1}\right)^{1/8} \, \left(\frac{n}{1}\right)^{1/8} \, \left(\frac{E_{iso}}{10^{53}}\right)^{-1/8} \, t_{break}^{3/8}
\end{equation}
where $\kappa$ is the radiation efficiency, $n$ is the number density of the ambient medium
($cm^{-3}$) and $t_{break}$ is the time of jet break observed in the afterglow emission 
measured in days.  
 In this work, the jet opening angle of these GRBs are estimated using the following 
methodology: for the unknown parameters, we adopted a uniform distribution for $\kappa$ 
between values of $0.01 - 0.99$, a lognormal distribution for $n$ with the mean value $0.009$ 
(mean of the median values estimated in \cite{Fong_etal_2015}) with a standard deviation of 
$2$ and a uniform distribution for $E_{iso}$ between the minimum and maximum values found in 
Table \ref{integ_analyzes}. Using these inputs, we found a normal distribution of possible 
$\theta_j$ values for these GRBs (for an example see the Figure \ref{theta_j_theta_0}a in appendix). The mean and standard deviation obtained from these 
distributions are used to report the possible range of $\theta_j$ values of these GRBs in 
Table \ref{geometry_inside}. We find these to be nearly consistent with the previous estimates 
reported in the literature. 

In short GRBs, only a handful of cases show a distinct jet break in their afterglow emission 
and most of the GRBs show a single power-law decline. In our study, we have $21$ GRBs that are 
viewed within the jet core, out of which $14$ GRBs do not possess a jet break in their 
afterglow emission. We estimate the jet opening angle in these cases using the condition of 
jet collimation as the jet drives itself out of the surrounding, mildly relativistic expanding
ejecta that is produced as a result of the NS-NS merger. \cite{Nagakura_etal_2014} conducted a
hydrodynamic study of this propagation of the jet and found that the jet undergoes collimation
at the least in the vicinity of the central engine where the ejecta is the densest. The 
density of the ejecta ($\rho_a$) in NS-NS merger events are found to decrease steeply with radius such 
that $\rho_{a} \propto r^{\gamma}$ where $\gamma \sim 3 - 4$ \citep{130603B_Hotokezaka_etal_2013_progenitor}. \cite{Bromberg_etal_2011} had done an analytical analysis of 
the interaction between the relativistic jet and the surrounding stellar mantle in case of 
long GRBs and found out that the condition for collimation\footnote{As shown in \cite{Bromberg_etal_2011}, for values of $\gamma > 2$ the jet is likely to be uncollimated. However, at radius, $r_{esc}$, the jet is likely to be collimated at the least once and later as the density decreases, the jet may not undergo any further collimation and in this work our aim is to find the maximum limiting value of $\theta_0$.} is ${\tilde L} \le \theta_0^{-4/3}$ 
where ${\tilde L}$ is the ratio of jet energy density ($L_j/\Sigma_j c)$, where $L_j$ is the jet 
luminosity and $\Sigma_j$ is the jet cross-section, and rest mass energy density of the ejecta
at the jet head ($\rho_a c^2$), and $\theta_0$ is the 
initial jet opening angle as the jet starts. As the jet interacts with the surrounding ejecta,
it heats up the ambient matter which in turn applies pressure on the jet which leads to 
collimation. The degree of collimation decreases as the density of the ambient ejecta 
decreases. Close to the nozzle radius of the jet, the density of the ejecta is high enough to
collimate the jet at least once and thus, the resultant jet opening angle, $\theta_j$ would be
lower than $\theta_0$ but is generally found to be larger than $\theta_0/5$
\citep{Mizuta_Ioka_2013}. 

The expression for ${\tilde L}$ is expanded by including the analytical model of the ejecta profile obtained from the
numerical simulations of \cite{130603B_Hotokezaka_etal_2013_progenitor} and is given in equation 6 in
\cite{Nagakura_etal_2014}. We thus find the jet opening angle, $\theta_0$, by the following
expression
\begin{equation}
\theta_0 = \left(5.63 \times 10^{-2} \, \frac{L_{iso}}{2\times10^{50}} \, \left(\frac{M_{ej}}{0.01}\right)^{-1} \, \left(\frac{t_i}{50\times10^{-3}}\right)^{2/3} \, \left(\frac{\epsilon_r}{1}\right)^{\gamma} \, \left(\frac{\epsilon_t}{1}\right)^{3-\gamma}\right)^{-3/4} \, \rm deg
\label{theta0}
\end{equation}
where $L_{iso}$ is the isotropic luminosity, $M_{ej}$ is the ejecta mass, $t_i$ is the time
taken to launch the jet, $\epsilon_r = r/r_{esc}$ where $r_{esc}$ is the radius from 
where the jet starts and $\epsilon_t = t/t_i$. We estimate the value $\theta_0$ at $r=r_{esc}$
which can be associated to the nozzle radius of the jet ($r_0$, see section \ref{outflow}) and
therefore in the above equation $\epsilon_r=1$ and $\epsilon_t=1$, which thus, makes the above
equation independent of $\gamma$. 

The time, $t_i$ can be splitted into three main parts \citep{Zhang2019_timedelay,Beniamini_etal_2020_CE}: time taken for the black hole to form
($\sim 0 \rm s$), time taken for the accretion disk to form and start accreting ($10 \, \rm 
ms$) and finally, the time taken to launch the jet ($r_0/\beta c$, where $\beta$ is the 
velocity of the jet at $r_0$ which is relatively very low and we assume a Lorentz factor of 
$1.1$). We consider a uniform distribution for mass of the ejecta between limits of $10^{-4} -
10^{-1} \, \rm M_{\sun}$. With these inputs into the equation \ref{theta0}, we find the 
possible distribution of $\theta_0$ such that $1 < \theta_0 <45^{\circ}$ and $\theta_0/5 \ge  
1$. 
The $\theta_0$ distribution is found to be uniform and thus, 
the upper limit of $\theta_j$ is taken as $max(\theta_0)$. The lower limit 
of $\theta_j$ is equal to the estimate made using the last time of observation in case of a single power law decline of 
afterglow emission and if no significant afterglow emission 
data is available, then the estimate is made from the 
distribution of $\theta_0/5$. The distribution of $\theta_0/5$ 
is found to be roughly Gaussian and thereby the mean and 
standard deviation are estimated (for an example see Figure 
\ref{theta_j_theta_0}b in appendix). Thus, in such cases 
$\theta_{j,min}$ equals the minimum value of $\langle 
\theta_0/5 \rangle$ including its error.
The hard upper limit of $\theta_0$ is considered as 
$45^{\circ}$ as in the simulations of \cite{Nagakura_etal_2014} it was found that the jet launched with $\theta_0=45^{\circ}$ 
was not able to penetrate through the ejecta. This is because a large cross-section leads to lower ${\tilde L}$ and thereby 
becomes unsuccessful to push aside the ejecta and eventually expands quasi-spherically. 

\begin{figure}
    \centering
    \includegraphics[scale=0.5]{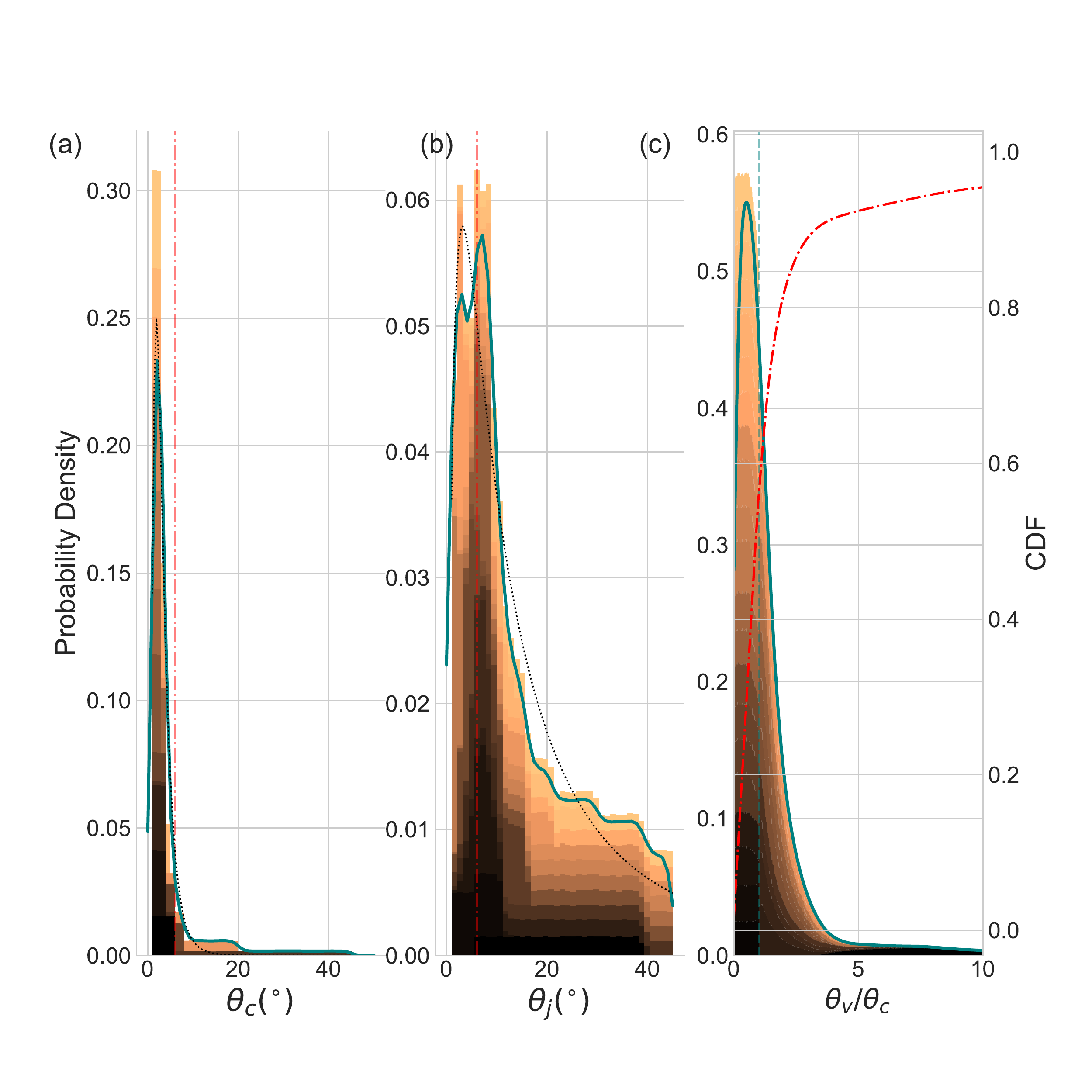}
    \caption{(a) The cumulative probability density of the possible range of $\theta_c$ determined for the sGRBs that are viewed outside the jet core in the sample is shown. The median of $\theta_c \sim 3^{\circ}$ is marked by the red vertical dash dot line. The lognormal fit to the kernel density curve (teal solid line) is shown in dotted black line. (b) The cumulative probability density of the possible range of $\theta_j$ determined for the sGRBs that are viewed within the jet core in the sample is shown. The median of $\theta_j \sim 10^{\circ}$ is marked by the red vertical dash dot line. The lognormal fit to the kernel density (teal solid line) is shown in dotted black line.  
    (c) The cumulative probability density of the ratio of the probable range of viewing angles to the jet opening angles of the sGRBs in the sample is shown. The cumulative probability distribution (right-hand y-axis) is shown in red dashed dot curve. The red dashed vertical line marks $\theta_v/\theta_c =1 $. We find the probability to observe sGRBs within the jet cone is $\sim 57\%$.  
    }
    \label{thetaj_thetav}
\end{figure}

The possible range of $\theta_j$ of GRBs, thus, observed  within the core of the jet are presented in the Table \ref{geometry_inside}. Figure \ref{thetaj_thetav}(b) shows the cumulative probability density distribution of the possible range of estimates of $\theta_j$.
We find the median of the $\theta_j$ distribution to be $\sim 10^{\circ}$ and the most probable $\theta_j$ is $4^{\circ} \pm 1^{\circ}$.


\section{ Discussion}
\label{discussion}
\subsection{Jet opening angle and viewing geometry}
The large isotropic energies estimated for the gamma ray burst requires that the outflow is emitted within a certain solid angle, thereby bringing the energetics in accord with known brightest 
explosions in the universe such as a supernova. Since, the outflow possesses very high Lorentz 
factors of the order of a hundred, the observer views only the emission within $1/\Gamma$ and
when $1/\Gamma \ll \theta_j$, the emission can be considered as isotropic. However, when the 
jet crashes into the afterglow emission and starts to decelerate, $1/\Gamma$ increases which 
eventually results in a scenario where $1/\Gamma > \theta_j$, thereby revealing the edge
of the jet emission. This results in a steeper decay of observed flux of the afterglow 
emission. With this understanding, we have been modeling the breaks observed in the decaying 
afterglow emission, where the emission transitions into a post break phase where the power law
index of the decay becomes steeper than $1.5$ and sometimes, even as steep as the accelerated electron energy power law index in case of sideways expansion of the jet \citep{sari1999_jets}. In cases, where no jet break was found, the last data point was considered to estimate the lower limit on the jet opening angle. This has been the standard procedure for estimating the opening angle of the GRB jet.  
 
The main difference of this work from the previous studies of \cite{Fong_etal_2013,Fong_etal_2015} is that
in their works they considered that all the observed sGRB jets are top-hat and thereby are viewed
within the jet core ($\theta_c \equiv \theta_j$). The possibility of a structured jet and a viewing geometry of the jet outside the jet core were not considered. In this work, while analysing the prompt emission of sGRBs we have taken into account both these above factors.  
The viewing geometry of the different sGRBs in our sample are inferred using the methodologies described in sections \ref{view_angle} and \ref{Thetaj} and the result is demonstrated in the Figure \ref{inter_1}.

The cumulative density distribution of the uniform distribution assumed between the minimum and maximum of the probable values of $\theta_c$ and $\theta_j$ estimated for the different sGRBs in the sample that are viewed outside and inside the jet core are plotted in the Figure \ref{thetaj_thetav}(a) and \ref{thetaj_thetav}(b) respectively.
We find the median jet core angle, $\theta_c$ 
is $\sim 3^{\circ}$ and the median of jet opening angle, $\theta_j$ is $\sim 10^{\circ}$. 
Thus, this study shows that the core of the jet 
of sGRBs can be very narrow 
(see also \citealt{Beniamini_etal_2019_sgrb_thetaj}).  This finding is consistent 
with the observational evidence found from the modeling of the afterglow emission of GRB 
170817A \citep{Mooley_etal_2018_170817A_jetsign} where the GRB jet core is found to 
be as narrow as $\le 3^{\circ}$. On the other hand, the median of jet opening angle, $\theta_j$ is largely consistent with the results of \cite{Fong_etal_2015}.

Taking into consideration the estimates of the possible viewing angles $\theta_v$ of the sGRBs
presented in the section \ref{view_angle}, we plot the cumulative distribution of 
$\theta_v/\theta_c$ and find that the probability to observe a sGRB within the jet core is 
$\sim 57\%$ and the probability to observe the GRB along the edge of the jet core ($1 < 
\theta_v/\theta_c \le 1.5$) is around $\sim 18\%$. 
Thus, we find that when the possibility of the jet having a structure is 
taken into consideration, the probability to observe the GRB jet along the edge is quite 
significant and we find this consistent with what has been anticipated by \cite{Lundman_etal_2013} (also see \citealt{Beniamini_etal_2019_jetstruc}). 

Based on the detection of GW 170817A by LIGO and VIRGO \citep{Abbott_etal_2017_170817A}, the rate of binary neutron star mergers are estimated to be ${\cal{R}}_{iso,NS-NS} = 1540^{+3220}_{-1220}\, \rm Gpc^{-3} yr^{-1}$. The brightest GRBs can be observed when the core of the jet is pointed towards the observer. 
The rate of sGRB events that would be viewed within the jet core is estimated by 
\begin{equation}
    {\cal{R}}_{j,NS-NS} = (1 - \rm cos\theta_c)\, {\cal{R}}_{\it iso,NS-NS}.
\end{equation}
Using the information of the $\theta_c$ from the current study, ${\cal{R}}_{j,NS-NS}$ is estimated to be $0.05 - 6.5$ and $1.8 - 26$ 
events per $\rm Gpc^{-3} \, \rm yr^{-1}$ for the most probable and median values of the $\theta_c$ distribution respectively. Considering the full LIGO sensitivity, the gravitational waves from sGRBs can be 
detected up to a luminosity distance of $200 \, \rm Mpc$ which would 
correspond to a co-moving volume of $1.1 \times 10^{-1} Gpc^{3}$. This predicts the rate of very bright sGRB events that would be detected 
with the jet core pointed towards the observer to be $0.005 - 0.72$ and $0.19 - 2.87$ events per year for the most probable and median values of the $\theta_c$ distribution respectively.


\subsection{Correlation between burst energetics and peak temperature}
\begin{figure}
    \centering
    \includegraphics[scale=0.5]{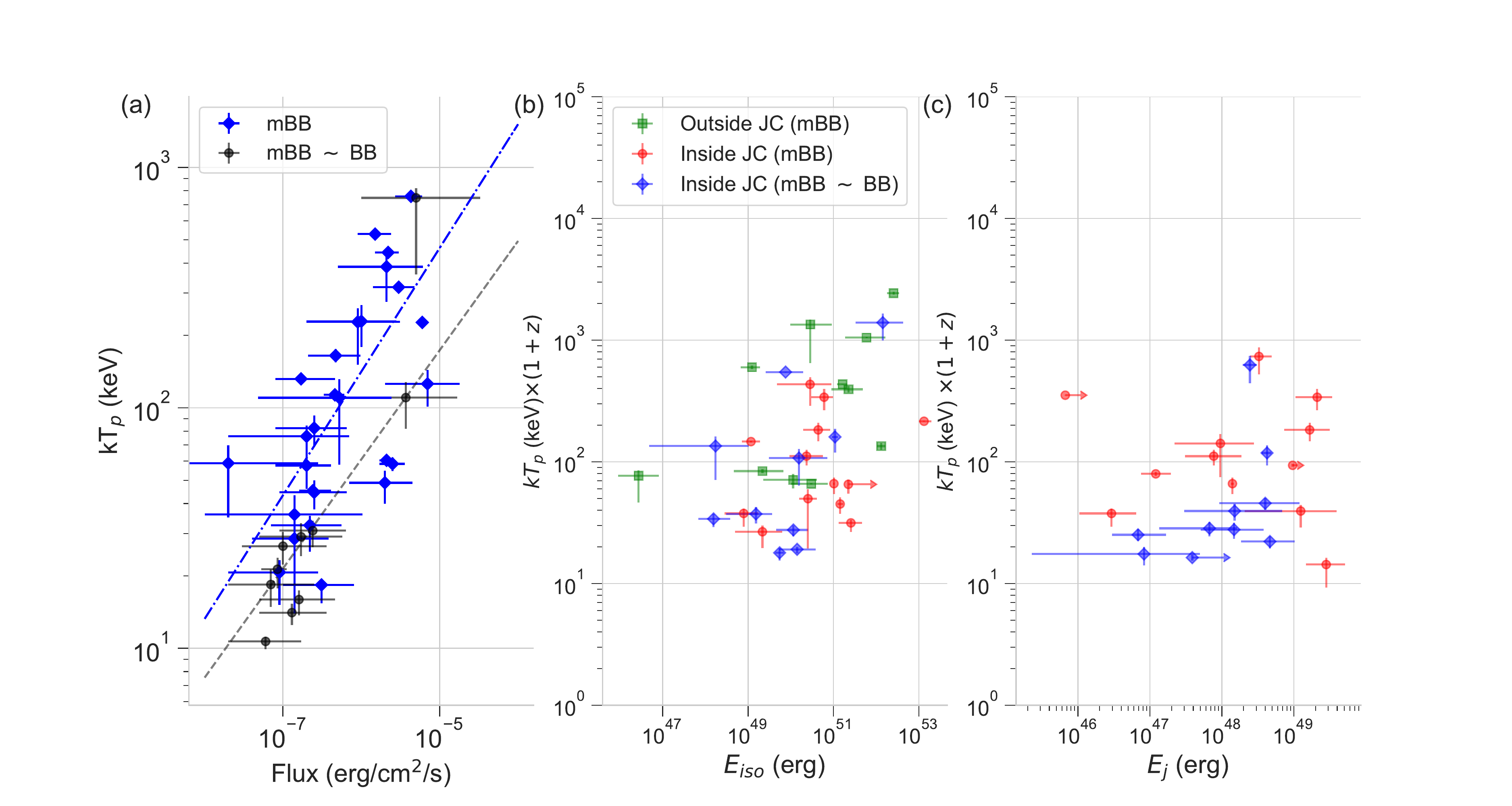}
    \caption{(a) The multi-color blackbody temperature, $kT_p$ versus energy flux is plotted. The power law correlation obtained between $kT_p$ and energy flux for the cases where {\tt mBB} $\sim$ {\tt BB} and the remaining other cases are shown by black dashed and blue dash dot curves. (b) The peak energy, $kT_p$ obtained in the peak count spectra versus the isotropic energies, $E_{iso}$, of the bursts are plotted. The bursts observed within (outside) the jet core are plotted in red circles (green squares) and for cases where {\tt mBB} $\sim$ {\tt BB} are plotted in blue diamonds. (c) $kT_p$ versus the bursts' energies corrected for the corresponding jet opening angles, $E_j$ for the cases where the jet is viewed within the jet cone are plotted.}
    \label{kT_Eiso_correlation}
\end{figure}

We first study the correlation between the obtained peak temperatures, $kT_{p}$ of the {\tt mBB} fits with the observed energy flux, $F_E$. In case of sGRBs whose {\tt mBB} fits are close to that of the blackbody, we find a power law correlation such that $kT_{p} \propto F_E^{0.45 \pm 0.24}$. Excluding these cases, the remaining sGRBs with {\tt mBB} fits exhibit a power law correlation of $kT_{p} \propto F_E^{0.51 \pm 0.11}$ (see Figure \ref{kT_Eiso_correlation}a). 
We note that both the correlations are consistent with each other. 
In a relativistically expanding outflow, the observed thermal flux ($F_{BB}$) and temperature ($T$) is expected to follow the correlation such as 
\begin{equation}
    F_{BB} \propto {\cal{R}}^2 \, T^4
\end{equation}
where ${\cal{R}} \propto r_{ph}/\Gamma$ which represents the transverse size of the emitting region (photosphere). It can vary from burst to burst and also, during a burst emission. Thus, the observed thermal flux is expected to vary from the expected emergent flux ($\sigma T^4$) from the photosphere \citep{ryde2004cooling,ryde2009quasi}.    

In this study, we find that $57\%$ of the sample is viewed within the jet core and is consistent with {\tt mBB} fit. This provides a strong assessment of the total intrinsic energy of the jet and we find that the isotropic energy of the bursts varies between $E_{iso} \sim 10^{48} - 10^{53} \, \rm erg$. Figure \ref{kT_Eiso_correlation}b shows that there is no strong correlation that can be ascertained between the peak temperature ($kT_p$) and $E_{iso}$. Using the average value of the possible $\theta_j$ estimated for each GRB (section \ref{Thetaj}), we also estimate the jet opening angle corrected burst energy, $E_j$ of the GRBs that are viewed within the jet core, which we find to vary between $10^{45} - 10^{50} \, \rm erg$. Again, we do not find any strong correlation between the $E_j$ and $kT_p$ (Figure \ref{kT_Eiso_correlation}b). 
These observations are found to be consistent with the sample study of sGRBs conducted by \cite{Husne_etal_2020}.

\begin{figure}
    \centering
    \includegraphics[scale=0.5]{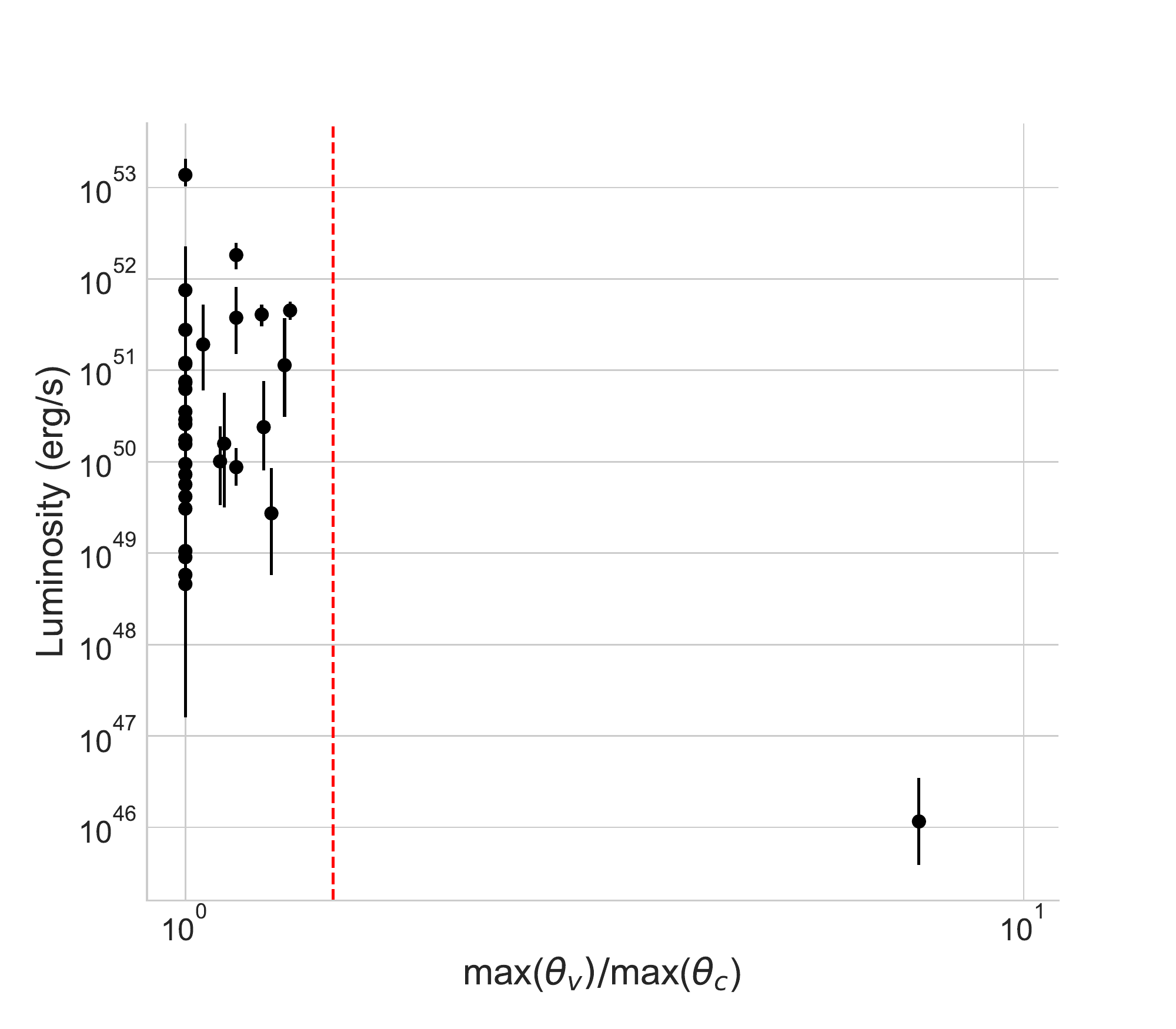}
    \caption{Observed burst luminosity versus $max(\theta_v)/max(\theta_c)$ is plotted. In the plot $max(\theta_v)/max(\theta_c) =1$ represents the cases where the bursts are viewed within the jet core and all else represents those that are viewed outside the jet core. The red dashed vertical line marks when $max(\theta_v)/max(\theta_c) =1.5$. Most of the off-axis sGRBs in the sample are observed close to the edge of the jet core.}
    \label{Luminosity_thetav_correlation}
\end{figure}

The sGRBs that are considered to be viewed outside the jet core, also do not show any particular correlation between $kT_{p}$ and $E_{iso}$. At the same time, we find that $E_{iso}$ estimates of these sGRBs vary between $10^{46} - 10^{52}$ erg. 
The brightest cases are those viewed close to the edge of the jet wherein a significant decrease in the Lorentz factor of the jet has not occurred. This is evident in the Figure \ref{Luminosity_thetav_correlation} where we find that by considering the upper limit of the possible range of $\theta_c$, most of the sGRBs viewed outside the jet core are likely to be viewed within $max(\theta_v)/max(\theta_c) =1.5$. However, we note that the actual $\theta_v/\theta_c$ is unknown as in most cases we do not have an absolute estimate of these parameters instead we only know the possible range of values.

\subsection{Jet Outflow parameters}
\label{outflow}
\begin{figure}
    \centering
    \includegraphics[scale=0.5]{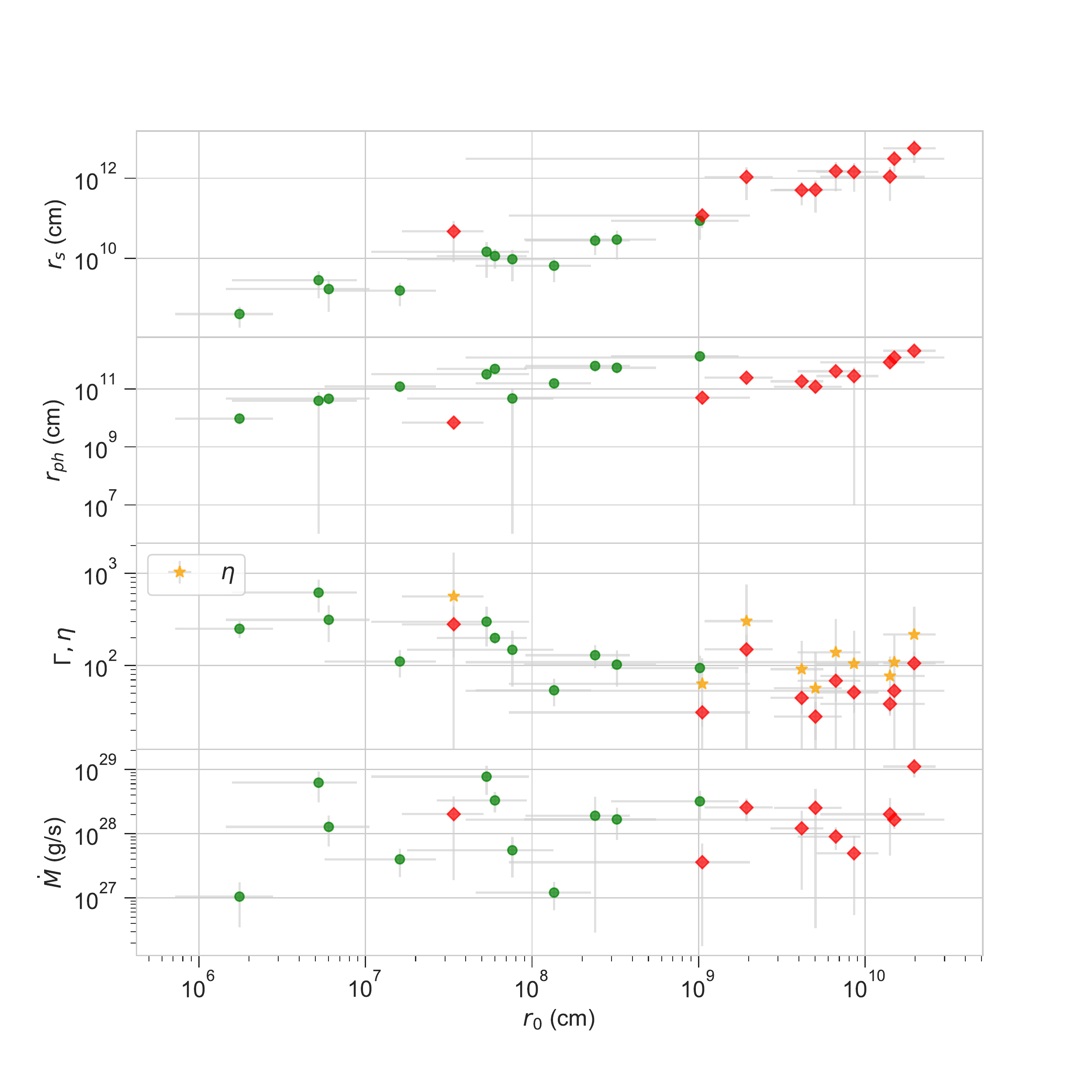}
    \caption{The outflow parameters: saturation radius,$r_s$, photospheric radius, $r_{ph}$, Lorentz factor at $r_{ph}$, $\Gamma$ and the mass ejection rate, $\dot M$ versus the nozzle radius, $r_0$ are plotted. The estimates for the scenario when the photosphere forms in the coasting phase and accelerating phase are shown in green circles and red diamonds respectively. The coasting Lorentz factor, $\eta_{acc}$ is plotted in yellow stars.}
    \label{outflow_param}
\end{figure}

Among the $21$ sGRBs that are viewed within the jet core, $10$ sGRBs possess $\alpha>+0.4$ and are nearly consistent with {\tt BB}. These cases are interpreted as the emission coming from the photosphere formed in the accelerating phase of the jet. The outflow parameters of these sGRBs are estimated using the methodology described in \cite{Begue_Iyyani2014}. The different outflow parameters such as the photospheric radius, $r_{ph}$, Lorentz factor of the outflow at the photosphere, $\Gamma(r_{ph})$, nozzle radius of the jet, $r_0$, saturation radius, $r_s$ and the maximum Lorentz factor attainable at $r_s$, $\eta = L/\dot M c^2$ are estimated and plotted in Figure\ref{outflow_param}. We find that the average values of $\langle r_{ph,acc} \rangle = 10^{11.4}\, \rm cm$, $\langle r_{s,acc} \rangle = 10^{11.9}\, \rm cm$, $\langle r_{0,acc} \rangle = 10^{9.6}\, \rm cm$, $\langle \Gamma_{acc} \rangle = 85$ and $\langle \eta_{acc} \rangle = 172$. 

The other $11$ sGRBs with softer $\alpha$ 
are considered as the emission from the photosphere formed in the coasting phase. 
In these cases of {\tt mBB} fits, we extract the blackbody component corresponding to the peak temperature, $T_p$ and estimate the respective blackbody flux which are then used in the calculations to estimate the outflow parameters of the jet using the methodology given in \cite{pe2007new,iyyani2013variable}. The obtained outflow parameters are shown in the Figure \ref{outflow_param}. We find that the average values of $\langle r_{ph,c} \rangle = 10^{11.2}\, \rm cm$, $\langle r_{s,c} \rangle = 10^{9.9}\, \rm cm$, $\langle r_{0,c} \rangle = 10^{7.7}\, \rm cm$ and $\langle \Gamma_c \rangle$\footnote{In the coasting phase, the Lorentz factor of the outflow at the photosphere, $\Gamma_c$ is equal to $\eta_c$.} $= 210$. We note that our estimates of Lorentz factors are relatively lesser than those reported in \cite{Husne_etal_2020} where they find the average value to be $775$. The major difference being that redshift is not known for most of the bursts in their sample and they had thus assumed a common redshift value of $z=1$ for the estimations of $\Gamma$. 

We note that in both the scenarios, the photosphere forms roughly at the similar radius of the order of $10^{11}\, \rm cm$ and the coasting Lorentz factor, $\eta_{acc}$ and the $\Gamma_c$ are also found to be consistent with each other. This is also reflected in the consistent mass ejection rate $\langle {\dot M} \rangle \sim 10^{28} g/s$ observed in both the scenarios (Figure \ref{outflow_param}). 
The major difference is observed in the outflow parameter of the nozzle radius, $r_0$ of the jet which in the case of accelerating phase, is found to be nearly two orders of magnitude larger than that in the case of the photosphere forming in the coasting phase. This large value of $r_0$ basically pushes the saturation radius to higher values, which eventually lets the formation of the photosphere in the accelerating phase. 
Such large values of $r_0$ have been previously estimated in case of long GRBs, where the $r_0$ has been found to increase from radius close to the vicinity of the central engine to a large radius comparable to the size of the stellar core \citep{iyyani2013variable,Iyyani_etal_2016}. The nozzle radius of the jet represents the 
radius from where the jet starts accelerating. As the jet drives through the surrounding 
ejecta, it is possible that the kinetic energy of the jet attained till then gets dissipated due to strong oblique shocks created by the interaction between the jet and the surrounding ejecta. Since the photon producing processes such as Bremsstrauhlung, double Compton etc are very efficient at these low radius and due to high optical depths, the dissipated energy is thermalized such that apparently a new fireball is formed at a larger radius which then starts accelerating more freely \citep{Thompson_etal_2007,iyyani2013variable,Iyyani_etal_2016}.  
With this understanding, we find that the sGRBs with photosphere forming in the accelerating phase tend to have strong oblique or collimation shocks as the jet interacts with the surrounding ejecta than in those sGRBs with their photospheres in the coasting phase.  


\subsection{Alternate interpretations}
\subsubsection{Synchrotron emission}
\label{sync}
The non-thermal nature of the GRB spectra has been inspirational in modeling the emission by synchrotron. The dissipation of the kinetic energy/ Poynting flux of the jet by internal shocks/ magnetic reconnections in the optically thin region in the outflow, heat electrons and in some cases, accelerate them to a power law distribution which eventually cools by radiating via synchrotron. If the magnetic fields are strong, then the electron cool on a timescale less than the dynamical time required by the electrons to cross the region of shock. This is referred as fast cooling synchrotron emission which can explain $\alpha\le -1.5$. If the magnetic field strength is weak, then the shocked electrons cool on a timescale greater than the dynamical time. This is referred as the slow cooling synchrotron emission which can explain spectra with $\alpha \le -0.67$. Synchrotron emission produced within its conventional microphysical parameters' space, cannot produce $\alpha >-0.67$ which is referred as the 'line of death' of synchrotron emission. 
However, in a recent study by \cite{burgess_etal_2019NatAs}
(also see \citealt{Oganesyan_etal_2019_sync}) have shown that synchrotron emission can produce consistent fits to even very hard spectra with $\alpha>-0.67$. Their study shows that synchrotron emission can explain GRB spectra with a wide range of $\alpha$ values. However, it has been noted that in such cases, the synchrotron emission models require extremely large emission radius which suggests very high ratio of total burst energy to the ambient matter density \citep{Ghisellini_etal_2020}.  
There have also been attempts to model the prompt emission spectra using doubly broken power law model which is interpreted as synchrotron emission \citep{Oganesyan_etal_2017,Ravasio_etal_2018,Ravasio_etal_2019}. In this interpretation, the double breaks observed in the GRB spectra are considered as the cooling frequency ($\nu_c$) and the minimum frequency ($\nu_m$) of synchrotron 
spectrum. To get these breaks consistently in the X-ray regime and a small ratio 
of $\nu_c/\nu_m$ would require a fine tuning in the microphysical parameters such that the synchrotron radiation remains in the marginally fast cooling regime \citep{Beniamini_etal_2018_marginalFC_sync}. Proton-synchrotron emission, on the other hand, seems to alleviate many of these issues \citep{Ghisellini_etal_2020}. 

The hard spectral slopes, $\alpha>0$, had been attempted to be explained by synchrotron self-absorption. However, in order to obtain the synchrotron self-absorption frequency, $\nu_{sa}$, in the X-ray regime, extreme conditions such as very high Lorentz factor ($\ge 10^4$) and magnetic fields ($\ge10^8 \, \rm Gauss$) are required to obtain the system to be optically thick to synchrotron emission even when it is optically thin to Thompson scattering and pair creation \citep{Granot_etal_2000,Lloyd_Petrosian2000}. In addition, if we consider the electrons to be cooling, then it is required that either $\nu_{sa} < \nu_{c} <\nu_m$ (fast cooling) or $\nu_{sa} < \nu_m < \nu_{c}$ (slow cooling). In this scenario, these extreme values of the Lorentz factor and magnetic fields push the synchrotron peak ($\nu_m$) to higher energies of the order of a few GeV which would severely violate the energy budget of the GRB. 
We also note that synchrotron emission spectrum from a relativistic Maxwellian electron distribution is found to have a spectral width of $SW = 25$ \citep{Axelsson_Borgonovo2015}. Most of the sGRB spectra in our sample are much narrower than this limit (also see \citealt{Burgess2019_spectralwidth}).  

\subsubsection{Subphotospheric dissipation} 
Photospheric emission is inherently anticipated in the classical fireball model \citep{meszaros2006gamma, Begue_etal_2013_phos_MC,Begue_etal_2014_phos_acc}. Even though photospheric emission can relieve many limitations of the synchrotron emission such as the narrow range of peak energy, high radiation efficiency etc, it was challenging to explain the GRB spectra which looked non-thermal in nature which were not very narrow as the blackbody emission expected from the photosphere. 
This led to the approach of considering the possibility of dissipation of kinetic/ Poynting flux happening below the photosphere. If the dissipation of the kinetic/ Poynting flux of the jet happens either continuously or at localized regions below the photosphere such that thermalization of the emission is not fully achieved, in such cases the emission from the photosphere is found to possess extremely non-thermal shapes \citep{Pe'er_Waxman2004,ryde2010identification,beloborodov2013regulation,iyyani2015extremely,Bjorn_etal_2015,Bhattacharya_etal_2018_photos_MC,Bjorn_etal_2019,Bhattacharya_Kumar2020_subphotos}. Such emission spectra are found to consistently fit the observed data with soft $\alpha$ values. Therefore, subphotospheric dissipation is one of the key alternate explanations for the sGRB spectra in our sample with soft $\alpha$.

\subsubsection{Cocoon emission}
Several numerical simulation studies \citep{Lazzati_etal_2017_cocoon} have shown that the sGRB jets emerging from NS-NS mergers are likely to be surrounded by a mildly relativistic cocoon. In the scenario where the GRB jet is viewed off-axis, the observed prompt emission can be composed of the emission coming from the cocoon. This transient X-ray emission from the cocoon is expected to be either thermal or a Comptonised spectrum. Below we discuss if the multicolor blackbody spectra presented in this work can be produced by the cocoon or not. 

\cite{Lazzati_etal_2017_cocoon} found that the emission from the cocoon is expected to have a peak photon energy $\ll 10 \, \rm keV$ which is in contrast to the high peak temperatures that are observed in our sample ($kT > 10 \, \rm keV$). Also, the isotropic energy from the cocoon is expected to be $< 10^{49}\, \rm erg$ which is again in contrast to the majority of the isotropic burst energies observed for the sGRBs in our sample.

We also note that sGRBs in which the spectra are found to be consistent with a blackbody, possess peak temperature, $kT > 10 \, \rm keV$ as well as $E_{iso} > 10^{48}\, \rm erg$ (Figure \ref{analyzes_results}b and \ref{analyzes_results}d). Thus, we find that the multi-color blackbody spectra found in the sGRB sample are less likely to be produced from a cocoon.

\section{Conclusions and summary}
\label{conclusion}
The detection of the gravitational waves along with the short GRB 170817A marked the onset of multi-messenger astronomy-astrophysics. In addition, extensive observations were carried out for the afterglow across various wavelengths such as radio, optical and X-rays. GRB 170817A was an atypical event in terms of both the prompt as well as afterglow emission because of its low luminosity and delayed onset and steadily rising afterglow emission respectively. The event affirmed certain aspects of gamma ray burst such as the presence of a jetted emission with a significant structure beyond its core and in addition that the event is viewed at an angle significantly away from the axis of the jet. This suggests that the observed prompt gamma ray emission of a GRB event depends on the above two aspects.
In addition, a large fraction of the time integrated sGRB gamma ray spectra are found to possess $\alpha>-0.67$, the line of death of synchrotron emission, along with steep indices for the high energy part of the spectra. These spectral features suggest that the observed emission is more likely from a photosphere. 

All the above factors, thus motivated us to analyze the sGRB prompt gamma ray bursts' spectra using the multi-color blackbody model which is interpreted as the non-dissipative photospheric emission from a jetted outflow with a power law decreasing Lorentz factor profile outside the jet core, accounting for different viewing angles. 
We find that $37/39$ sGRBs in our sample (sGRBs with known redshifts) are consistent with this model interpretation. From the spectral fits, we find that $16/37$ sGRBs in the sample are observed outside the jet core whereas out of the remaining sGRBs that were viewed within the jet core, $11$ cases had emission consistent from that of the photosphere produced in the coasting phase and other $10$ cases where the emission was consistent with photosphere formed in the accelerating phase. 
From the spectral fits of the cases where the burst was viewed off jet core and close to the edge of the jet core, we infer the power law index of the Lorentz factor jet profile to be most probably between $1.3 - 2.2$. 
The study of the outflow dynamics of the sGRBs viewed on-axis shows that the average Lorentz factor $\langle \Gamma \rangle = 210$ $(85)$ and the nozzle radius of the jet is deduced to be $\langle r_0 \rangle = 10^{7.7} \, (10^{9.6}) \,  \rm cm$ in case of photosphere forming in the coasting (accelerating) phase. 

The opening angle of the jet in cases where the jet break was observed is estimated
using the standard formula, however, wherever the break was not observed we estimated
the possible range of the jet opening angle using the condition of jet confinement as
the jet drives out through the surrounding ejecta. The median value of $\theta_j$ distribution is deduced to be $10^{\circ}$.  
In case of GRBs observed outside the jet 
core, the jet core angle is estimated using the condition $\theta_c < \theta_v$. With these estimates, the most probable (median) value 
of the distribution of the $\theta_c$ is inferred to be $2^{\circ} \pm 1^{\circ}$ 
($2.6^{\circ}$). Thus, our study shows that sGRBs tend to possess very narrow jet cores consistent with the observations of GRB 170817A.  
The 
predicted rate of sGRB events that can be detected by LIGO with jet core pointed towards the observer is inferred to be $0.19 - 2.87$ events 
per year for $\theta_c = 2.6^{\circ}$. 


In this study, we, thus, demonstrate that analyzing the sGRB spectra using a simple non-dissipative
photospheric emission model accounting for the jet structure and viewing geometry allows us to assess the viewing angle as well as the power law behaviour of the angle
dependent Lorentz factor profile of the jet. 

We also note that the inferred values of the different parameters of the GRB outflow are physically reasonable and viable. This justifies our consideration of non-dissipative photospheric emission model including the geometrical broadening for analyzing the prompt emission of sGRBs, as well as enhances the probability of this physical model in giving rise to the observed short GRB spectra. This conclusion entitles for further investigation of sGRB spectra using photospheric emission models including different Lorentz factor jet profile structures via numerical simulations. In addition, the viability of this physical interpretation of the spectral data and the limitations faced by the synchrotron model on the other hand, allows us to speculate that the outflow is likely to be baryon dominated which implies that the remnant of the merger of the binary neutron stars maybe a black hole rather than a magnetar. In the era of multi-messenger astronomy, this notion can be confirmed with future observations and analysis of both electromagnetic and gravitational waves detected from sGRBs.


We would like to thank Prof. Felix Ryde, Dr. Christoffer Lundman and Prof. Dipankar Bhattacharya for insightful discussions and comments on the manuscript. We would also like to thank the anonymous referee for the very insightful and valuable suggestions which have considerably improved the manuscript.
This work has made use of data supplied by the UK Swift Science Data center at the University of Leicester and {\it Fermi} data obtained through High Energy Astrophysics Science Archive Research Center Online Service, provided by the NASA/ Goddard Space Flight Center.

\begin{table*}
	\centering
	\caption{List of short GRBs detected by {\it Fermi} and {\it Niel Gehrels Swift} with known redshift until 31st December 2018.}
	\label{sample_def}
	\begin{tabular}{llllllll} 
		\hline
		No:& GRB name & RA & Dec & {\it Fermi} & BAT & Redshift & References \\
		& & & & T$_{\rm 90}$ (s) & T$_{\rm 90}$ (s) & &    \\
		\hline
		1 & 170817A &197.45 & -23.38 & 2.048 &  & 0.009  &\citet{170817A_FermiGBM,170817A_redshift} \\
		2 & 170428A & 330.05 & +26.91&  & 0.200 & 0.454 &\citet{170428A_SwiftBAT,170428A_redshift} \\
		3 & 161104A & 77.90 & -51.45 &  & 0.100 & 0.788 & \citet{161104A_refinedBAT,161104A_redshiftgcn}\\
		4 & 160821B &279.98 & +62.39  &1.088 & 0.480  & 0.160  & \citet{160821B_FermiGBM,160821B_SwiftBAT}\\
		& & & & & & & \citet{160821B_redshift}\\
		5 & 160624A &330.21 &+29.66 & 0.400 &0.192 & 0.483  & \citet{160624A_FermiGBM,160624A_SwiftBAT}\\
		& & & & & & &\citet{160624A_redshift} \\
		6 & 150423A &221.60 &+12.27 & &0.216 & 1.394  &
		\citet{150423A_SwiftBAT,150423A_redshift}\\
		7 & 150120A &10.33 &+33.98 & 3.328 &1.196 & 0.460  &\citet{150120A_FermiGBM,150120A_swiftBAT}  \\
		& & & & & & &\citet{150120A_redshift}\\
		8 & 150101B &188.04 &-10.98 &0.080 & 0.012 & 0.134  &\citet{150101B_FermiGBM,150101B_SwiftBAT, 150101B_redshift} \\
		9 & 141212A  &39.17 &+18.16 & &0.288 & 0.596  & \citet{141212A_SwiftBAT,141212A_redshift} \\
		10 & 140903A &238.02 &+27.61 & &0.296 & 0.351  & \citet{140903A_SwiftBAT,140903A_redshift} \\
		11 & 140622A &317.15 &-14.41 & &0.132 & 0.959  & \citet{140622A_SwiftBAT,140622A_redshift} \\
		12 & 131004A &296.11 &-2.95 &1.152 &1.536 & 0.717  &\citet{131004A_FermiGBM,131004A_SwiftBAT}\\ 
		& & & & & & &\citet{131004A_redshift1,131004A_redshift2,131004A_redshift3} \\
		13 & 130603B &172.22 &+17.06 & &0.176 & 0.356  & \citet{130603B_SwiftBAT,130603B_redshift1,130603B_redshift2} \\
		 & & & & & & & \citet{130603B_redshift3}\\
		14 & 120804A & 233.95 & -28.768 &  & 0.810 & 1.300  & \citet{120804A_SwiftBAT,120804A_redshift} \\
		15 & 111117A &12.71 &+23.00 &0.432 &0.464 & 2.211 &\citet{111117A_FermiGBM,111117A_SwiftBAT}  \\
		& & & & & & &\citet{111117A_redshift}\\
		16 & 101219A &74.59 &-2.53 & &0.828 & 0.718  &\citet{101219A_SwiftBAT,101219A_redshift} \\
		17 & 100724A &194.57 &-11.09 & &1.388 & 1.288  & \citet{100724A_SwiftBAT,100724A_redshift} \\
		18 & 100628A &225.94 &-31.65 & &0.036 & 0.102  & \citet{100628A_SwiftBAT,100628A_redshift} \\
		19 & 100625A &15.79 &-39.09 &0.240 &0.332 & 0.452 & \citet{100625A_FermiGBM,100625A_SwiftBAT,100625A_redshift} \\
		20 & 100206A &47.16 &+13.16 &0.176 &0.116 & 0.407 & \citet{100206A_redshift1,100206A_redshift2} \\
		21 & 100117A & 11.26 &-1.59 & 0.256 &0.292 & 0.915  & \citet{100117A_FermiGBM,100117A_SwiftBAT,100117A_redshift} \\
		22 & 090927 & 343.95 &-70.98 & 0.512 &2.160 & 1.370 & \citet{090927_FermiGBM,090927_SwiftBAT,090927_redshift} \\
		23 & 090515 &164.15 &+14.44 & &0.036 & 0.403  &\citet{090515_SwiftBAT,090515_redshift} \\
		24 & 090510 &333.55 &-26.58 & 0.960 &5.664 & 0.903  &\citet{090510_FermiGBM,090510_SwiftBAT,090510_redshift} \\
		25 & 090426 &189.07 &+32.98 & &1.236 & 2.609  & \citet{090426_SwiftBAT,090426_redshift} \\
		26 & 090417A &34.99 &-7.15 & &0.068  & 0.088  &\citet{090417A_SwiftBAT,090417A_redshift} \\
		27 & 080905A &287.67 &-18.88 &0.960 &1.016 & 0.122 & \citet{080905A_FermiGBM,080905A_SwiftBAT}\\
		& & & & & & &\citet{080905A_redshift} \\
		28 & 071227 &58.13 &-55.98 & &1.8 & 0.383  &\citet{071227_SwiftBAT,071227_redshift} \\
		29 & 070923 &184.64  &-38.28 & &0.040 & 0.076  & \citet{070923_SwiftBAT,070923_redshift} \\
		30 & 070729 &56.31 &-39.32 & &0.988 & 0.800  & \citet{070729_SwiftBAT, Fong_etal_2013} \\
		31 & 070724A &27.81 &-18.59 & &0.432 & 0.457  & \citet{070724A_SwiftBAT,070724A_redshift} \\
		32 & 070429B &328.02 &-38.83 & &0.488 & 0.904 &\citet{070429B_SwiftBAT,070429B_redshift}  \\
		33 & 061217 &160.42 &-21.148 & &0.224 & 0.827  &\citet{061217_SwiftBAT,061217_redshift} \\
		34 & 061201 &332.08 &-74.57 & &0.776 & 0.111 & \citet{061201_SwiftBAT,061201_redshift} \\
		35 & 060801 &212.98 &+16.99 & &0.504 & 1.130 & \citet{060801_SwiftBAT,060801_redshift} \\
		36 & 060502B &278.96 &+52.63 & &0.144 & 0.287  &\citet{060502B_SwiftBAT,060502B_redshift} \\
		37 & 051221A &328.71 &+16.89 & &1.392 & 0.547 &\citet{051221A_SwiftBAT,051221A_redshift} \\
		38 & 050813 &242.01 &+11.23 & &0.384 & 0.722  &\citet{050813_SwiftBAT,050813_redshift} \\
		39 & 050509B &189.12 &+29.00 & &0.024 & 0.225  &\citet{050509B_SwiftBAT,050509B_redshift} \\
		
				\hline
	\end{tabular}
\end{table*}

\begin{table*}
	\centering
	\caption{Spectral results of the time integrated analysis over the duration of the bursts in the sample.}
	\label{integ_analyzes}
	\begin{tabular}{*{8}{l}}
		\hline
		GRB name & T$_{\rm start}$ & T$_{\rm stop}$ & $\alpha$/ $\zeta$ & $\rm E_{peak}$/kT &  Flux ($10^{-8}$) & $E_{\it iso}$ ($10^{48}$)  & Chosen  \\
		& (s)& (s) & & (keV) & (erg/cm$^2$/s)& (erg) & Best fit model\\
		\hline
		170817A &-0.37 &2 &$1.45_{-0.29}^{+1.19}$ &$20_{-5}^{+1}$ &$6_{-4}^{+12}$ &$0.012_{-0.007}^{+0.023}$ &mBB \\
		170428A & -0.1 & 0.5 &$0.82_{-0.07}^{+0.07}$ &$73_{-8}^{+7}$ & $90_{-50}^{+80}$ & $733_{-407}^{+652}$ & mBB\\
		161104A &-0.3 &0.1 &$-1.29_{-0.17}^{+0.17}$ & - &$11.4_{-7.1}^{+20}$ & $353_{-220}^{+610}$ & PL \\
		160821B &-0.15 &0.6 &$0.57_{-0.02}^{+0.02}$ & $39_{-6}^{+5}$ & $14_{-9}^{+ 17}$ &$11_{-7}^{+13} $ & mBB\\
		160624A &-0.4 &0.4 &$0.83_{-0.06}^{+0.06}$ & $277_{-10}^{+10}$ & $80_{-40}^{+50}$ & $756_{-378}^{+472}$ & mBB \\
		150423A &-0.3 &0.3 &$0.83_{-0.09}^{+0.09}$ & $63_{-9}^{+8}$ &$30_{-18}^{+35}$ & $3764_{-2259}^{+4392}$ & mBB \\
		150120A &-0.9 &2.1 &$0.55_{-0.01}^{+0.01}$ & $95_{-17}^{+12}$ & $12_{-8}^{+ 17}$ & $101_{-67}^{+143}$& mBB \\
		150101B &-0.2 &0.02 &$0.62_{-0.02}^{+0.02}$ & $518_{-1}^{+1}$ & $110_{-50}^{+60}$ & $56_{-26}^{+31}$& mBB \\
		141212A  &-0.12 &0.6 &$0.63_{-0.03}^{+0.08}$ & $32_{-11}^{+3}$ & $10_{-8}^{+ 26}$ & $158_{-126}^{+409}$ & mBB \\
		140903A &-0.7 &0.8 &$0.52_{-0.01}^{+0.01}$ &$46_{-7}^{+7}$ & $35_{-21}^{+50}$ & $156_{-93}^{+222}$& mBB \\
		140622A &-0.2 &0.2 &$+1.0$ &$11_{-2}^{+2}$ &$5.8_{-3.5}^{+7.0}$ &$289_{-174}^{+348}$ & BB \\
		131004A &-0.4 &1.8 &$-0.86_{-0.17}^{+0.01}$ &$44_{-6}^{+6}$ &$48_{-23}^{+40}$ &$1165_{-558}^{+ 971}$ & Band \\
		130603B &-0.12 &0.25 &$0.75_{-0.01}^{+0.01}$ & $134_{-0.03}^{+0.03}$ & $603_{-27}^{+27}$ & $2774_{-124}^{+124}$ &mBB \\
		120804A &-0.3 &2.6 &$0.66_{-0.02}^{+0.02}$ & $42_{-1.40}^{+1.39}$ &$43_{-9}^{+11}$ & $4513_{-945}^{+1155}$ &mBB\\
		111117A &-0.3 &1.1 &$0.75_{-0.04}^{+0.04}$ &$165_{-3}^{+3}$ &$47_{-14}^{+17}$ & $18260_{-5439}^{+6604}$ & mBB \\
		101219A &-0.4 &1 &$-0.75_{-0.06}^{+0.06}$ & - & $32_{-10}^{+13}$ &$770_{-235}^{+316}$ &PL \\
		100724A &-0.4 &1.56 &$0.56_{-0.03}^{+0.03}$& $35_{-3}^{+3}$ &$11_{-8}^{+25}$ & $1139_{-828}^{+2588}$ & mBB \\
		100628A &-0.3 &0.04 &$+1.0$ &$26_{-3}^{+3}$ &$15_{-8}^{+ 11}$ &$4.3_{-2.3}^{+3.2}$ &BB \\
		100625A &-0.2 &0.7 &$0.68_{-0.01}^{+0.01}$& $256_{-0.03}^{+0.03}$ & $150_{-24}^{+23}$ &$1209_{-194}^{+185}$ & mBB \\
		100206A &-0.1 &0.2 &$0.79_{-0.02}^{+0.02}$ & $183_{-22}^{+21}$ & $260_{-130}^{+210}$ & $1638_{-819}^{+1323}$ & mBB \\
		100117A &-0.1 &0.3 &$0.73_{-0.03}^{+0.03}$ & $163_{-2}^{+2}$ &$92_{-24}^{+26}$ & $4086_{-1066}^{+1155}$ &mBB \\
		090927 &-0.3 &2.8 &$0.55_{-0.02}^{+0.02}$ & $67_{-15}^{+13}$ & $16_{-11}^{+28}$ &$1920_{-1320}^{+3360}$ & mBB \\
		090515 &-0.4 &0.1 &$+1.0$ &$22_{-4}^{+4}$ &$5_{-4}^{+7}$ & $31_{-25}^{+43}$&BB \\
		090510 &-0.565 &0.395 &$0.705_{-0.02}^{+0.04}$ &$ 1833_{-380}^{+210}$ &$3300_{-2100}^{+5000}$ &$137600_{-34410}^{+68810}$ & mBB+PL \\
		090426 &-0.4 &1.5 &$0.58_{-0.02}^{+0.04}$ & $29_{-8}^{+3}$ & $13_{-10}^{+26}$ & $7532_{-5794}^{+15060}$ & mBB \\
		090417A &-0.2 &0.07 &$+1$ &$15_{-2}^{+2}$ & $11_{-5}^{+7}$ & $2.3_{-1.1}^{+1.5}$ &BB \\
		080905A &-0.11& 1.21 &$0.76_{-0.03}^{+0.03}$ & $297_{-0.33}^{+0.33}$ & $100_{-21}^{+25}$ & $42_{-9}^{+10}$ &mBB \\
		071227 &-0.7 &2.2 &$0.66_{-0.04}^{+0.05}$ &$92_{-0.23}^{+0.21}$ &$16_{-6}^{+ 10}$ & $87_{-33}^{+55}$ &mBB \\
		070923 &-0.3 &0.1 &$0.65_{-0.04}^{+0.04}$ & $157_{-0.02}^{+0.02}$ & $36_{-16}^{+19}$ & $6_{-2}^{+3}$ &mBB \\
		070729 &-0.2 &1 &$-0.95_{-0.19}^{+0.19}$ &- & $7.53_{-4.98}^{+16.6}$ &$239_{-158}^{+527}$ &mBB \\
		070724A &-0.2 &0.6 &$1.12_{-0.25}^{+0.96}$ & $15_{-3}^{+1}$ & $3.3_{-2.6}^{+7}$ & $27_{-22}^{+58}$ &mBB \\
		070429B &-0.4&0.7&$0.68_{-0.05}^{+0.10}$ & $19_{-5}^{+2}$ & $6_{-5}^{+13}$ & $258_{-215}^{+559}$ &mBB \\
		061217 &-0.3 &0.6 &$0.71_{-0.05}^{+0.08}$ & $51_{-26}^{+7}$ & $5_{-4}^{+ 18}$ & $173_{-139}^{+623}$ &mBB \\
		061201 &-0.2 &1.1 &$-0.99_{-0.08}^{+0.09}$ & - & $26.4_{-10.3}^{+16.6}$ &$9_{-4}^{+6}$ & PL \\
		060801 &-0.2 &0.5 &$0.83_{-0.08}^{+0.07}$ & $134_{-6}^{+6}$ & $41_{-18}^{+26}$ & $3067_{-1347}^{+1945}$ & mBB \\
		060502B &-0.1 &0.2 &$0.83_{-0.09}^{+0.07}$ & $72_{-22}^{+15}$ & $26_{-20}^{+ 50}$ & $73_{-56}^{+139}$ & mBB \\
		051221A &-0.2 &2.1 &$-1.43_{-0.03}^{+0.03}$ & - & $48.7_{-6.75}^{+7.79}$ &$624_{-87}^{+99}$ & PL \\
		050813 &-0.3 &0.5 &$+1$ & $15_{-3}^{+2}$ & $3.8_{-2.5}^{+6}$ & $95_{-62}^{+150}$  & BB \\
		050509B &-0.2 &0.1 &$0.59_{-0.04}^{+0.08}$ & $66_{-45}^{+9}$ & $3.6_{-3.5}^{+18}$ & $5.8_{-5.6}^{+29}$ & mBB \\
		
				\hline	
				\end{tabular}
      \small
      \\
      Note: For {\tt mBB}, {\tt BB} and {\tt Band} fits, the energy flux is reported for the energy range $0.001$ keV - $100$ MeV, however for {\tt PL} fits, the energy flux is reported for the energy range $15 - 150 \, \rm keV$ for {\it Niel Gehrels Swift} BAT detected GRBs.

\end{table*}

\begin{table*}
	\centering
	\caption{Spectral results of the analyses of the peak spectra of the bursts in the sample.}
	\label{peak_analysis}
	\begin{tabular}{*{8}{l}}
		\hline
		GRB name & T$_{\rm start}$ & T$_{\rm stop}$ & $\zeta$ & $\alpha$ & $\rm E_{peak}/kT$ &  Flux ($10^{-8}$) & Chosen  \\
		& (s)& (s) & & &(keV) & (erg/cm$^2$/s) & Best fit model\\
		\hline
		170817A &-0.32 &0.256 &$0.72_{-0.04}^{+0.11}$ & $-0.79 \pm 0.29$ &$76_{-30}^{+8}$ & $20_{-18}^{+50}$ & mBB \\
		170428A &0.23 &0.26 &$0.98_{-0.12}^{+0.003}$ & $-0.1 \pm 0.12$ & $126_{-25}^{+18}$ & $700_{-500}^{+1100}$ & mBB \\
		161104A$^{*}$ &-0.13 &0.03 &($83_{-59}^{+4}$) &$+1.0$ ($+0.74 \pm 0.07$) & $9_{-0.5}^{+0.5}$ ($11_{-0.76}^{+0.48}$) &$5.9_{-2.2}^{+2.5}$ ($6_{-4}^{+11}$) & BB (mBB) \\
		160821B &-0.08 &0.04 &$0.75_{-0.06}^{+0.13}$ & $-0.7 \pm 0.33$ &$33_{-7}^{+3}$ & $22_{-15}^{+34}$ &mBB\\
		160624A &-0.1489 &0.1294 &$0.92_{-0.09}^{+0.05}$ & $-0.22 \pm 0.16$ & $229_{-50}^{+39}$ & $100_{-80}^{+180}$ &mBB \\
		150423A &-0.0713 &0.1644 &$0.68_{-0.05}^{+0.06}$ & $-0.98 \pm 0.24$ & $165_{-10}^{+11}$ & $47_{-26}^{+50}$ &mBB \\
		150120A &0.22 &0.392 &$0.59_{-0.03}^{+0.03}$ & $-1.42 \pm 0.19$ & $45_{-0.64}^{+0.65}$ & $24_{-8}^{+17}$ & mBB \\
		150101B &-0.0914 &0.0186 &$0.61_{-0.02}^{+0.02}$ & $-1.29 \pm 0.13$ & $528_{-17}^{+18}$ & $150_{-60}^{+90}$ &mBB\\
		141212A  &-0.0336 &0.3264 &$0.62_{-0.03}^{+0.04}$ & $-1.22 \pm 0.19$ & $45_{-7}^{+5}$ &$25_{-16}^{+40}$ &mBB\\
		140903A &-0.326 &0.0544 &$0.55_{-0.02}^{+0.03}$ & $-1.63 \pm 0.16$ & $82_{-13}^{+11}$ & $25_{-17}^{+40}$ & mBB \\
		140622A$^{*}$ &-0.0520 &0.1480& ($70_{-45}^{+15}$) &$+1.0$ ($+0.75 \pm 0.05$) &$11_{-1}^{+1}$ ($14_{-2}^{+1}$) &$14_{-70}^{+11}$ ($13_{-8}^{+23}$) &BB (mBB) \\
		131004A &0.041 &0.0974 &$1.54_{-0.39}^{+1.06}$ & $+0.07 \pm 2.12$ & $18_{-3}^{+1}$ & $31_{-21}^{+50}$ & mBB\\
		130603B &-0.05 &0.023 &$1.09_{-0.11}^{+0.16}$ & $+0.03 \pm 0.22$ & $49_{-9}^{+6}$ & $200_{-130}^{+250}$ & mBB\\
		120804A &0.3 &0.4 &$0.76_{-0.04}^{+0.04}$ & $-0.66 \pm 0.13$ & $59_{-4}^{+4}$ & $250_{-80}^{+111}$ & mBB \\
		111117A &0.35 &0.47 &$0.72_{-0.03}^{+0.03}$ & $-0.78 \pm 0.12$ & $758_{-18}^{+18}$ & $430_{-160}^{+170}$ &mBB \\
		101219A &-0.18 &0.06 &-&$-0.48_{-0.15}^{+0.14}$ & - & $38_{-22}^{+54}$ & PL \\
		100724A &0.0031 &0.486 &$0.69_{-0.07}^{+0.13}$ & $-0.95 \pm 0.42$ & $29_{-5}^{+3}$ & $14_{10}^{+24}$ &mBB\\
		100628A$^{*}$ &-0.14 &0.04 & $(66_{-40}^{+19})$ &$+1.0$ ($+0.79 \pm 0.03$) &$23_{-3}^{+3}$ ($31_{-5}^{+3}$) & $20_{-9}^{+15}$ ($24_{-15}^{+40}$) & BB (mBB) \\
		100625A$^{*}$ &0.17 &0.2 &($82_{-8}^{+291}$) &$+1.0$ ($+0.81 \pm 0.04$) &$82_{-17}^{+13}$ ($110_{-28}^{+18}$) & $370_{-250}^{+500}$ ($370_{-300}^{+1300}$) & BB (mBB)  \\
		100206A &-0.03 &0.05 &$0.77_{-0.02}^{+0.02}$ & $-0.61 \pm 0.06$ & $227_{-0.05}^{+0.05}$ & $600_{-90}^{+80}$  & mBB \\
		100117A &-0.02 &0.11 &$0.66_{-0.03}^{+0.02}$ & $-1.06 \pm 0.11$ & $443_{-4}^{+4}$ & $220_{-70}^{+80}$ &mBB \\
		090927 &0.03 &0.46 &$0.59_{-0.02}^{+0.02}$ &$-1.41 \pm 0.11$ & $113_{-0.96}^{+0.92}$ & $46_{-13}^{+15}$ & mBB \\
		090515$^*$ &-0.19 &0.04 & ($34_{-14}^{+48}$) &$+1.0$ ($+0.76 \pm 0.22$) & $20_{-3}^{+3}$ ($27_{-4}^{+4}$) & $10_{-5}^{+9}$ ($10_{-7}^{+26}$) &BB (mBB) \\
		090510 &-0.27 &0.028&$0.68_{-0.02}^{+0.02}$&$-0.87_{-0.14}^{+0.13}$ & $386_{-110}^{+73}$ & $210_{-160}^{+400}$ &mBB \\
		
		090426$^{*}$ &0.43 &0.53 & ($70_{-42}^{+16}$) &$+1.0$ ($+0.76 \pm 0.06$) &$13_{-2}^{+1}$ ($16_{-2}^{+1}$) & $16_{-9}^{+13}$ ($16_{-11}^{+30}$) &BB (mBB) \\
		090417A &-0.1 &0.04 &$0.59_{-0.05}^{+0.09}$ & $-1.46 \pm 0.44$ & $36_{-22}^{+7}$ & $14_{-13}^{+90}$ & mBB \\
		080905A$^{*}$ &0.9565  &0.9781 &($5_{-0.51}^{+22}$)&$+1.0$ ($+0.79 \pm 0.25$) &$557_{-164}^{+104}$ ($747_{-388}^{+73}$) &$1300_{-1000}^{+2500}$ ($500_{-400}^{+2800}$) & BB (mBB) \\
		071227 &0.226 &0.465 &$0.84_{-0.17}^{+0.09}$ & $-0.48 \pm 0.42$ & $58_{-2}^{+2}$ & $20_{-12}^{+21}$ &mBB \\
		070923 &-0.14 &0.04 &$0.69_{-0.03}^{+0.06}$ & $-0.92 \pm 0.19$ & $228_{-76}^{+32}$ & $90_{-70}^{+220}$ &mBB \\
		070729 &0.08 &0.2 &$0.51_{-0.07}^{+0.04}$ & $-1.99 \pm 0.46$ & $59_{-24}^{+11}$ & $2_{-15}^{+26}$  & mBB \\
		070724A &0.0064 &0.2736 &$0.53_{-0.02}^{+0.04}$ & $-1.77 \pm  0.20$ & $132_{-2}^{+2}$ & $17_{-9}^{+29}$ & mBB \\
		070429B &0.02 &0.21 & $0.86_{-0.17}^{+0.09}$ &$-0.41 \pm 0.35$ &$21_{-6}^{+3}$ &$9_{-7}^{+19}$ &mBB \\
		061217$^{*}$ &-0.081 &0.155 & ($39_{-19}^{+45}$) &$+1.0$ ($+0.77 \pm 0.19$) & $22_{-4}^{+3}$ ($29_{-5}^{+4}$) & $14_{-9}^{+16}$ ($17_{-12}^{+40}$) & BB (mBB) \\
		061201 &0.38 &0.55 &$0.73_{-0.05}^{+0.06}$ & $-0.76 \pm 0.19$ & $317_{-0.93}^{+0.97}$ & $300_{-160}^{+180}$ & mBB \\
		060801 &-0.1 &0.1 & - &$-0.96 \pm 0.4$ & - & $7.8_{-7.1}^{+92}$ & PL \\
		060502B &-0.06 &0.04 &$0.94_{-0.11}^{+0.04}$ & $-0.17 \pm 0.15$ & $110_{-52}^{+22}$ &$52_{-48}^{+190}$& mBB \\
		051221A &-0.07 &0.1 &$0.70_{-0.02}^{+0.02}$ & $-0.84 \pm 0.09$ & $61_{-2}^{+2}$ & $210_{-40}^{+50}$ & mBB \\
		050813$^{*}$ &-0.07 &0.27 & ($66_{-41}^{+23}$) &$+1$ ($+0.77 \pm 0.06$) &$16_{-3}^{+2}$ ($21_{-4}^{+3}$) & $8_{-5}^{+9}$ ($8.5_{-3.2}^{+2.7}$) & BB (mBB) \\
		050509B$^{*}$ &-0.08 &0.02 & ($35_{-14}^{+49}$) &$+1$ ($+0.76 \pm 0.19$) & $14_{-3}^{+2}$ ($18_{-4}^{+3}$) & $6_{-4}^{+8}$ ($7_{-5}^{+18}$) & BB (mBB) \\
\hline
\end{tabular}\\
$^*$ The GRBs whose spectra are consistent with {\tt BB}.
\end{table*}

    \begin{table*}
	\centering
	\caption{The possible range of values estimated for $\theta_c$ ,$\theta_v$, $p$ and $\Gamma_0$ for the GRBs that are viewed outside the jet core ($\theta_c$). In these cases, the jet opening angle, $\theta_c \le \theta_j < 45^{\circ}$.}
	\label{geometry_outside}
	\begin{tabular}{|c|c|c|c|c|c|}
		\hline
		 \multirow{1}{*}{GRB name} &
		 \multicolumn{1}{c|}{$\theta_c (^{\circ})$ }&\multicolumn{1}{c|}{$\theta_v (^{\circ})$}&
		 \multicolumn{1}{c|}{$\alpha$} &
		 \multicolumn{1}{c|}{$p$} & \multicolumn{1}{c|}{$\Gamma_0$}\\
		\hline
		170817A &  $1 -6$ & $20 - 45$ &$-0.79 \pm 0.29$ &$1-2.9$& $-$ \\
		150423A &$1 - 4$ &$\theta_c - (1.5 - 4.6)$& $-0.98 \pm 0.24$ &$1-1.5$ &$>12-39$ \\
		150120A &$1 - 4$& $\theta_c - (1.4 - 4.4)$& $-1.42 \pm 0.19$ &$1-3$ &$>13-41$ \\
		150101B &$1 - 45$ & $\theta_c - (6 - 45)$ & $-1.29 \pm 0.13$ &$1-3$ & $-$ \\
		141212A &$1 - 8$ &$\theta_c - (2.8 - 8.9)$ &$-1.22 \pm 0.19$ &$1-3$ &$>7-21$\\
		120804A &$1 - 3$ &$\theta_c - (1.28 - 4)$& $-0.66 \pm 0.13$ &$1.3 - 2.4$ &$> 14 - 45 $\\ 
		111117A &$1 - 4$ &$\theta_c - (1.46 - 4.6)$ &$-0.78 \pm 0.12$ &$1.1 - 1.7$ &$> 12 - 39$\\ 
		100724A & $1 - 3$ &$\theta_c - (1.2 - 3.94)$& $-0.95 \pm 0.42$ & $1 - 3.2$ &$> 15 - 46$\\ 
		100206A & $ - $& $ - $ &$-0.61 \pm 0.06$ &$1.8 - 2.6$ & $ - $\\ 
		100117A &$1 - 3$&$\theta_c - (1.18 - 3.7)$ &$-1.06 \pm 0.11$ &$1-3$ &$>15-48$\\ 
		090927 &$1 - 20$ &$\theta_c - (6.62 -21)$& $-1.41 \pm 0.11$ &$1-3$ &$> 3-9$\\
		090417A & $-$ & $ - $& $-1.46 \pm 0.44$ &$1-3$ & $ - $\\
		071227 &$1 - 4$ &$\theta_c - (1.44 - 4.6)$& $-0.48 \pm 0.42$ &$1 - 3$ &$>13-40$\\ 
		070923 & $-$ & $ - $ & $-0.92 \pm 0.19$ &$1 - 1.7$ & $ - $\\  
		070729 &$1 - 5$ &$\theta_c - (1.97 - 6.2)$& $-1.99 \pm 0.46$ &$1-3$ & $>9-29$ \\ 
		070724A &$1 - 3$ & $\theta_c - (1.21 - 3.8)$& $-1.77 \pm 0.20$ &$1-3$ &$>15-47$\\ 
		
		\hline
	\end{tabular}

    \end{table*}
    
    \begin{table*}
	\centering
	\caption{The possible range of values estimated for $\theta_j$, $p$ and $\Gamma_0$ for GRBs that are viewed within the jet core ($\theta_c$). In these cases, the jet core angle, $\theta_c \le \theta_j$ and the viewing angle, $ 0 \le \theta_v \le \theta_c$.} 
	\label{geometry_inside}
	\begin{tabular}{|c|c|c|c|c|}
		\hline
		 \multirow{1}{*}{GRB name} &
		  \multicolumn{1}{c|}{$\theta_j (^{\circ})$ }& \multicolumn{1}{c|}{$\alpha$ }&
		 \multicolumn{1}{c|}{$p$} & \multicolumn{1}{c|}{$\Gamma_0$} \\
		\hline
		170428A & $5.5 - 39$ &$-0.1 \pm 0.12$ & - & $162-436$\\
		161104A &$1 - 11$ &$+0.74\pm 0.07$& - & $30-47$ \\
		160821B & $5.05 - 8.88$ $^{a}$ &$-0.7 \pm 0.33$ & $1 -6$ &$36-72$\\
 		160624A &$2.8 - 40$ &$-0.22 \pm 0.16$& - & $179-447$\\
		140903A & $4.72 - 8.44$ $^{b}$ &$-1.63 \pm 0.16$ & $1 - 3$ &$75 - 147$\\
		140622A &$1 - 45$ & $+0.75 \pm 0.05$ & - & $37-70$ \\
		131004A &$8 - 16$ & $+0.07 \pm 2.12$ & - &$60 - 128$ \\ 130603B & $3.07 - 5.40 $ $^{c}$ & $+0.03 \pm 0.22$ & - &$93 - 166$\\ 
		100628A &$3 - 45$ & $+0.79 \pm 0.03$ & - & $22-40$\\ 
		100625A &$5.3 - 9$ & $+0.81 \pm 0.04$ & - & $83-216$\\ 
		090510 & $0.41 - 0.73^{d}$ & $-0.87^{+0.13}_{-0.14}$ & $1.04 - 1.29$ & $379 - 854$\\
		090515 &$3 - 45$ & $+0.76 \pm 0.22$ & - & $30-60$\\ 
		090426 &$1.36 -  2.47$ $^e$ &$+0.76 \pm 0.06$ & - &$71-141$\\ 
		080905A &$3.4 - 45$ & $+0.79 \pm 0.25$ & - &$123-438$\\ 
		070429B &$3 - 45$ & $-0.41 \pm 0.35$ & - &$50 - 100$ \\ 
		061217 & $1.5 - 21$ & $+0.77 \pm 0.19$ & - &$41-96$ \\ 
		061201 &$1.97 - 3.49$ $^f$ & $-0.76 \pm 0.19$ & $1 - 2$ & $199 - 301$\\ 
		060502B & $3 - 45$ & $-0.17 \pm 0.15$ & - &$59 - 237$\\ 
		051221A &$ 6.80 - 11.96$ $^{g}	$ & $-0.84 \pm 0.09$ & $1.2 - 1.6$ &$171 - 227$\\  
		050813 & $2 - 30$ & $+0.77 \pm 0.06$ & - &$31-72$\\ 
		050509B &$5 - 45$ & $+0.76 \pm 0.19$ & - &$59 - 145$\\ 
		
		\hline
	\end{tabular}
	\\
        $^a$ \cite{Jin_etal_2018} has reported a $\theta_j = 5.72^{\circ}$.\\
        $^b$ \cite{140903A_Troja_etal_2016} has reported a $\theta_j = 5.2^{\circ} \pm 0.69^{\circ}$.\\
        $^c$ \cite{130603B_progenitor_Fan_etal_2013} and \cite{Fong_etal_2014} have reported $\theta_j = 4.87^{\circ}$ and $4^{\circ} - 8^{\circ}$.\\
        $^d$ \cite{Corsi_etal_2010} and \cite{Fraija_etal_2016} have estimated a $\theta_j = 0.1^{\circ} - 0.7^{\circ}$. \\
        $^e$ \cite{Guelbenzu_etal_2011} has reported a $\theta_j = 6.5^{\circ} \pm 0.4^{\circ}$ (n is assumed to be $10 \, \rm cm^{-3}$).\\
        $^f$ \cite{Stratta_etal_2007} has reported a $\theta_j = 1.2^{\circ} - 1.9^{\circ}$. \\
        $^g$ \cite{Soderberg_etal_2006} has reported a $\theta_j = 5.7^{\circ} - 7.3^{\circ}$.

    \end{table*}

\newpage
\bibliographystyle{aasjournal}
\bibliography{sGRB_project1_ref}

\appendix
\section{Spectral Models}
\label{spec_models}
We have analyzed the spectra with 5 models. These include
\\ 
(a) Power law ({\tt PL})\\
\begin{equation}
    N(E) = K \left(\frac{E}{1 \, \rm keV}\right)^{\alpha} 
\end{equation}
where $K$ is the amplitude, $\alpha$ is the power law spectral index. 
\\
(b) Power law with exponential cutoff ({\tt CPL})\\
\begin{equation}
    N(E) = K \left(\frac{E}{1 \, \rm keV}\right)^{\alpha} e^{-E/Ec}
\end{equation}
where $K$ is the amplitude, $\alpha$ is the power law spectral index and $E_c$ is the cutoff break energy such that the $E_{peak}$ in $\nu F_{\nu}$ spectrum is given by $(2+\alpha) E_c$.
 \\
(c) Band function ({\tt Band})\\
\begin{equation}
    N(E) =  K \left(\frac{E}{100 \, \rm keV}\right)^{\alpha}  e^{\left(\frac{-(\alpha+2)E}{E_{peak}}\right)} \, \, {\rm if} \, E < (\alpha - \beta) \frac{E_{peak}}{(\alpha+2)} 
    \end{equation}
\begin{equation}
    = K \left(\frac{E}{100 \, \rm keV}\right)^{\beta}  e^{(\beta - \alpha)} \, \left(\frac{(\beta-\alpha) Epeak}{100\, \rm keV (\alpha+2)}\right)^{(\alpha -\beta)} \, \, {\rm if} \, E > (\alpha - \beta) \frac{E_{peak}}{(\alpha+2)} \nonumber
\end{equation}
where $K$ is the amplitude, $\alpha, \, \beta$ are the low and high energy power law spectral indices respectively, and $E_{peak}$ is the peak energy in the $\nu F_{\nu}$ spectrum. 
\\
(d) Blackbody ({\tt BB})
\begin{equation}
    N(E) =  K \frac{E^2}{e^{\left(\frac{E}{kT}\right)} -1}
\end{equation}
where $K$ is the amplitude, $kT$ is temperature in $keV$. 
\\
(e) Multicolor Blackbody ({\tt mBB})
\begin{equation}
    N(E) = \frac{4 \pi E^2}{h^2 c^2} \left(\frac{K}{\zeta}\right) T_{p}^{(2/\zeta)} \int_{T_{min}}^{T_{p}} \frac{T^{\frac{-(2+\zeta)}{\zeta}}}{e^{(E/T)} - 1} dT
\end{equation}

where $K$ is the amplitude, $\zeta$ is power law index of the radial dependence of temperature ($ T(r) \propto r^{-\zeta}$), $T_{p}$ is the peak temperature in $keV$ and $T_{min}$ is the minimum temperature of the underlying blackbodies and is considered to be well below the energy range of the observed data. 

\begin{figure}
    \centering
    \includegraphics[width=.3\linewidth]{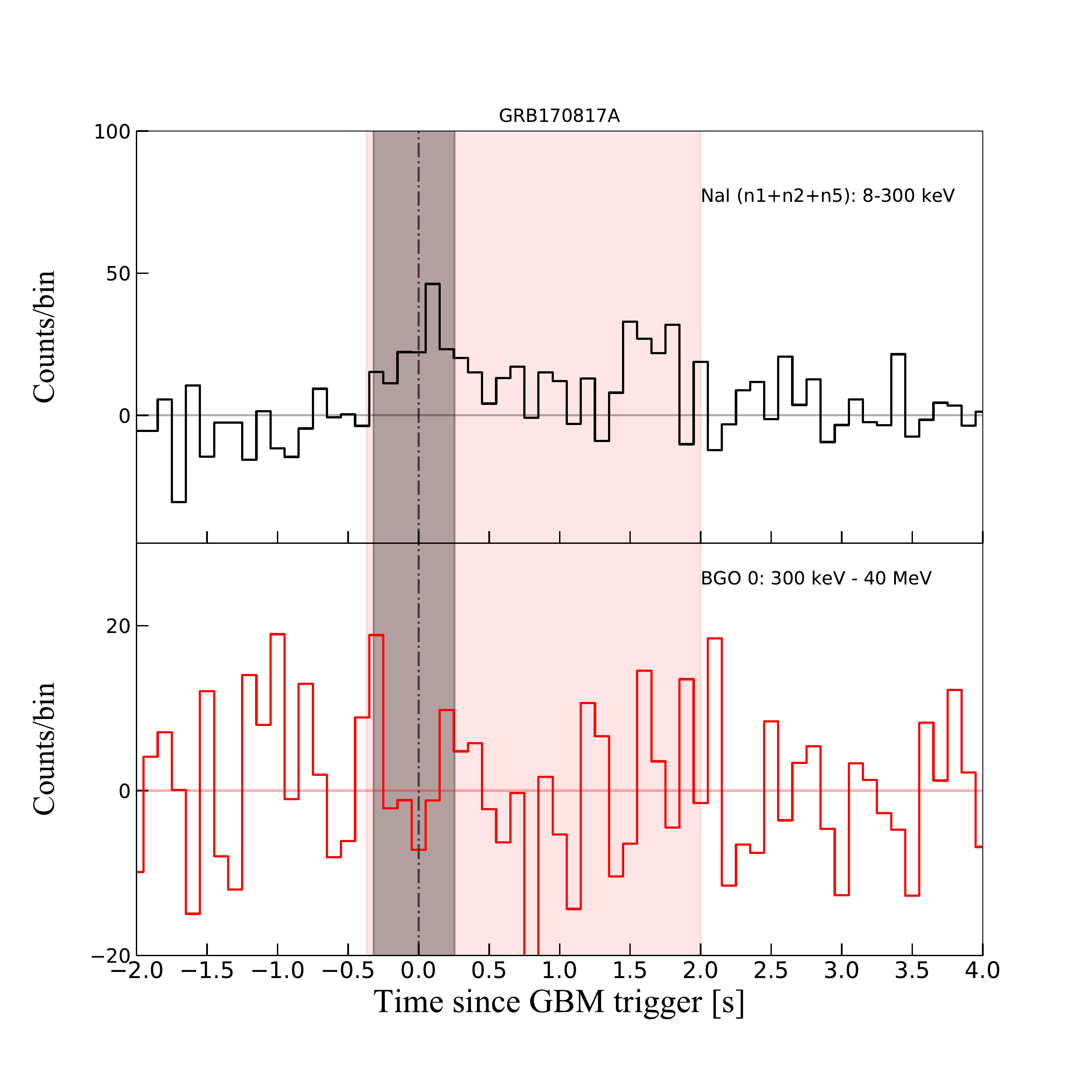}
    \includegraphics[width=.3\linewidth]{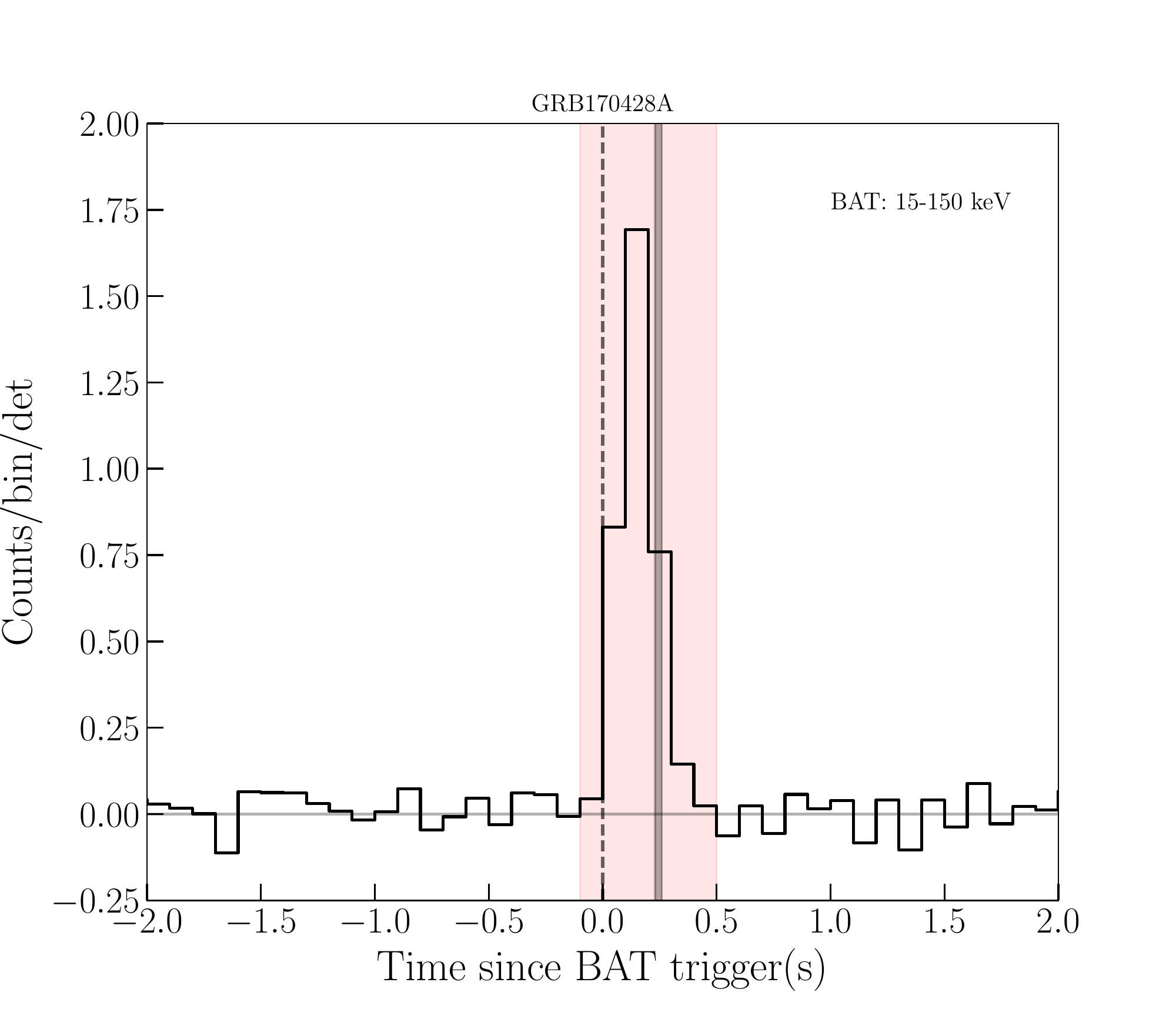}
    \includegraphics[width=.3\linewidth]{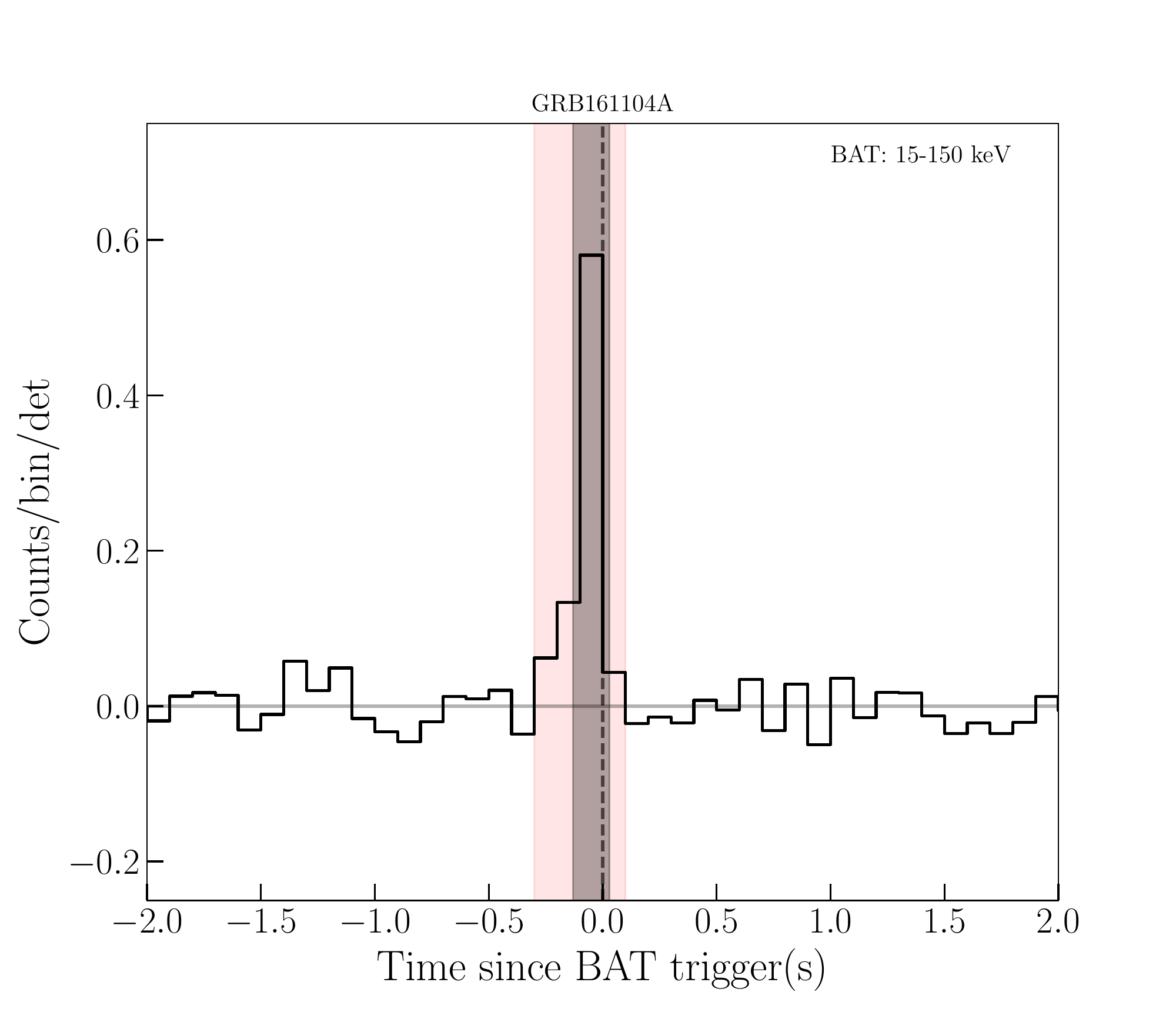}
    \includegraphics[width=.3\linewidth]{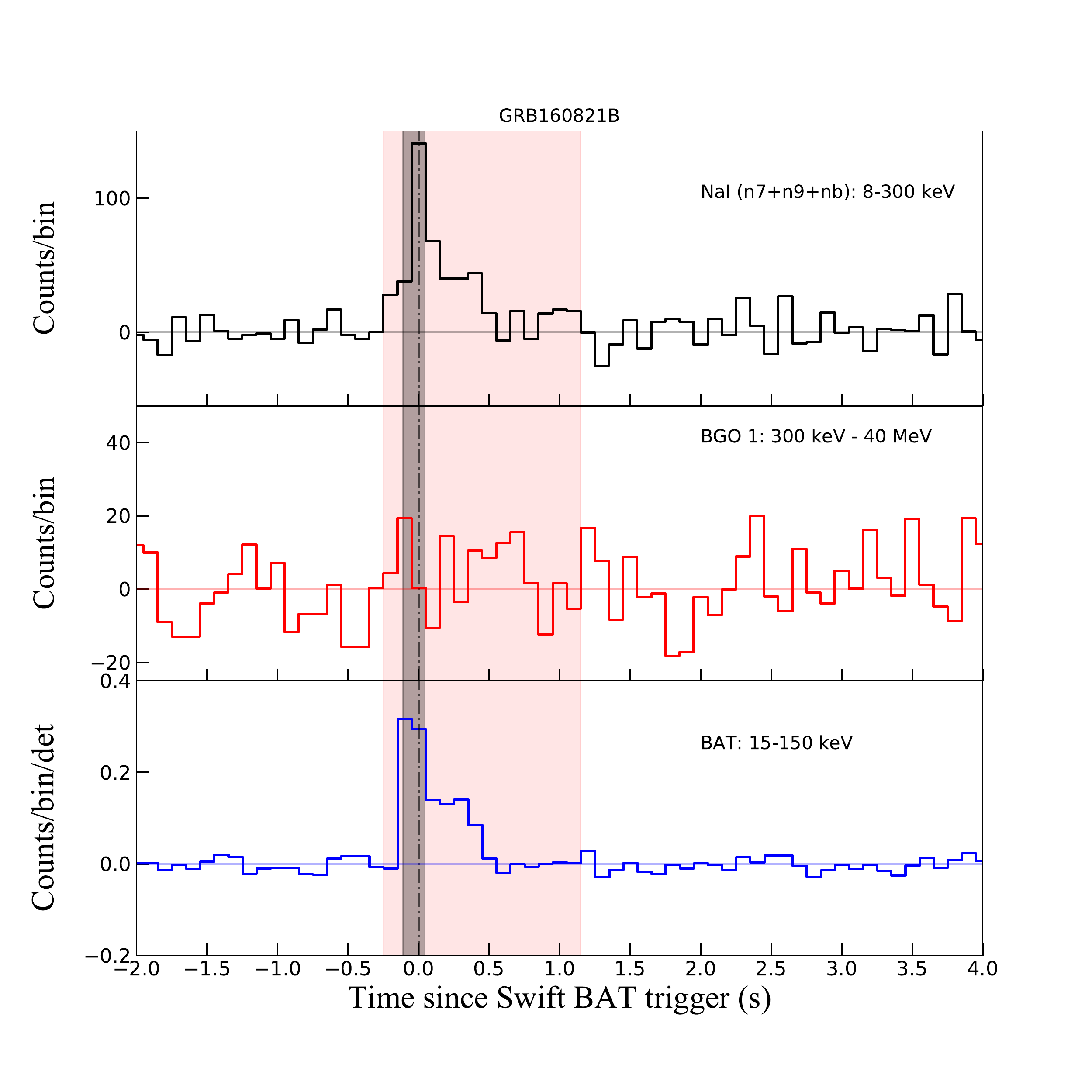}
    \includegraphics[width=.3\linewidth]{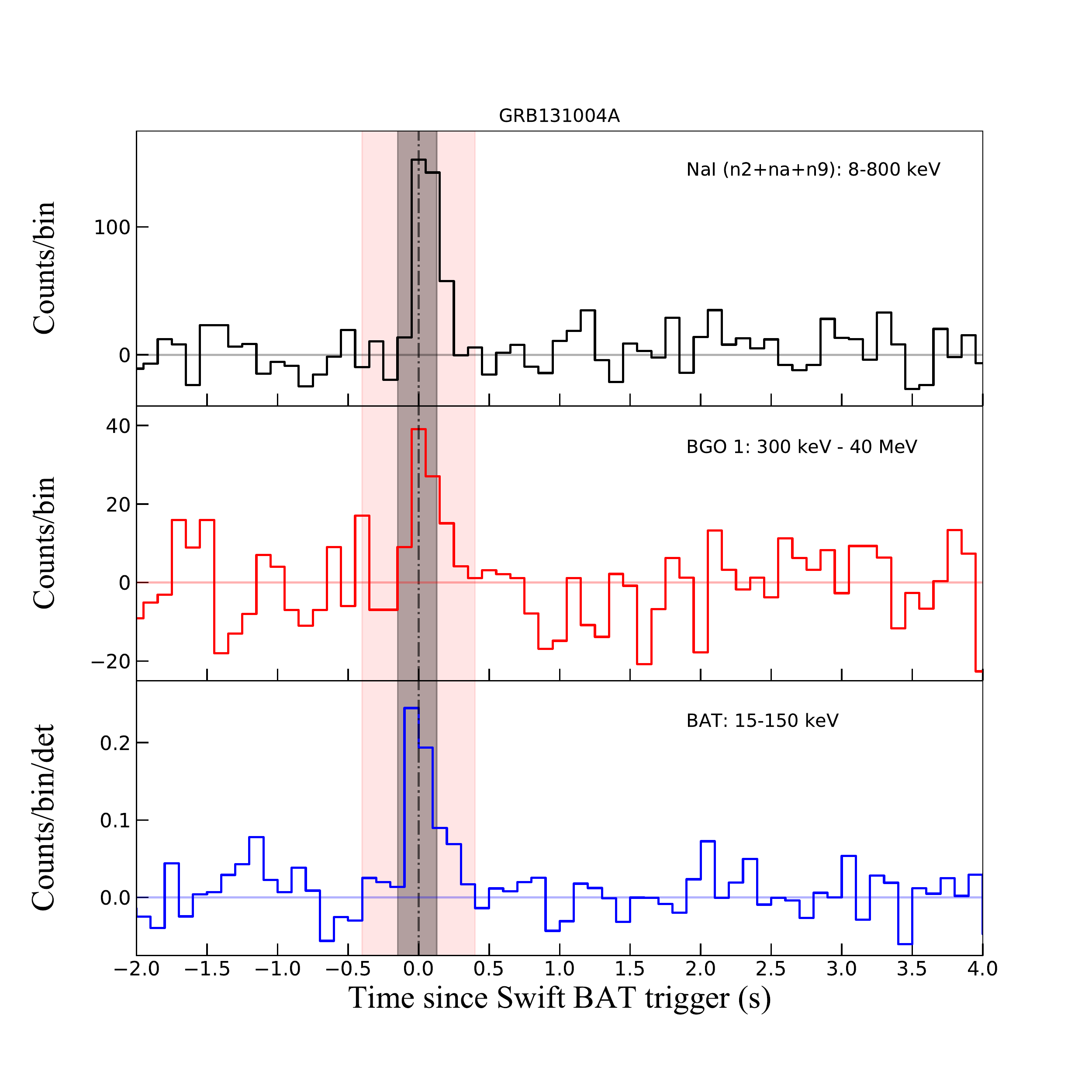}
    \includegraphics[width=.3\linewidth]{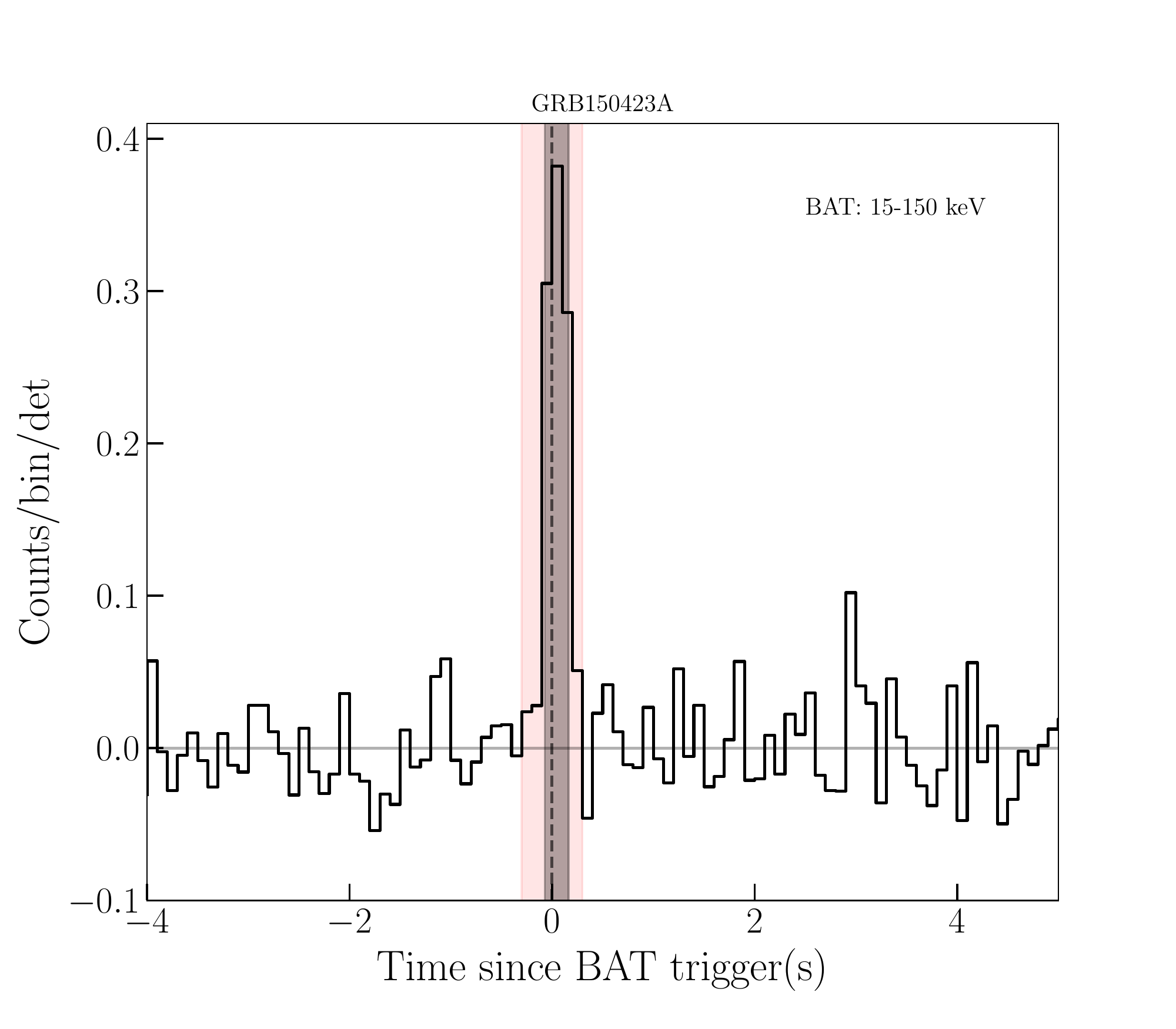}
    \includegraphics[width=.3\linewidth]{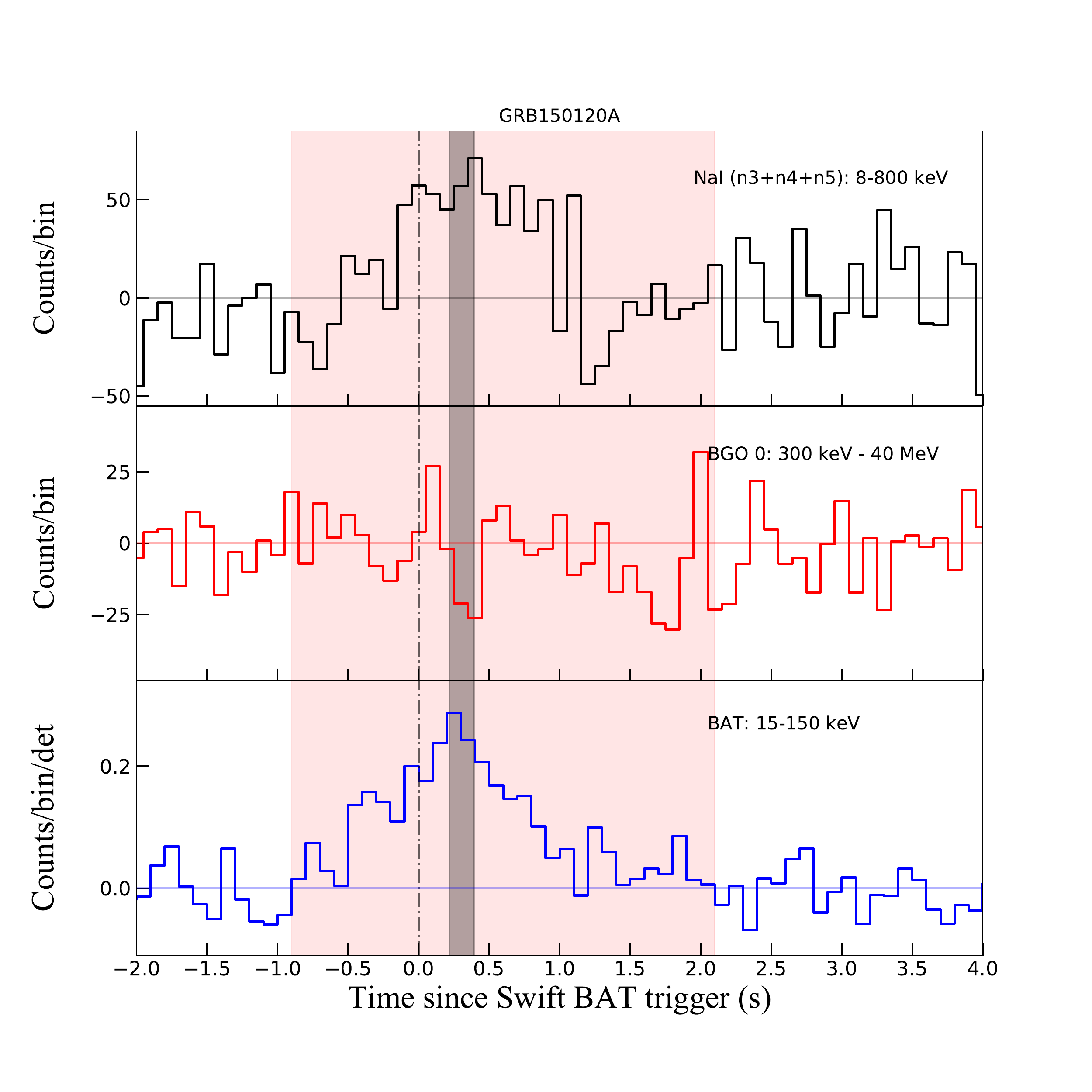}
    \includegraphics[width=.3\linewidth]{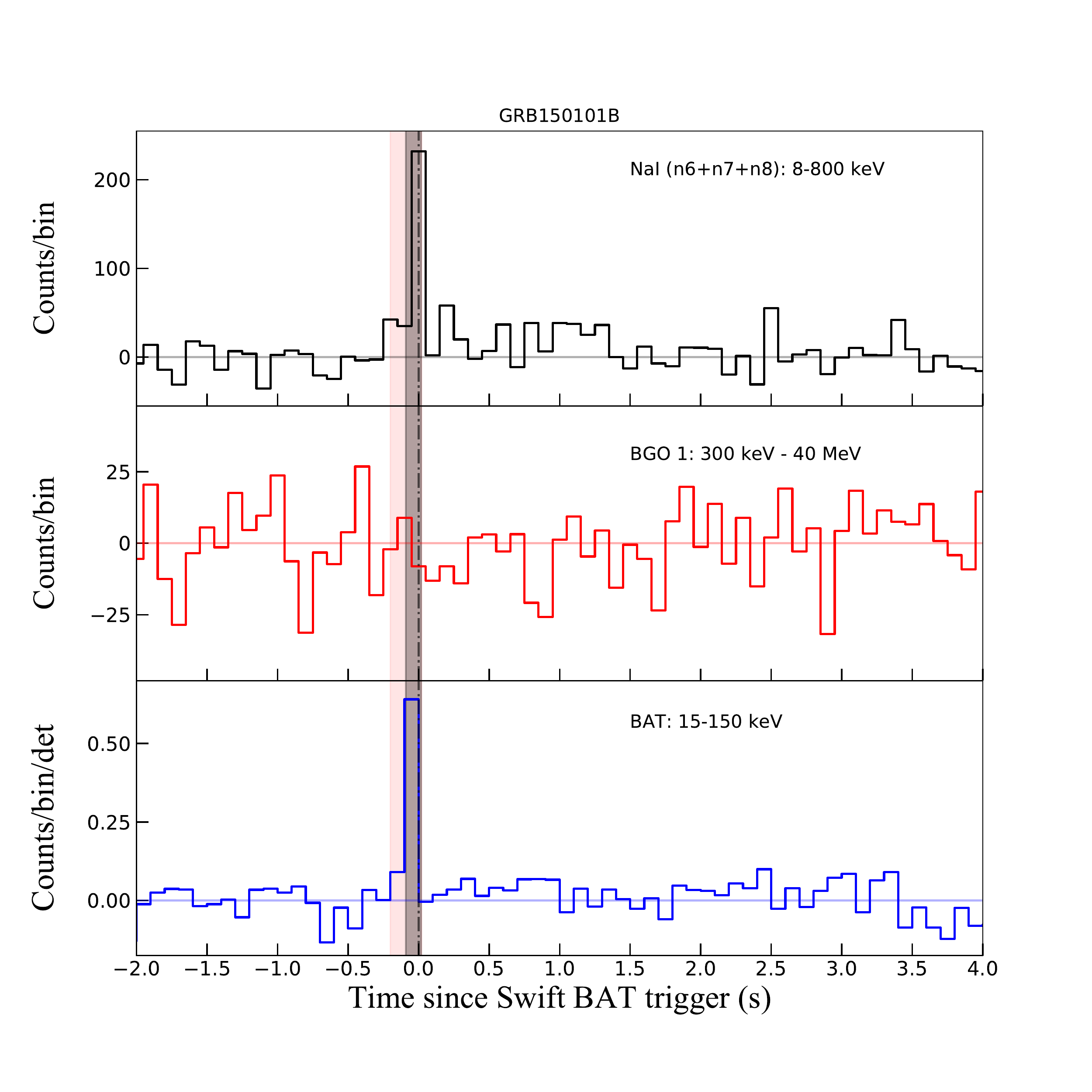}
    \includegraphics[width=.3\linewidth]{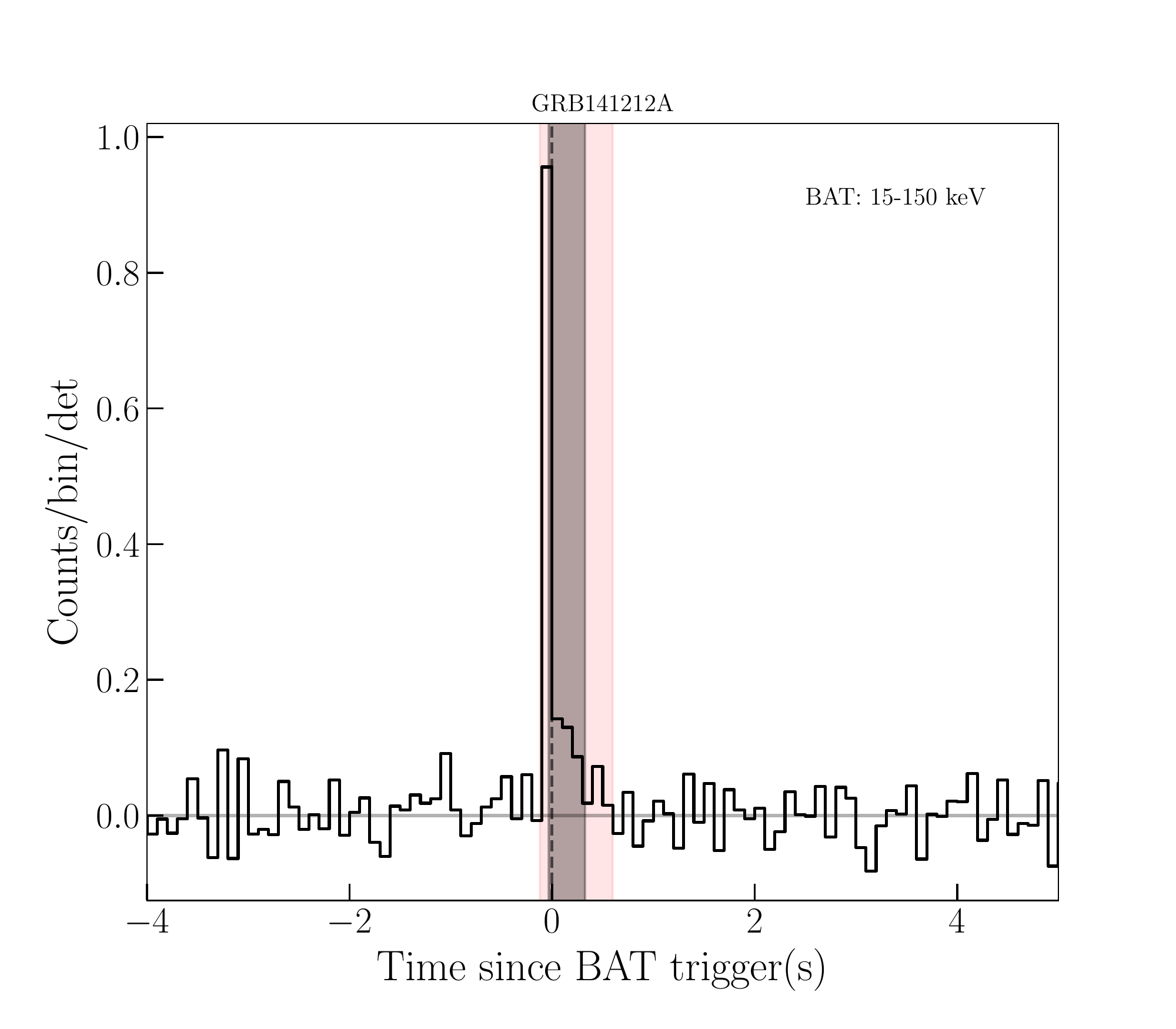}
    \includegraphics[width=.3\linewidth]{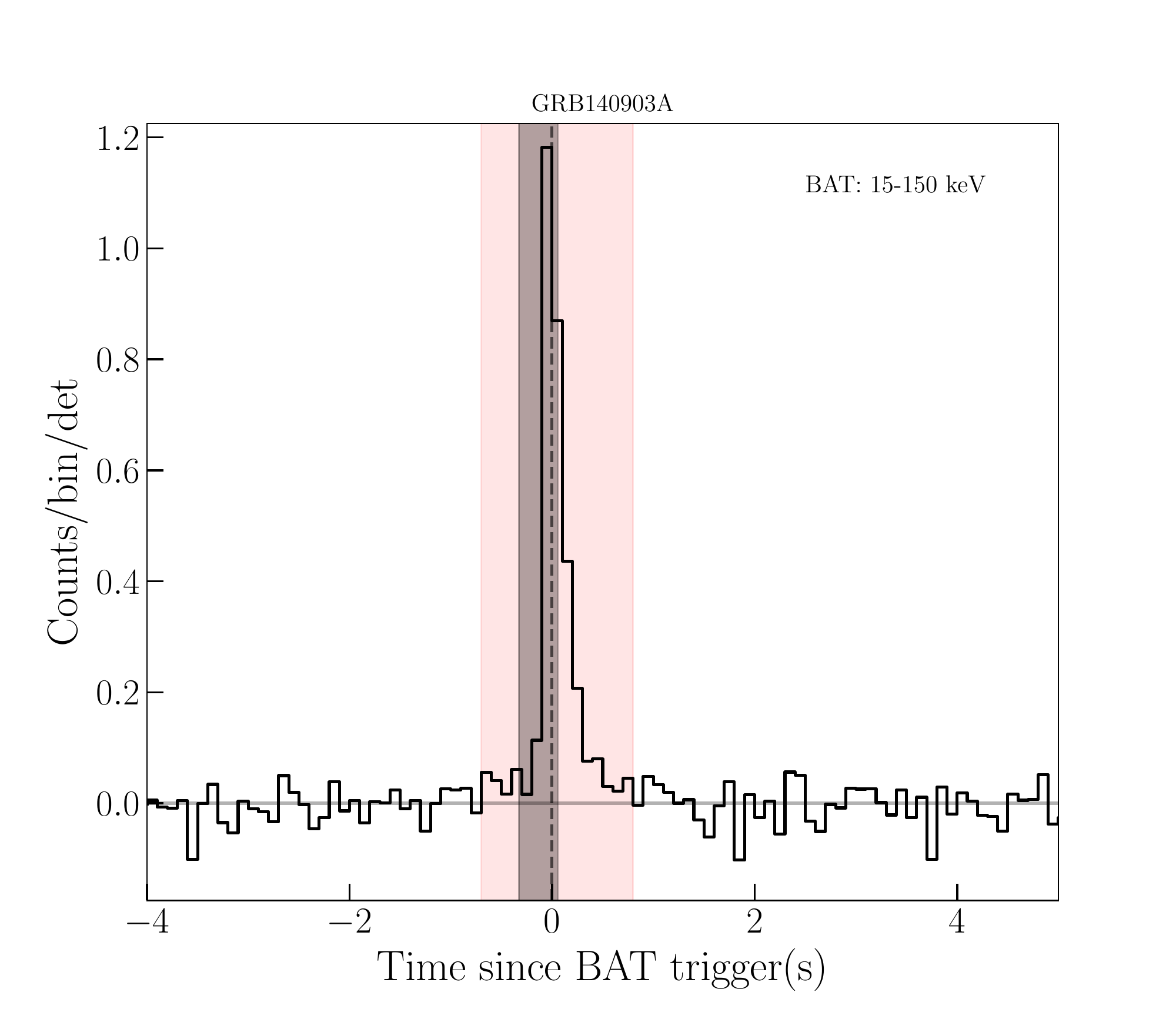}
    \includegraphics[width=.3\linewidth]{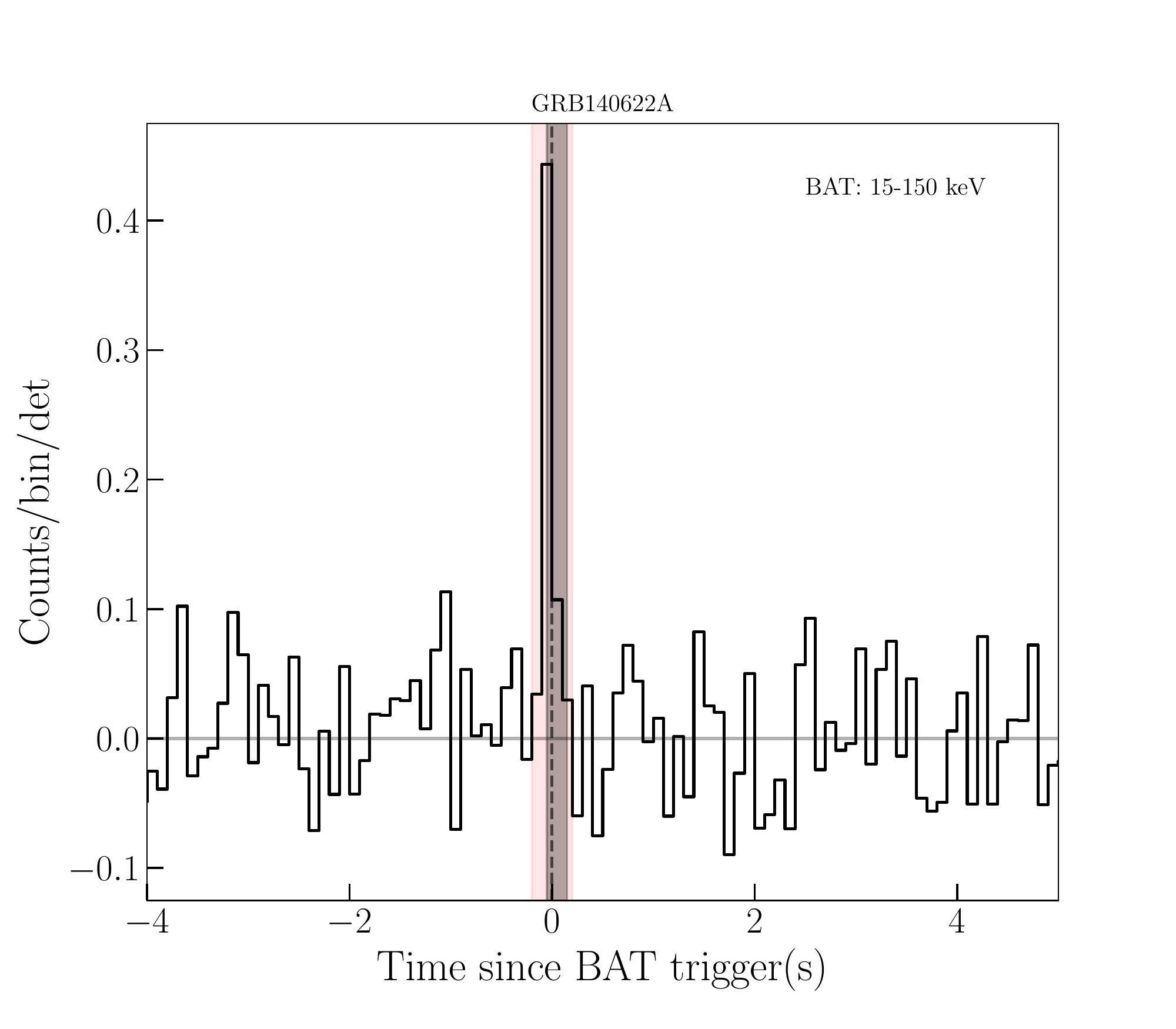}
    \includegraphics[width=.3\linewidth]{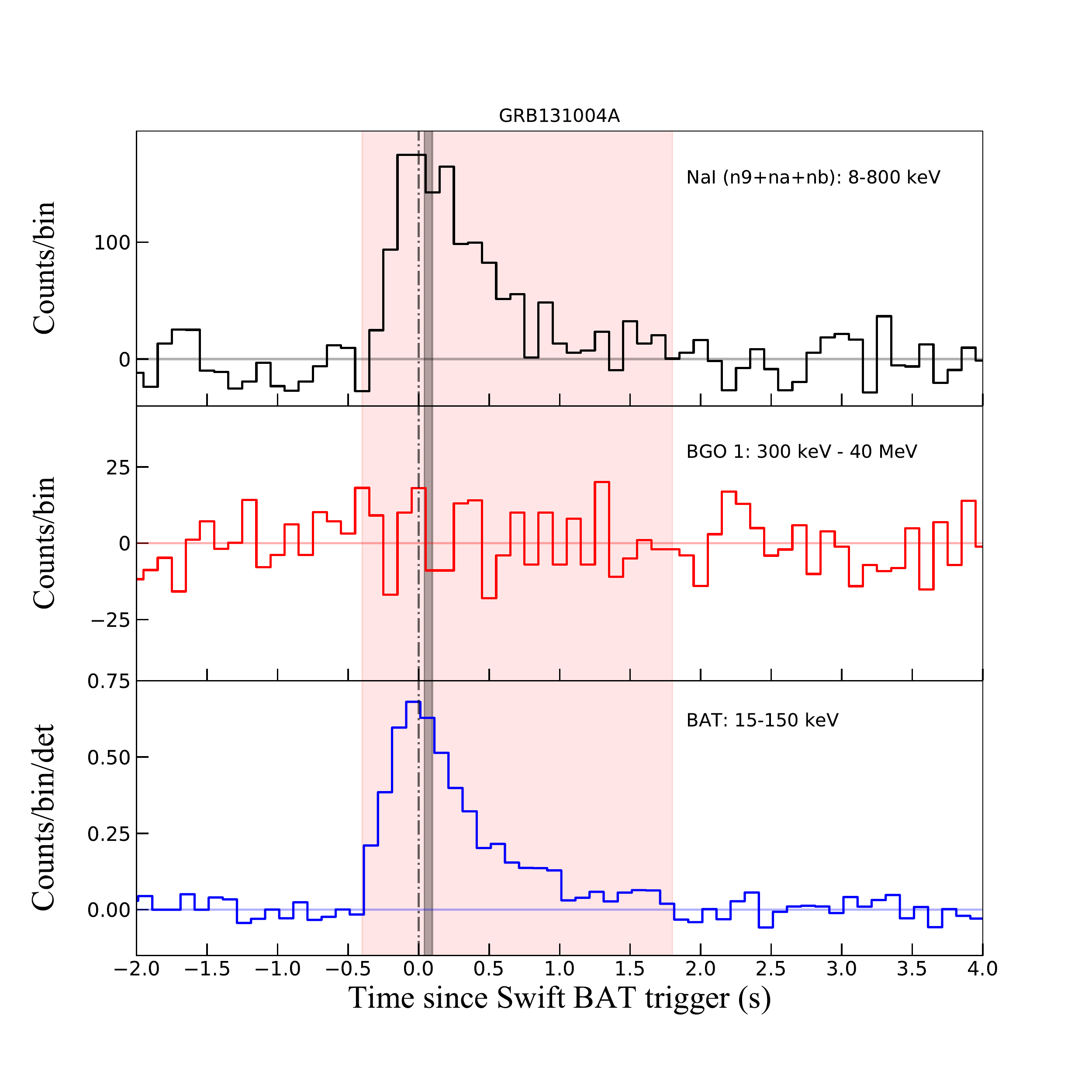}
    \caption{Light curves of the short GRBs in our sample, binned in $0.1$ s are shown. The pink and grey shaded regions mark the time intervals used for time integrated and peak count spectral analyzes respectively.}
    \label{lcs_1}
\end{figure}

\begin{figure}
    \centering
    \includegraphics[width=.3\linewidth]{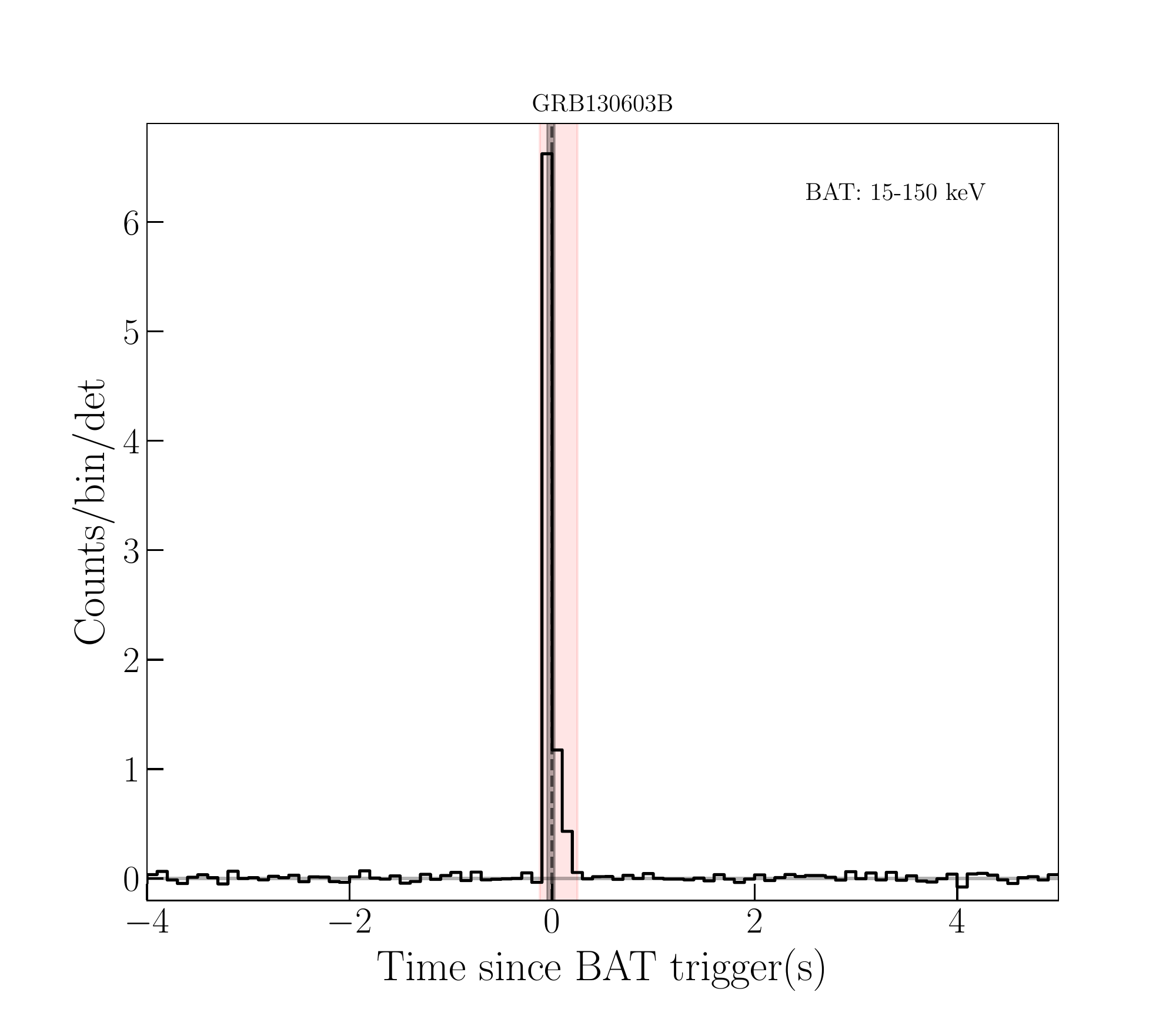}
    \includegraphics[width=.3\linewidth]{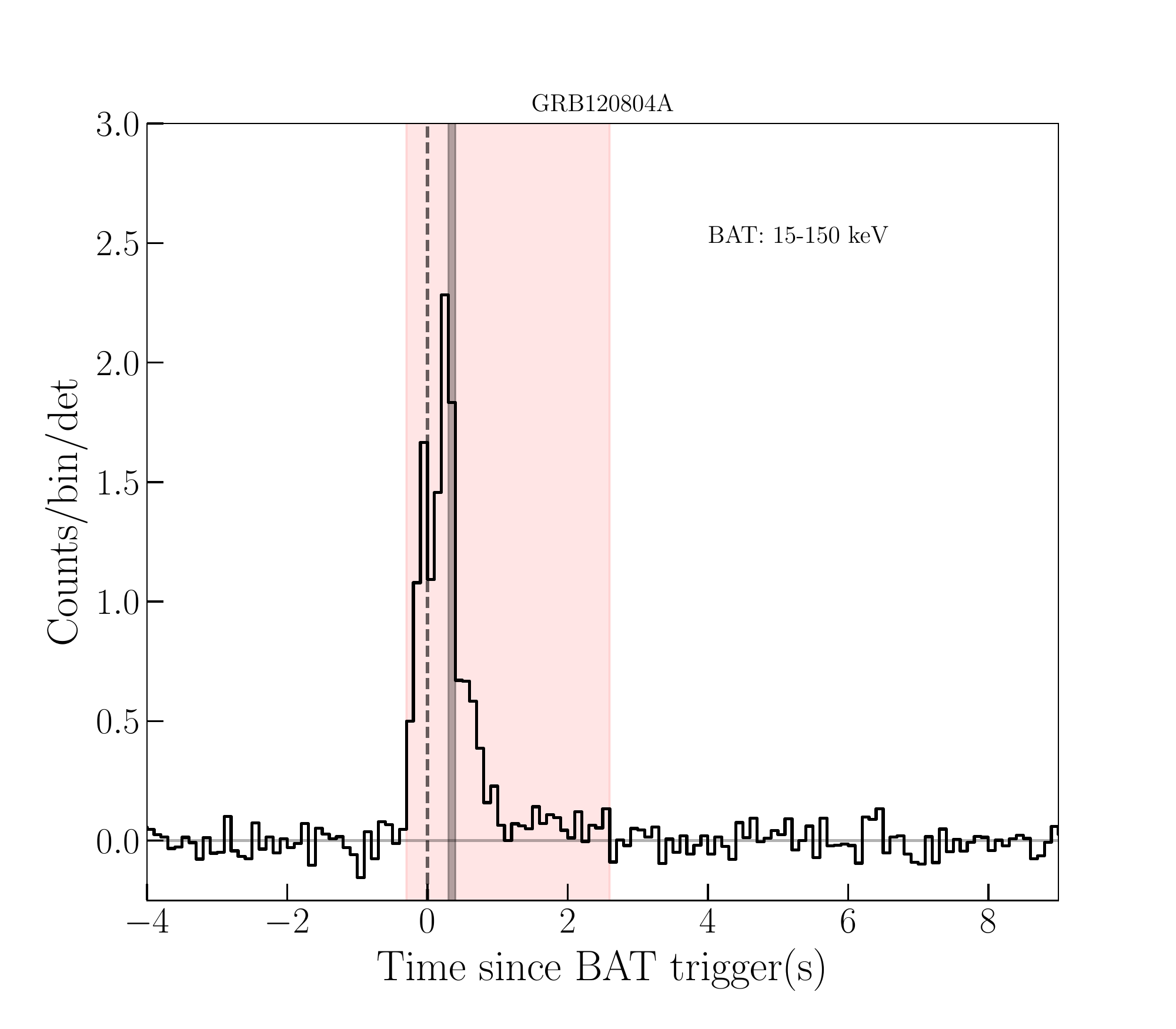}
    \includegraphics[width=.3\linewidth]{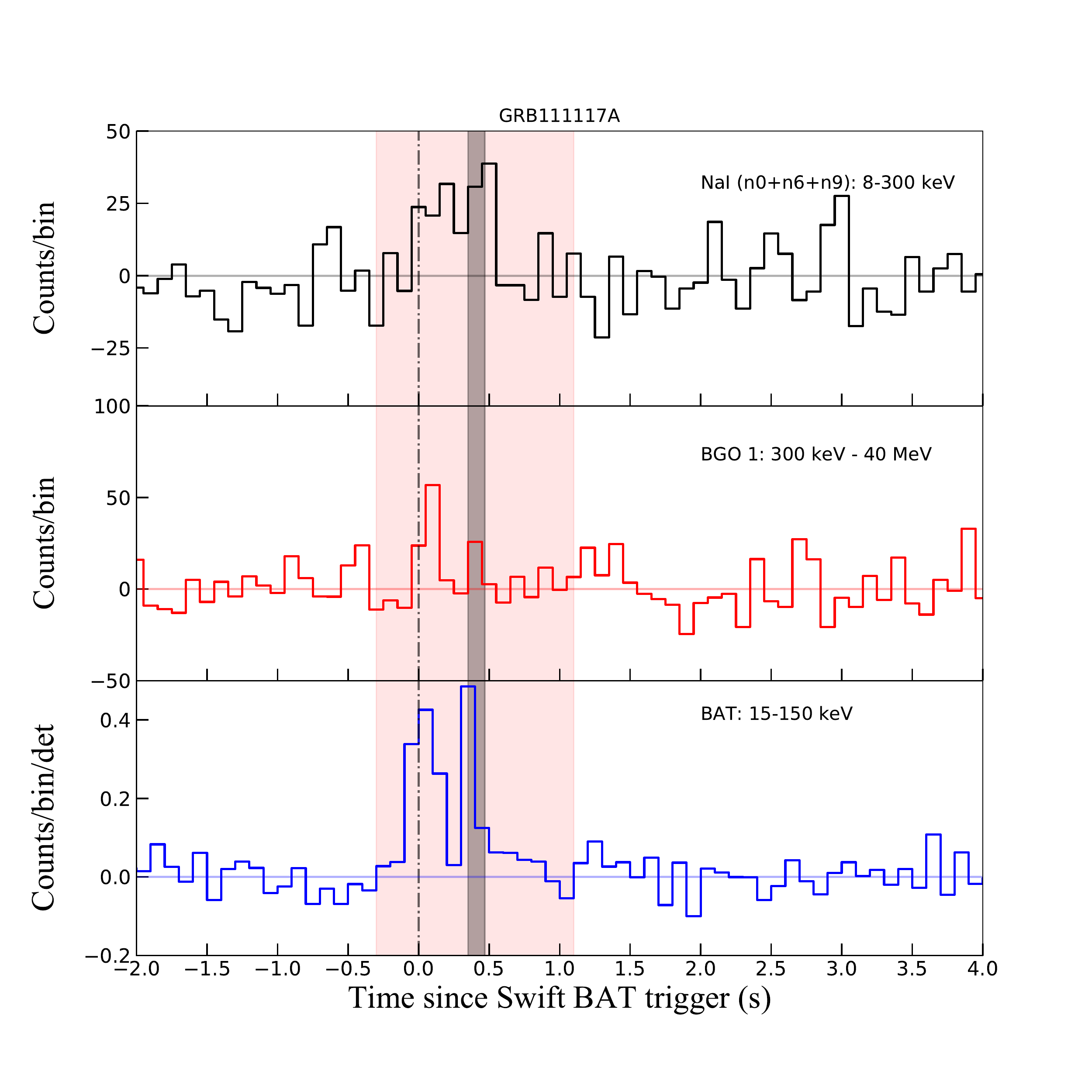}
    \includegraphics[width=.3\linewidth]{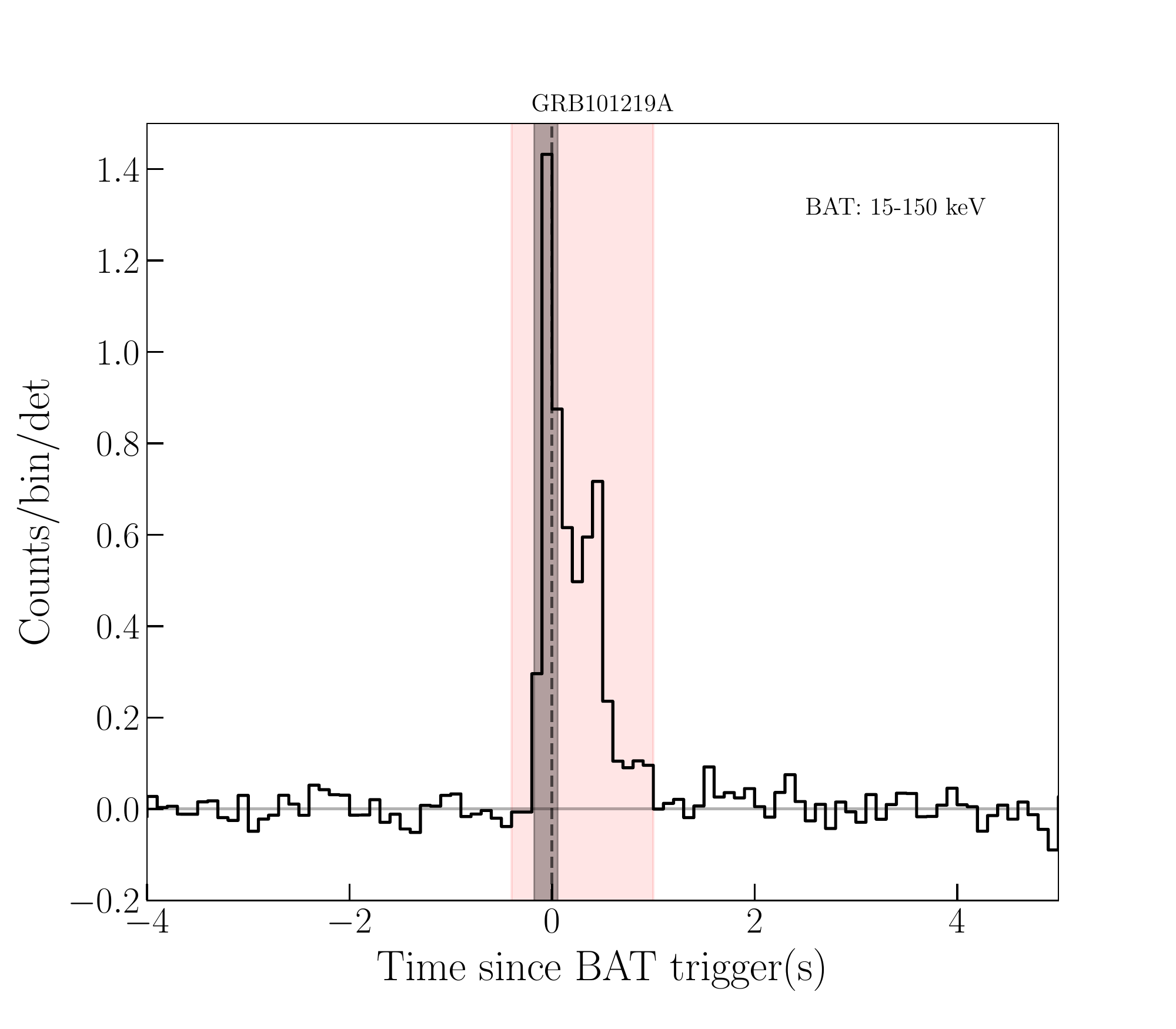}
    \includegraphics[width=.3\linewidth]{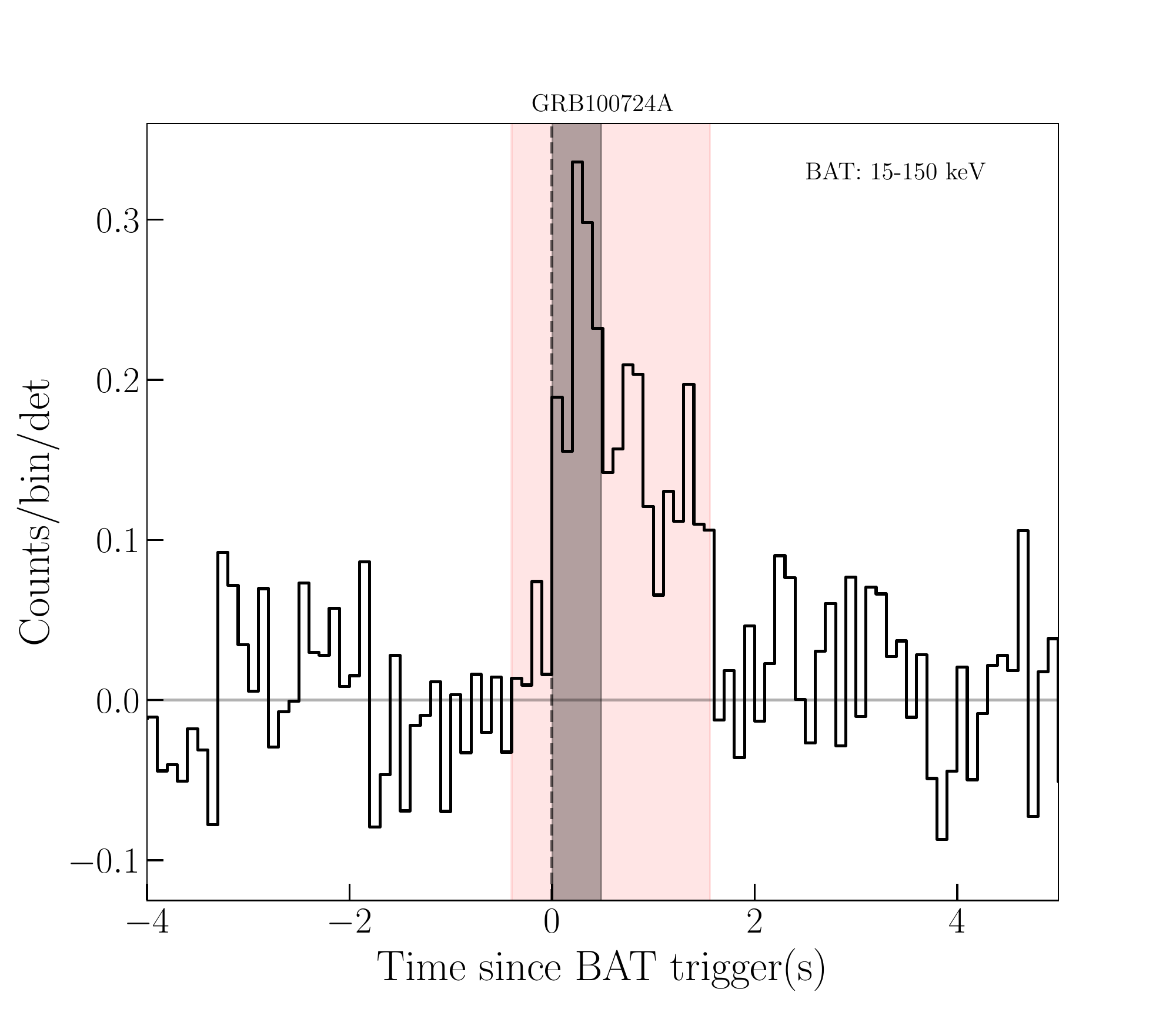}
    \includegraphics[width=.3\linewidth]{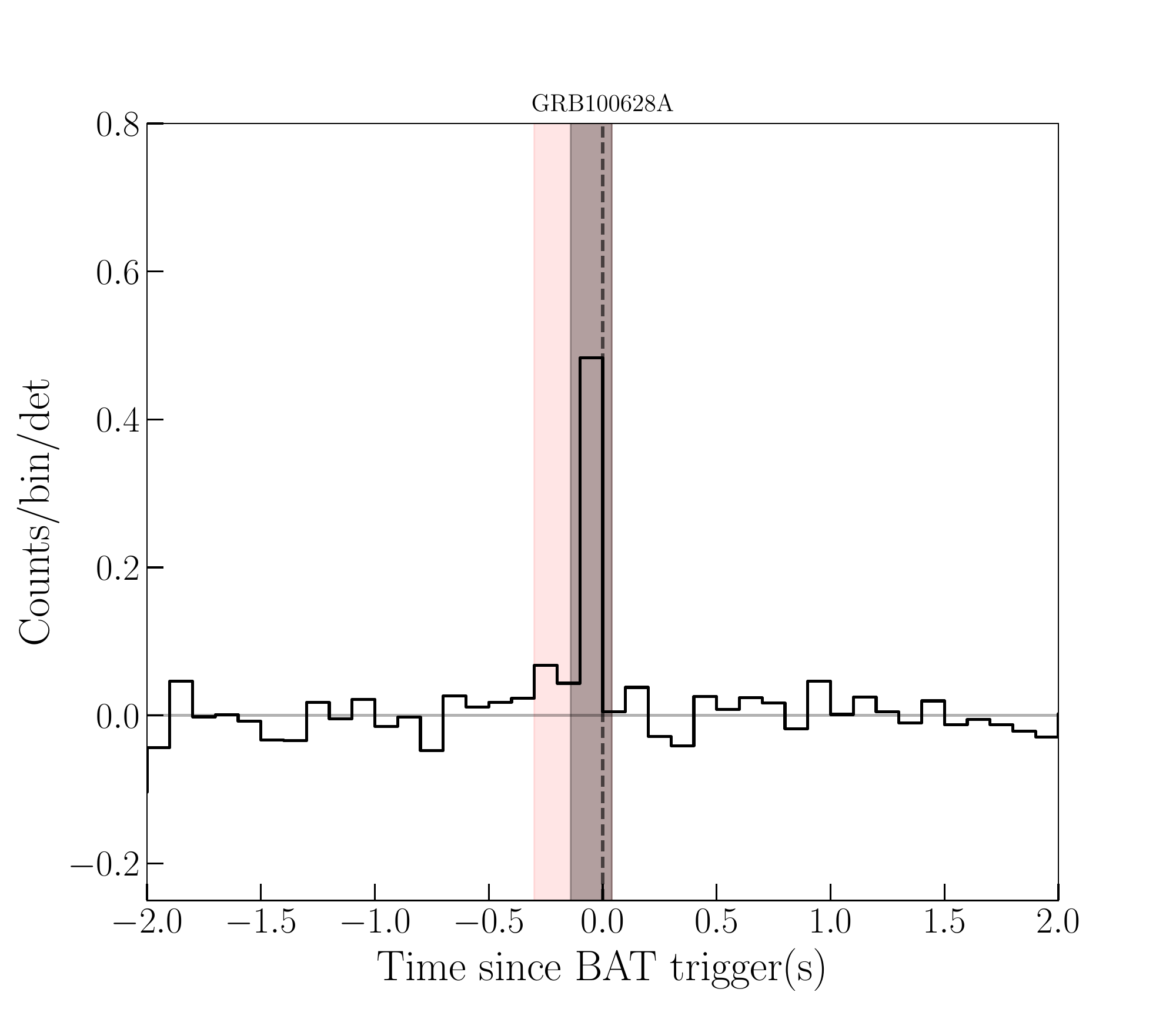}
    \includegraphics[width=.3\linewidth]{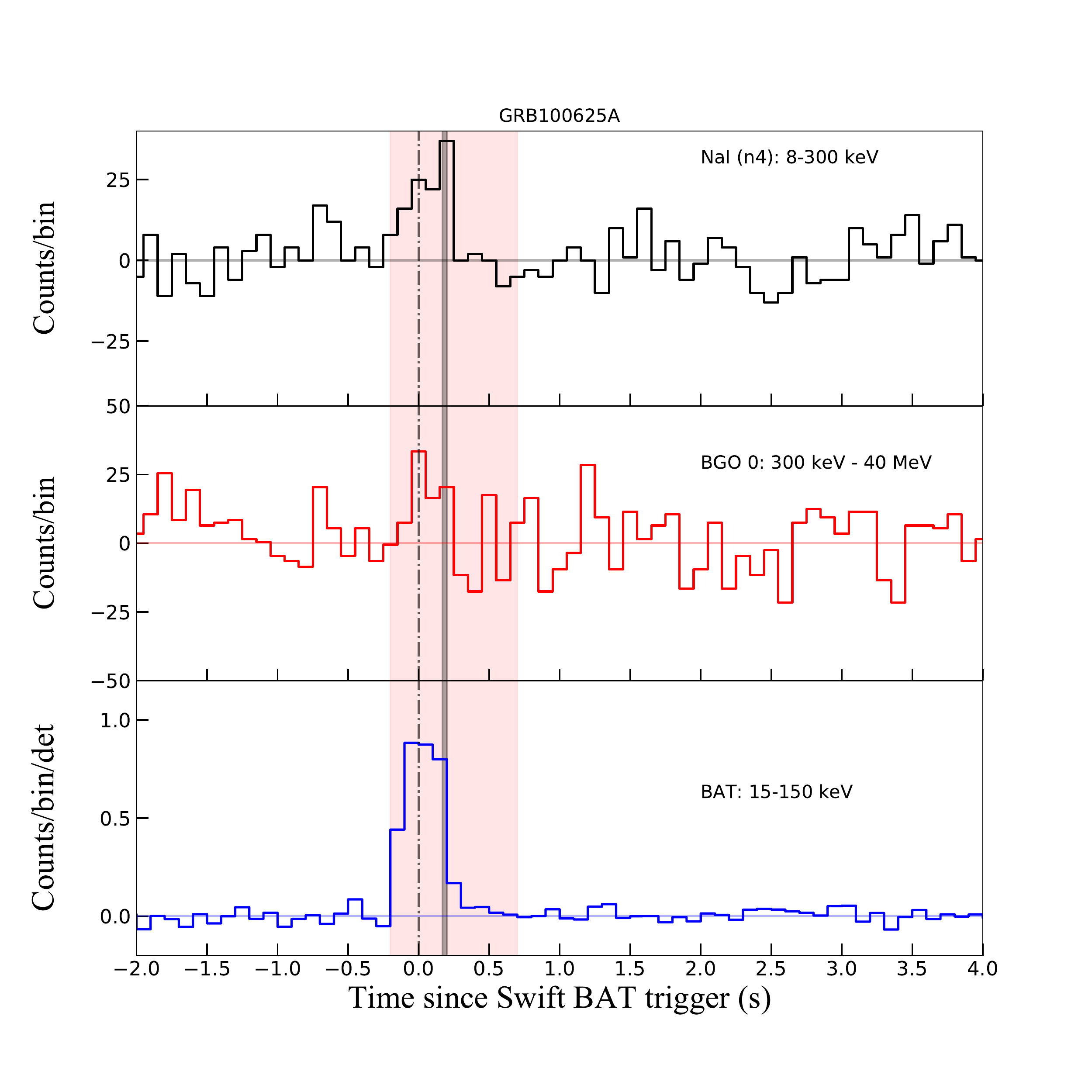}
    \includegraphics[width=.3\linewidth]{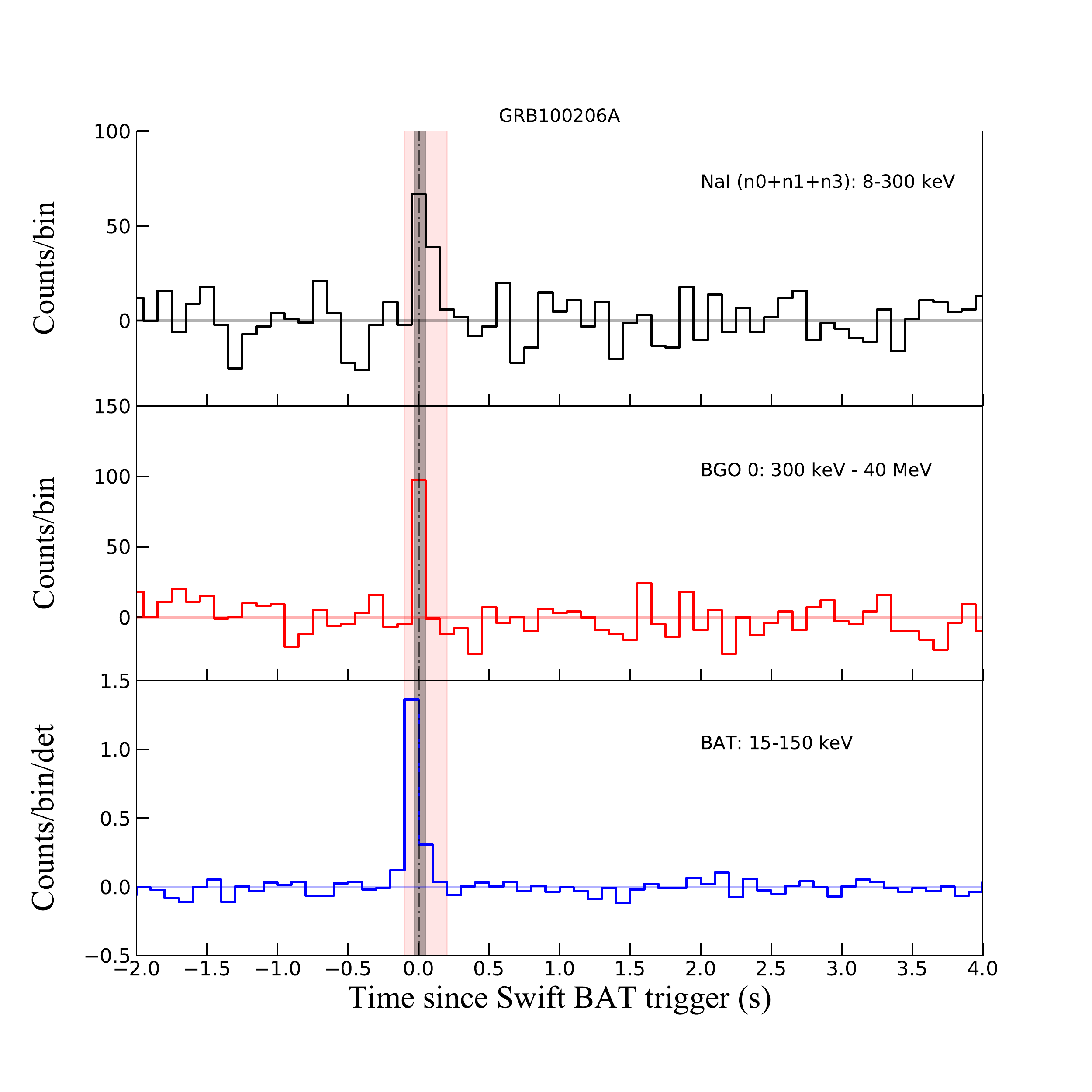}
    \includegraphics[width=.3\linewidth]{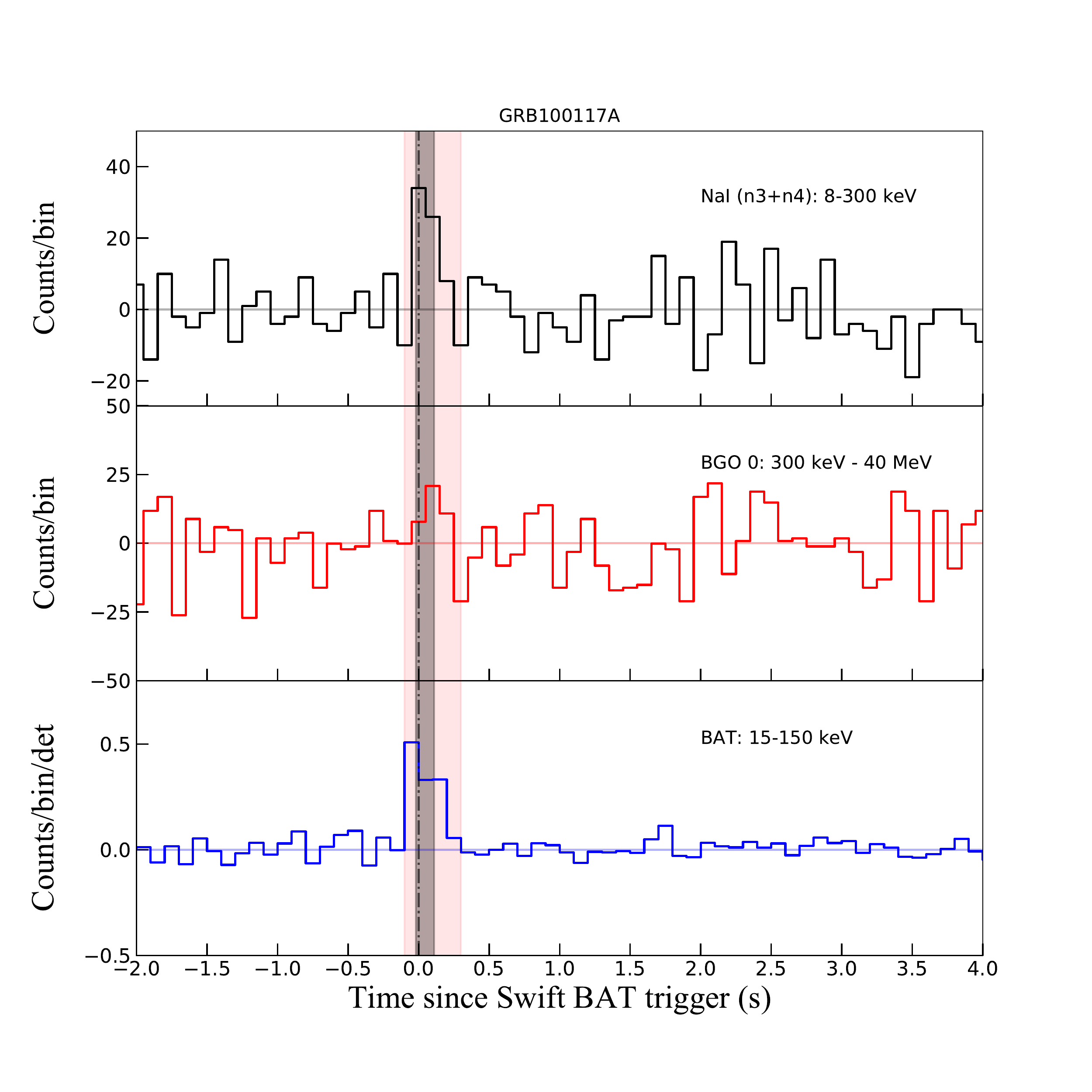}
    \includegraphics[width=.3\linewidth]{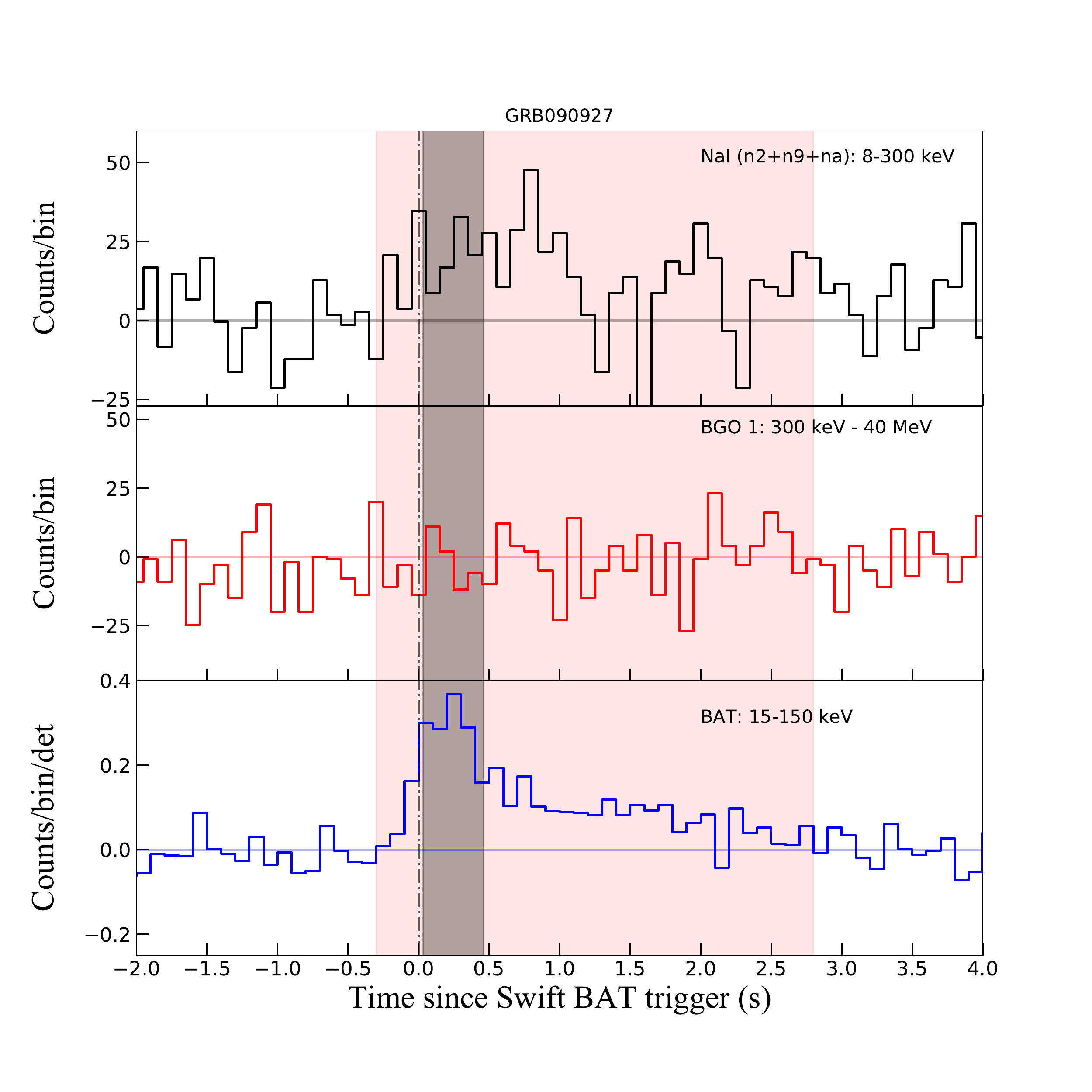}
    \includegraphics[width=.3\linewidth]{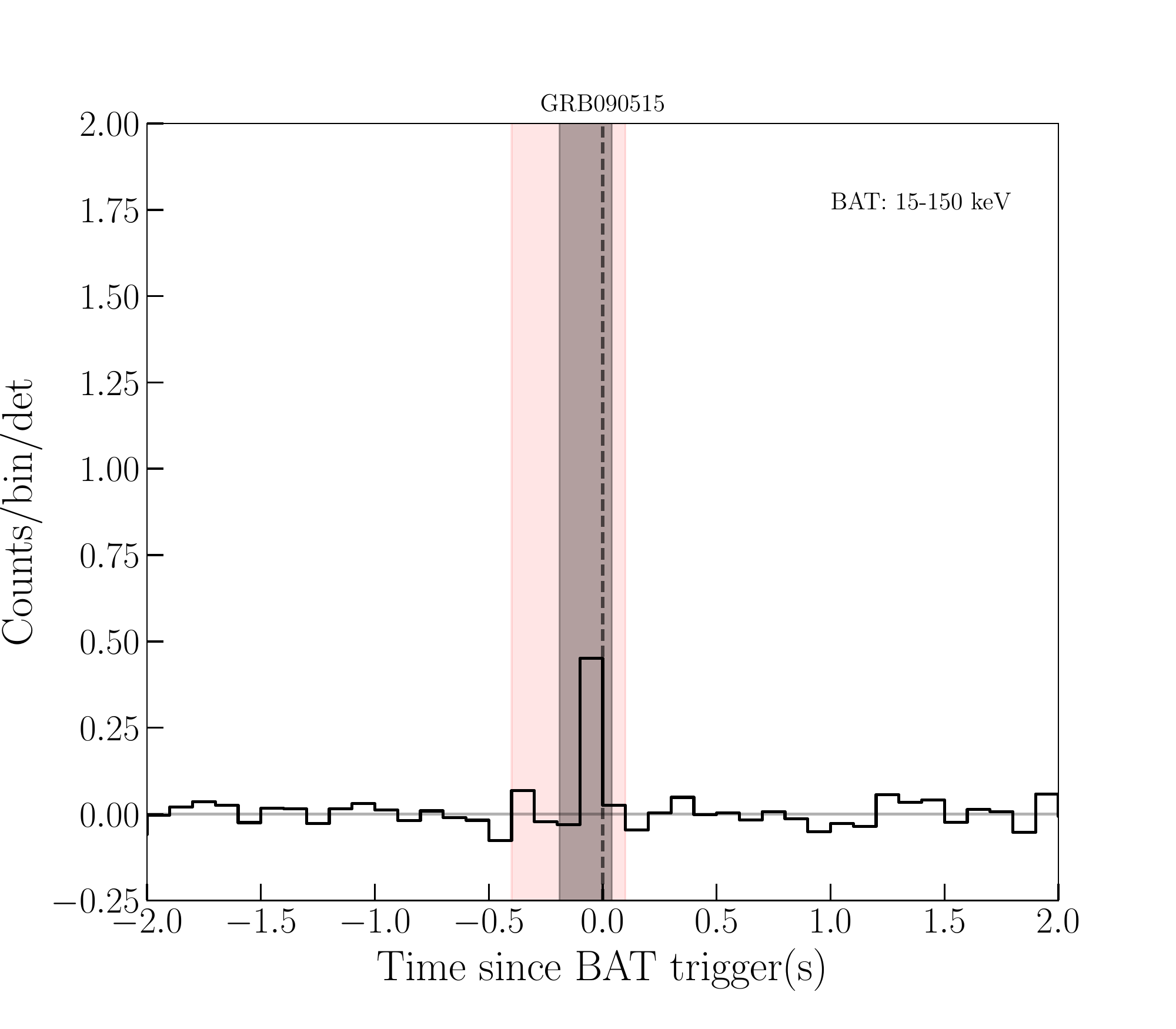}
    \includegraphics[width=.3\linewidth]{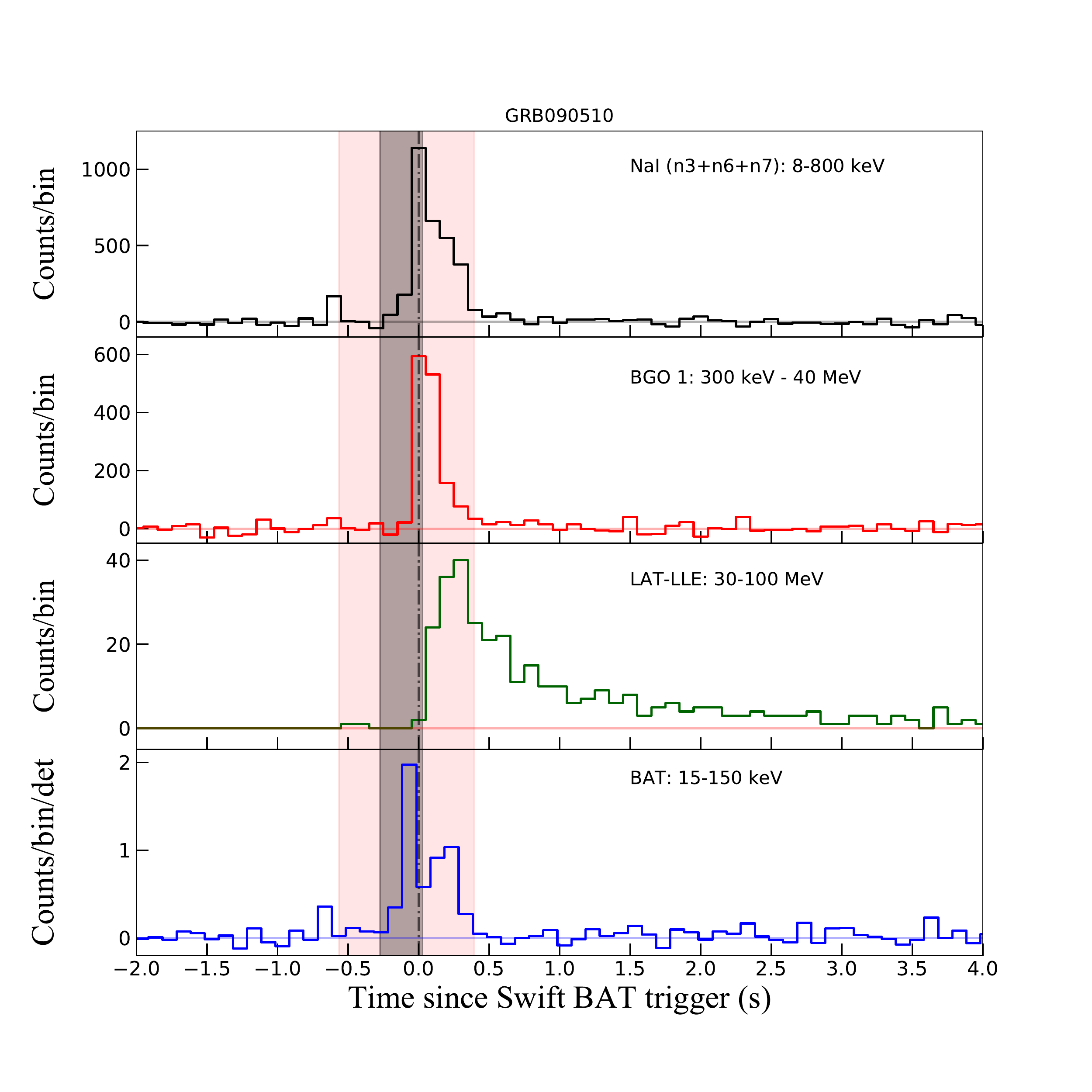}
    \caption{Figure \ref{lcs_1} continued.}
    \label{lcs_2}
\end{figure}

\begin{figure}
    \centering
    \includegraphics[width=.3\linewidth]{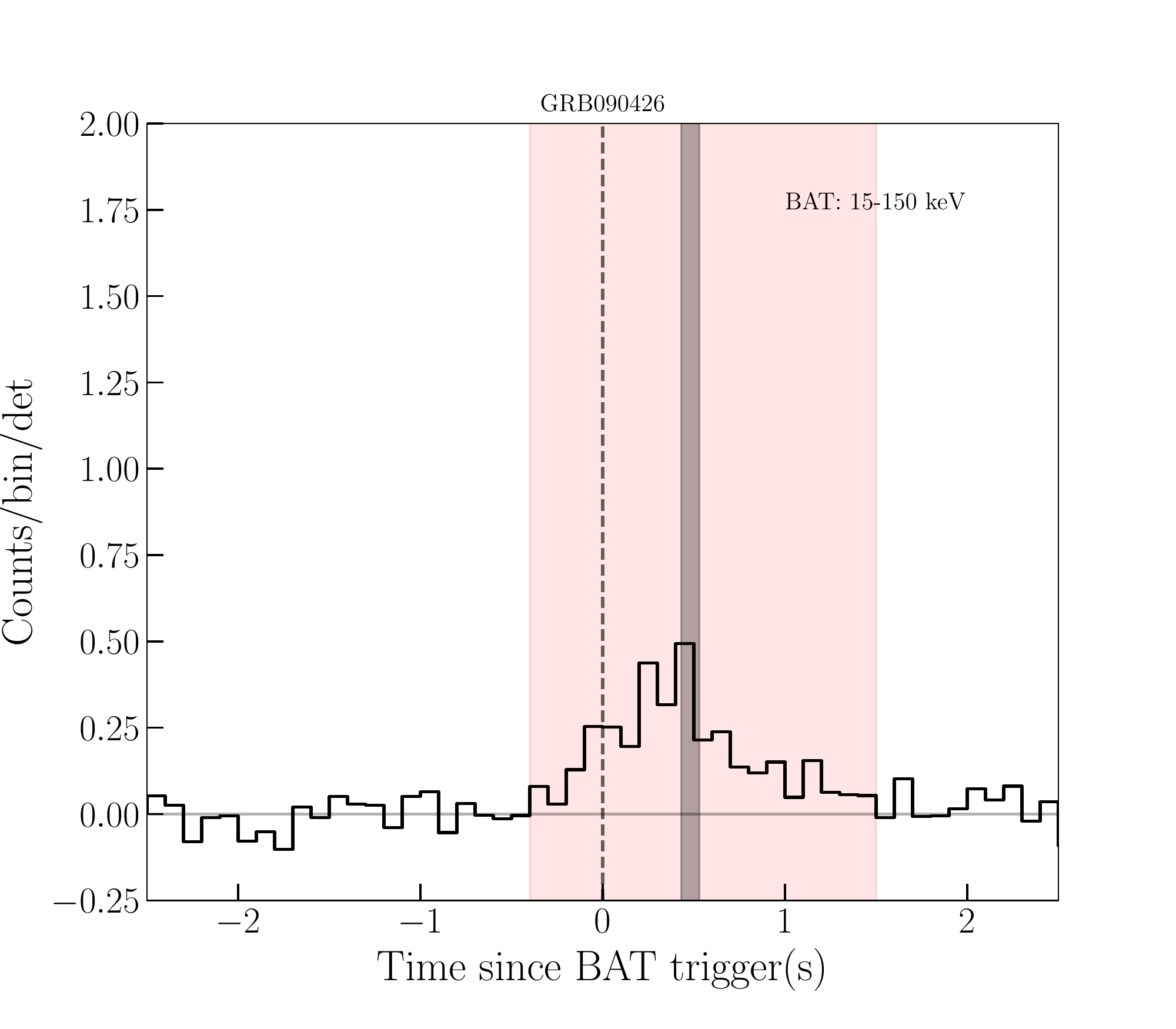}
    \includegraphics[width=.3\linewidth]{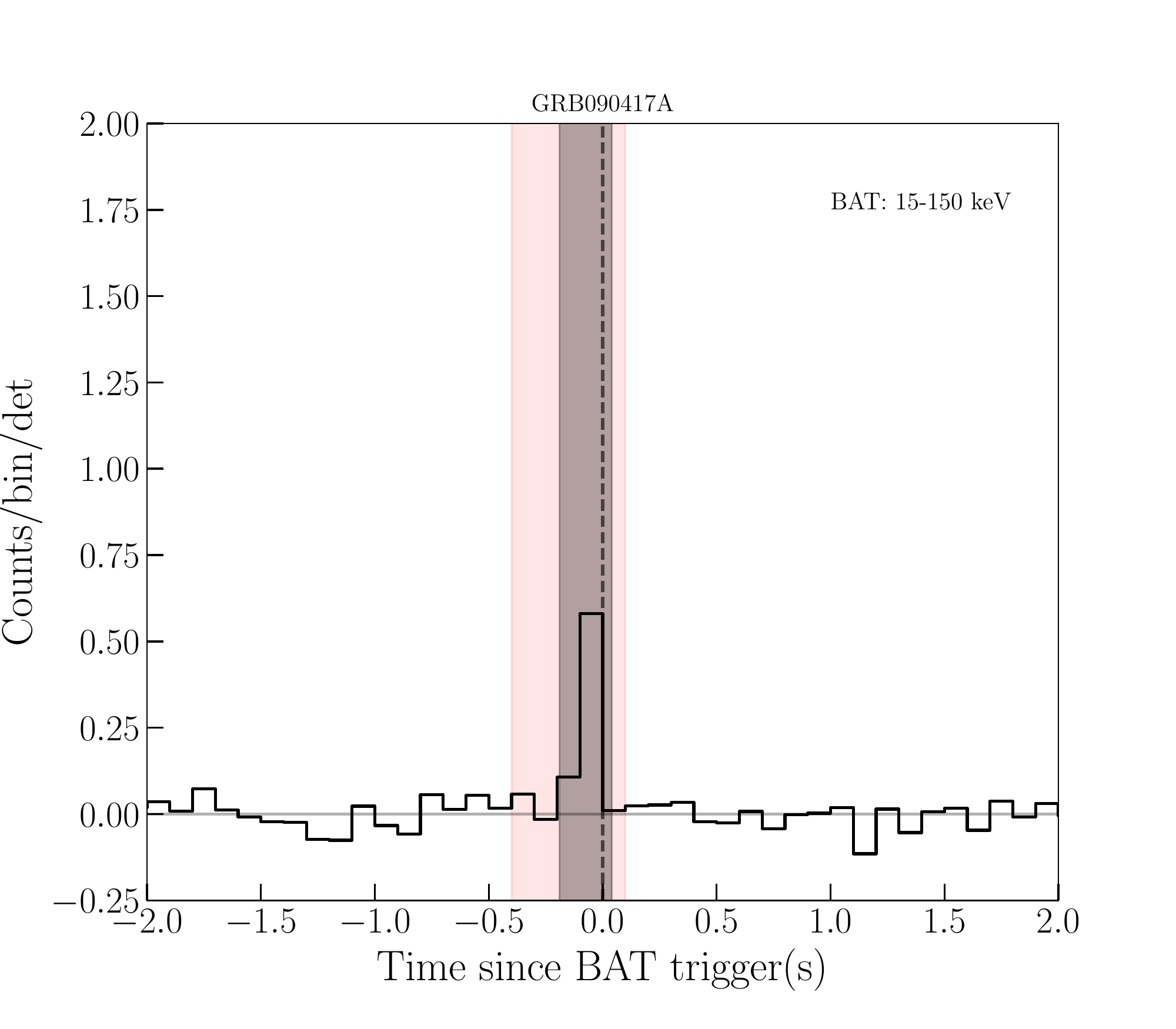}
    \includegraphics[width=.3\linewidth]{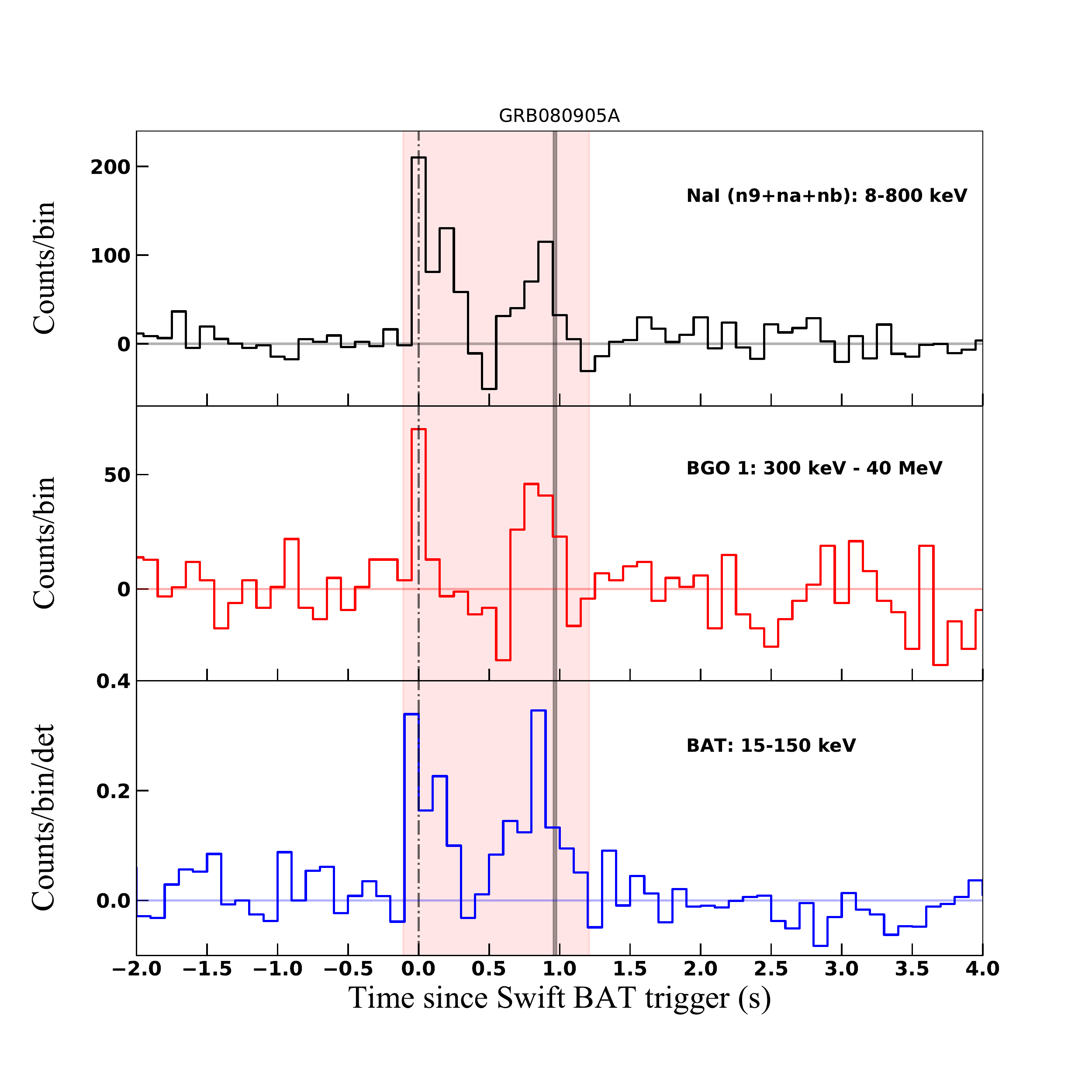}
    \includegraphics[width=.3\linewidth]{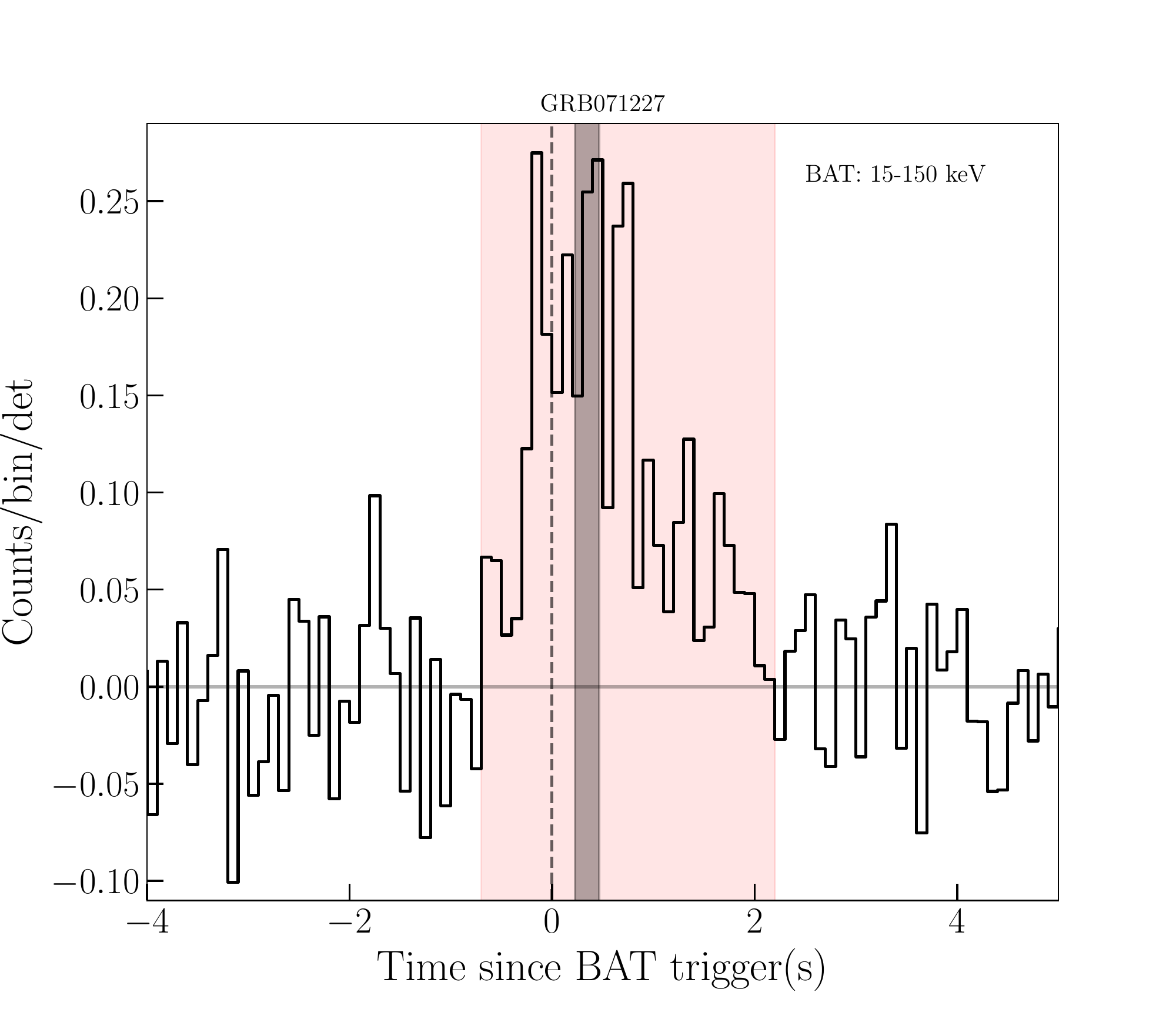}
    \includegraphics[width=.3\linewidth]{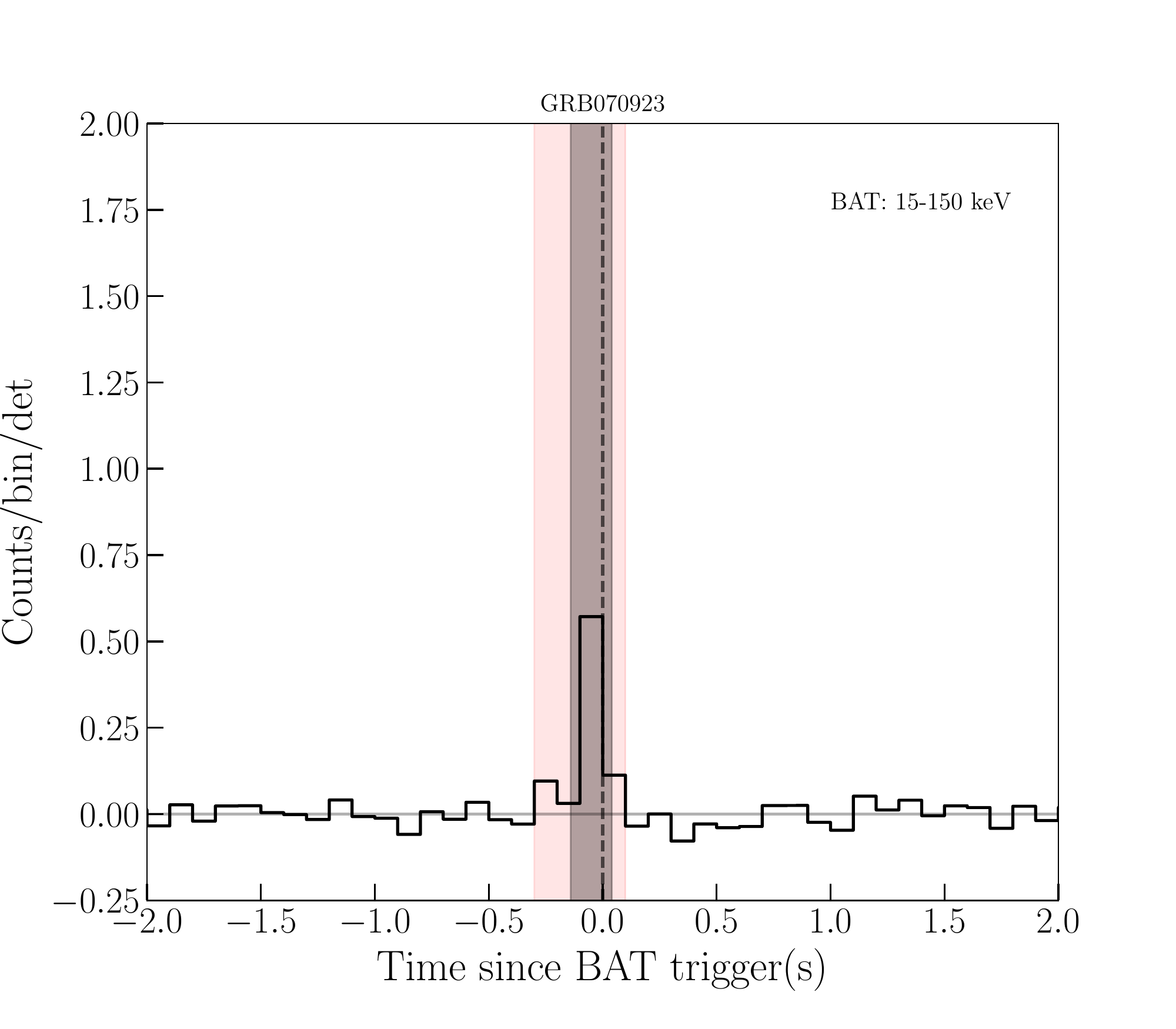}
    \includegraphics[width=.3\linewidth]{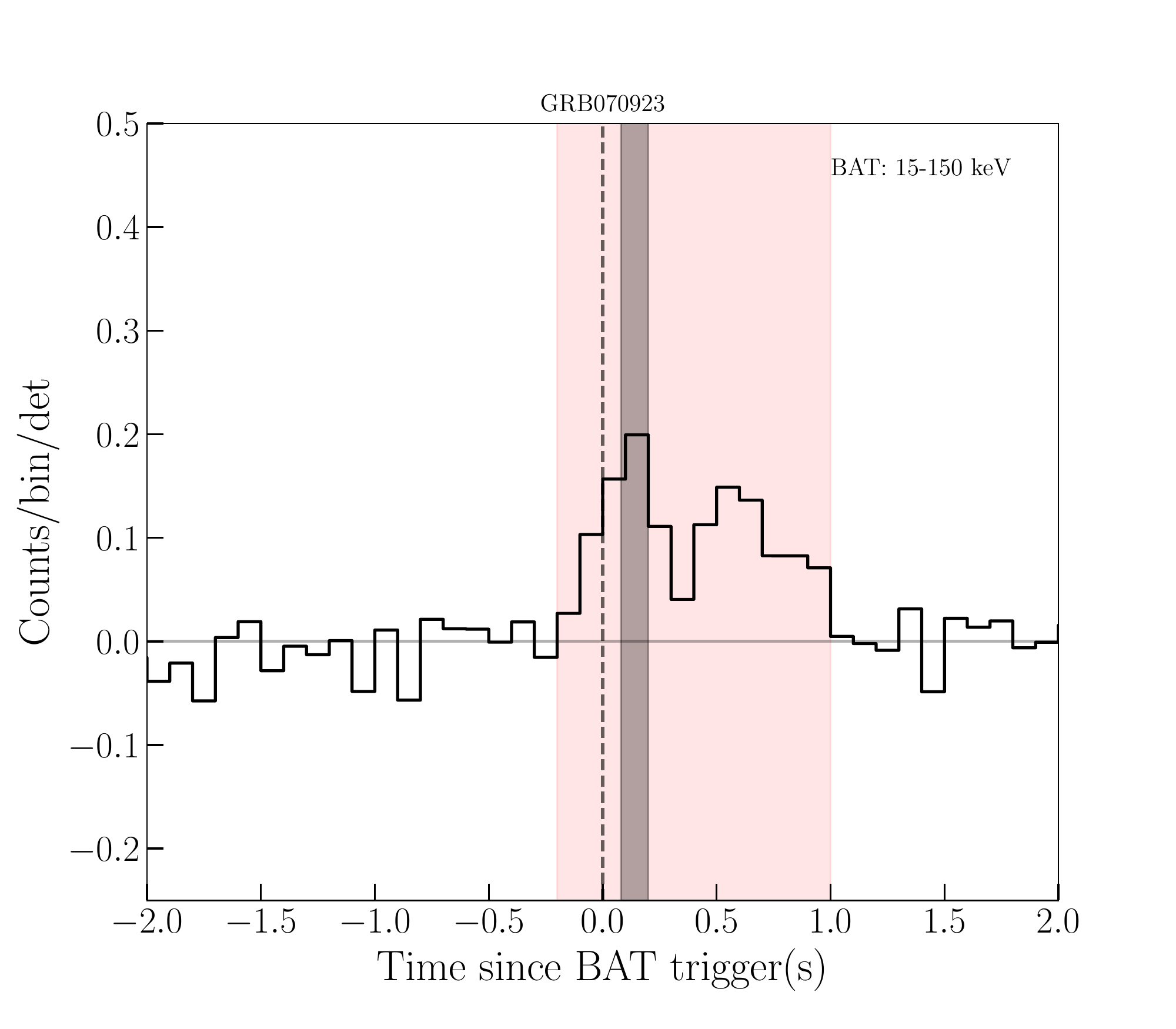}
    \includegraphics[width=.3\linewidth]{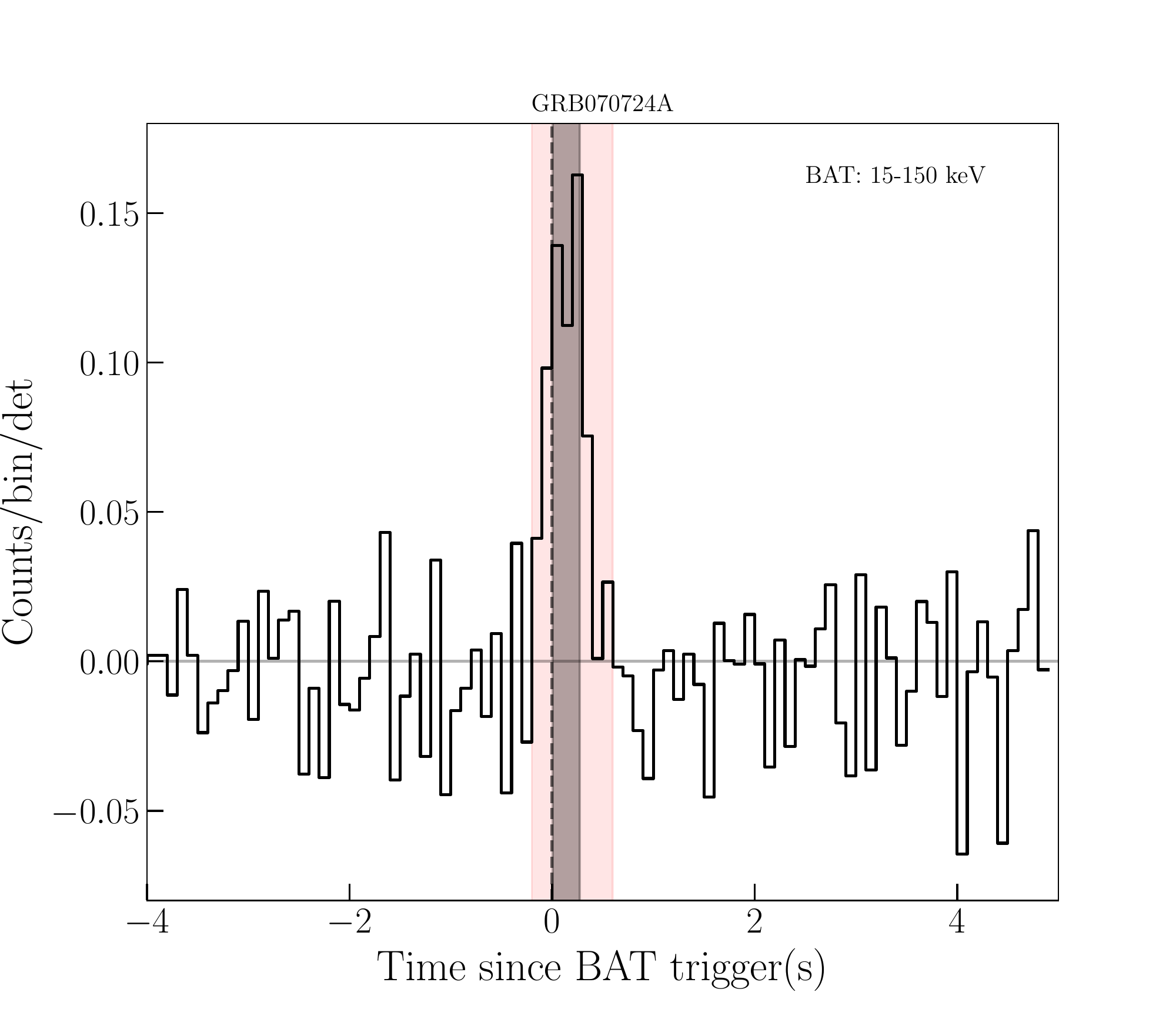}
    \includegraphics[width=.3\linewidth]{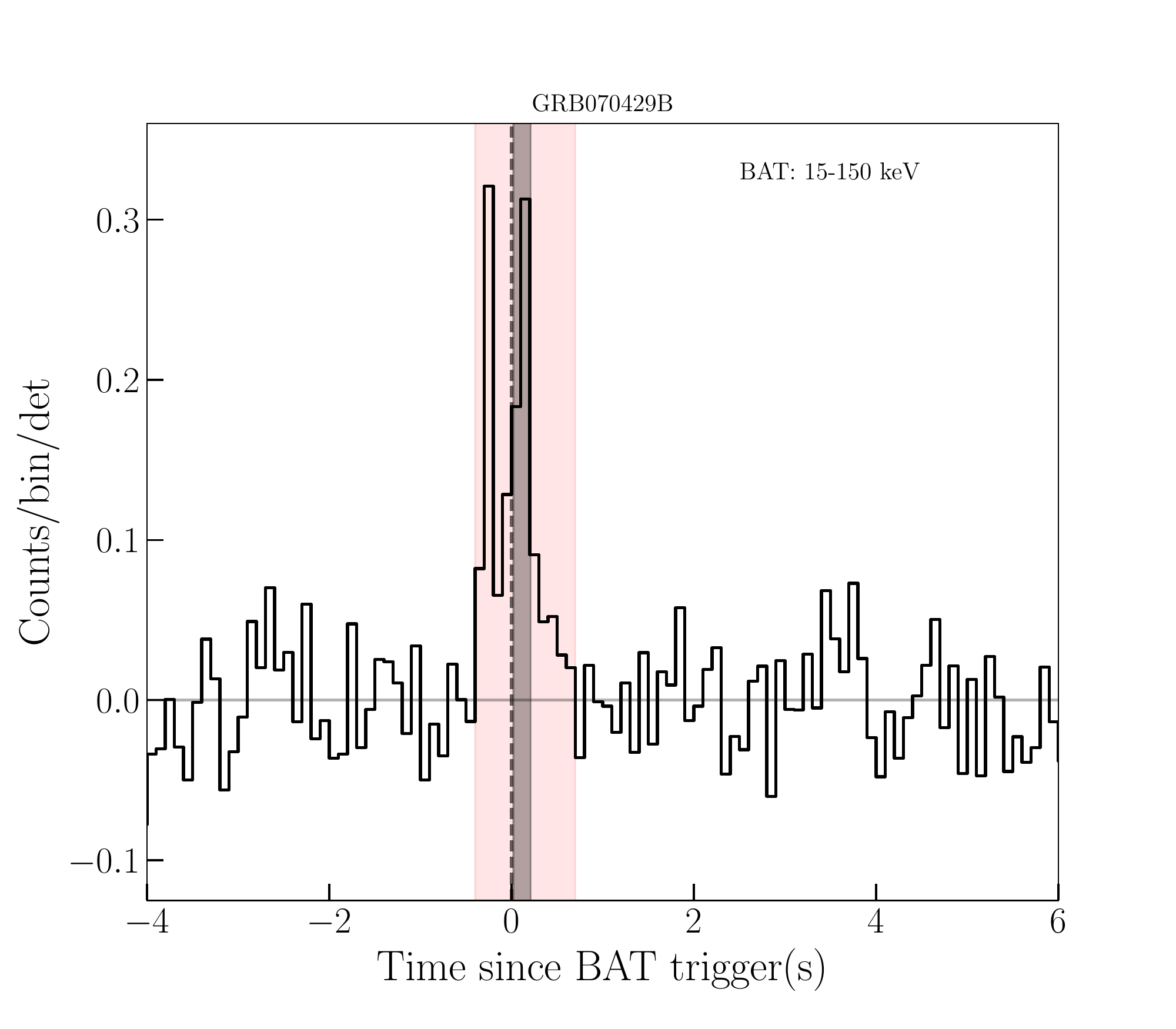}
    \includegraphics[width=.3\linewidth]{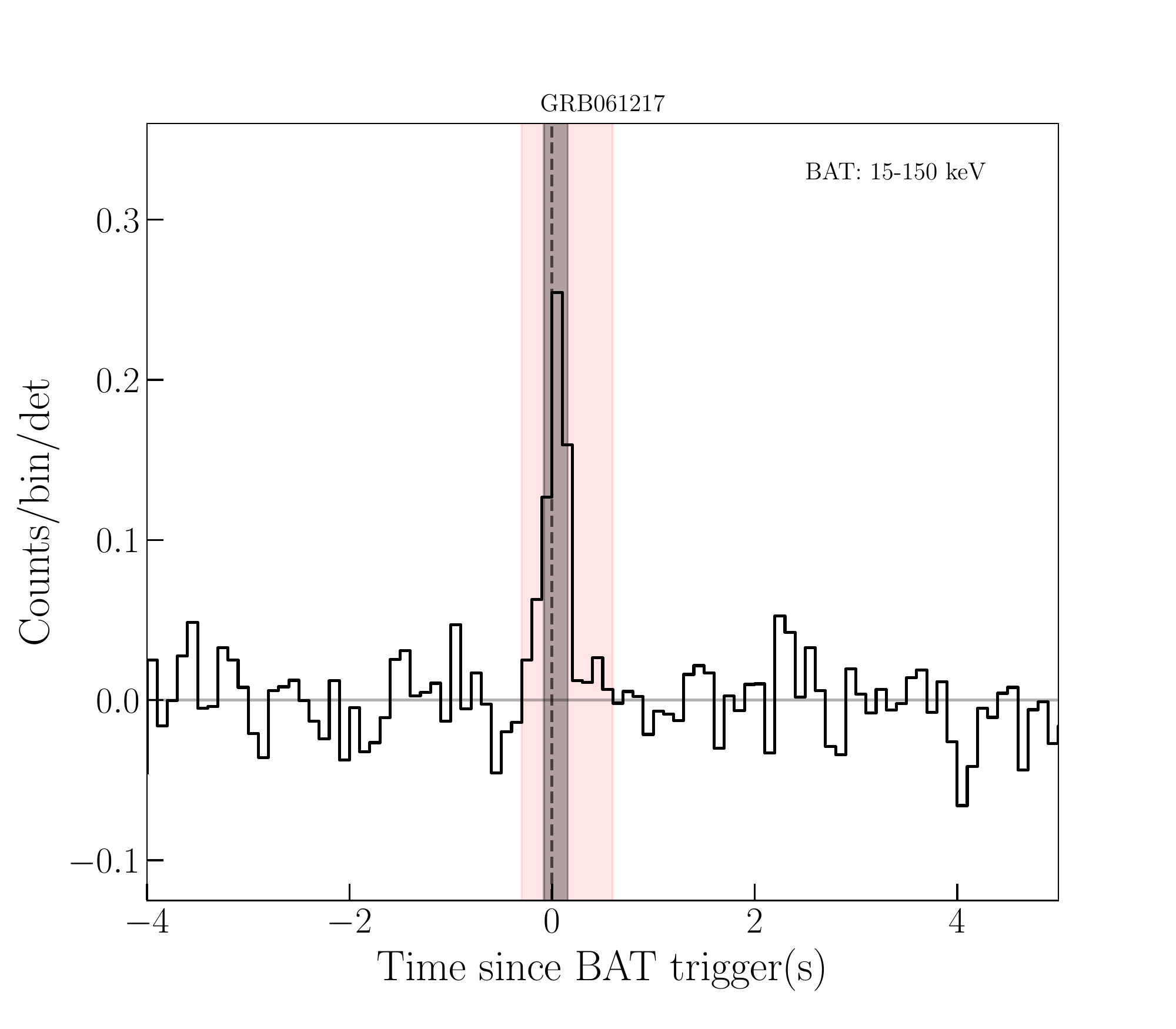}
    \includegraphics[width=.3\linewidth]{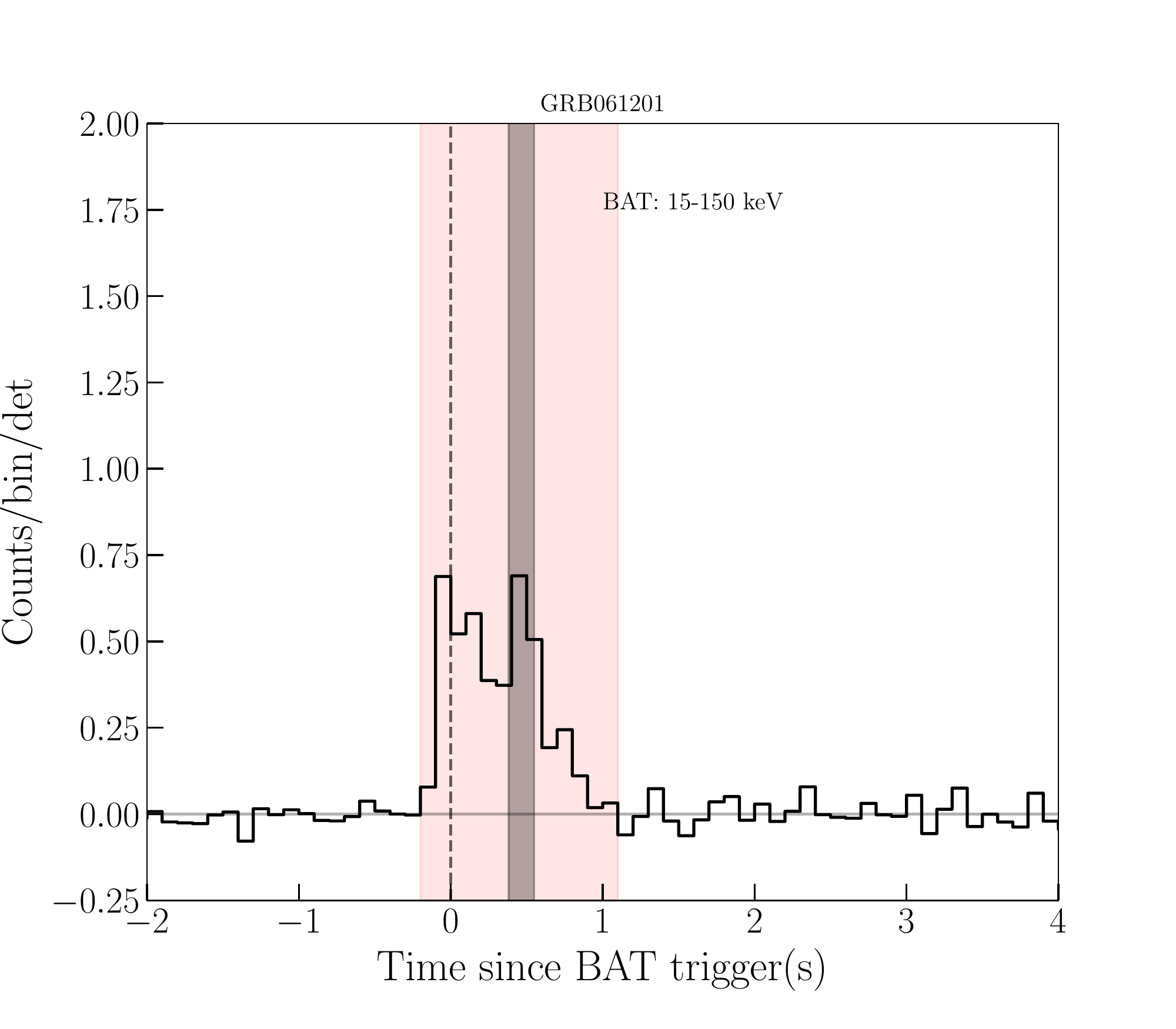}
    \includegraphics[width=.3\linewidth]{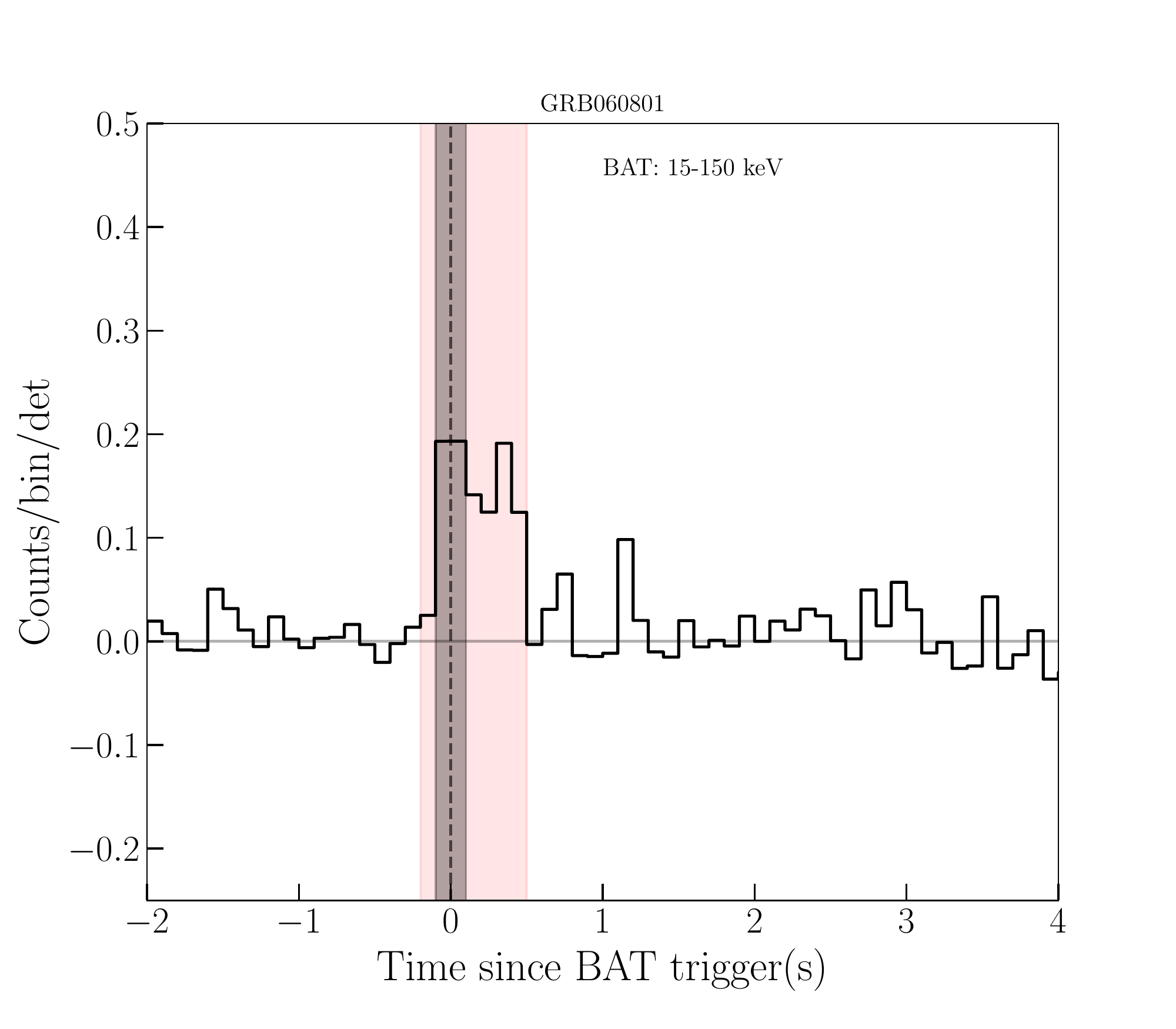}
    \includegraphics[width=.3\linewidth]{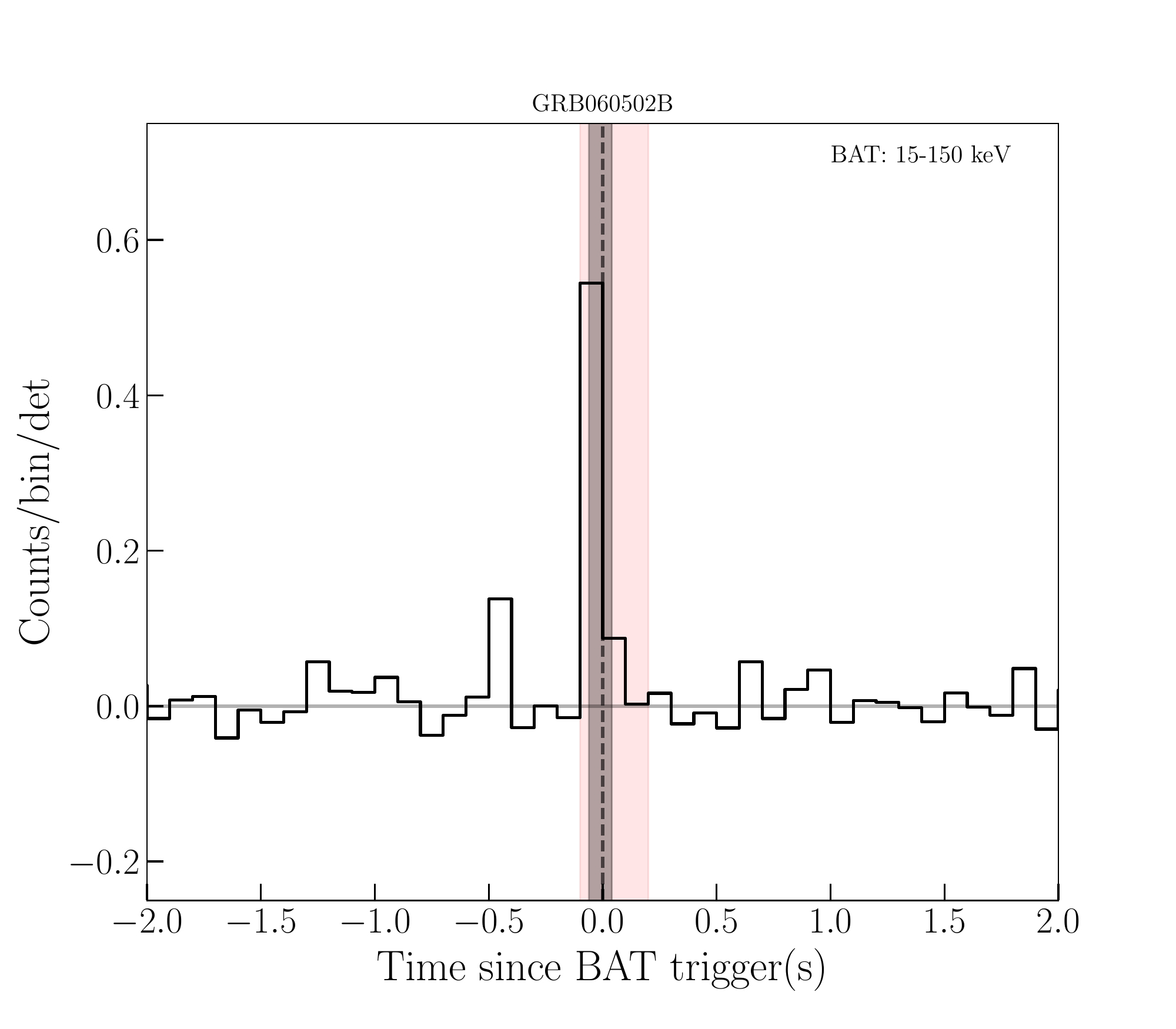}
     \caption{Figure \ref{lcs_1} continued.}
    \label{lcs_3}
\end{figure} 

\begin{figure}
    \centering
    \includegraphics[width=.3\linewidth]{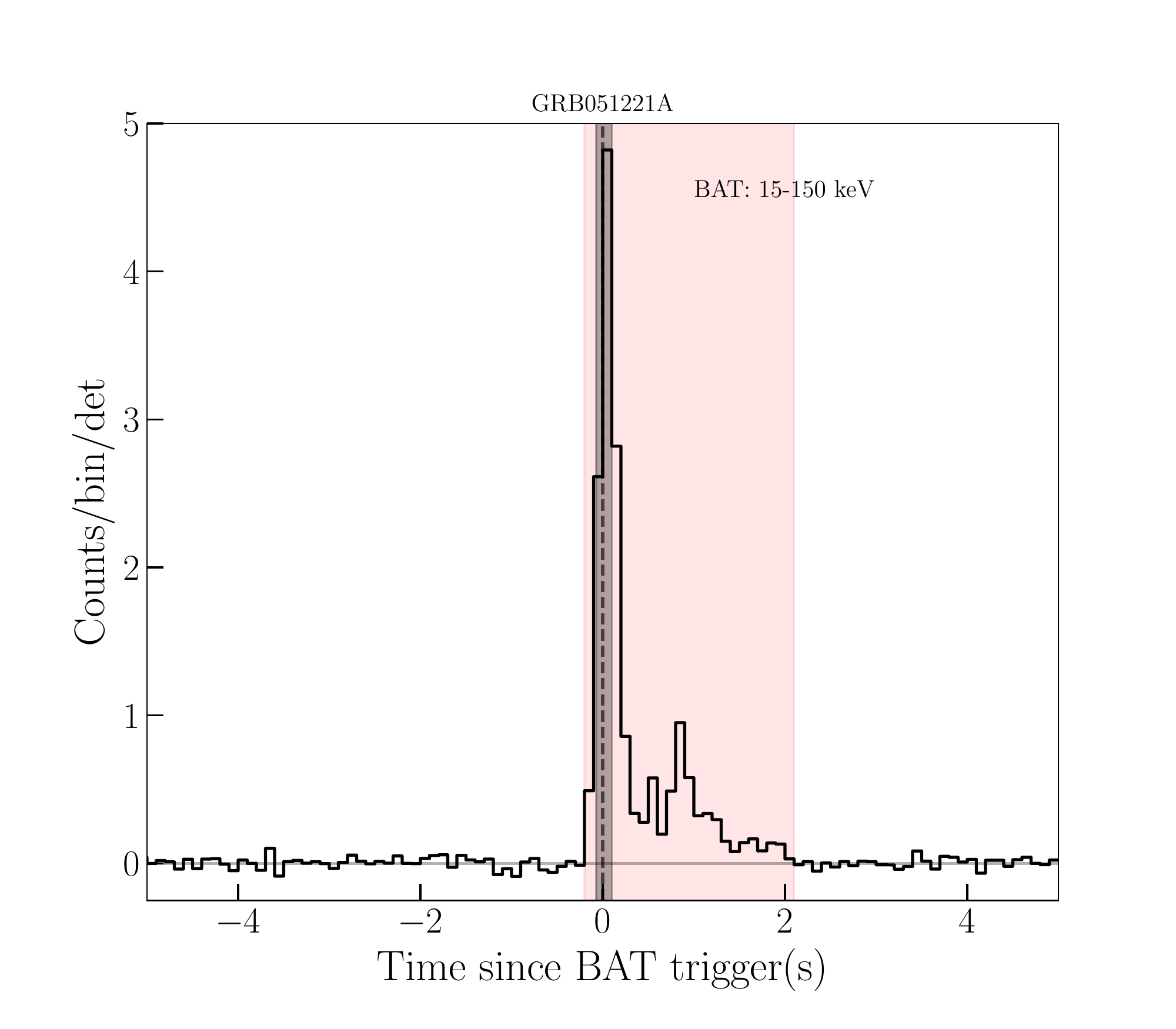}
    \includegraphics[width=.3\linewidth]{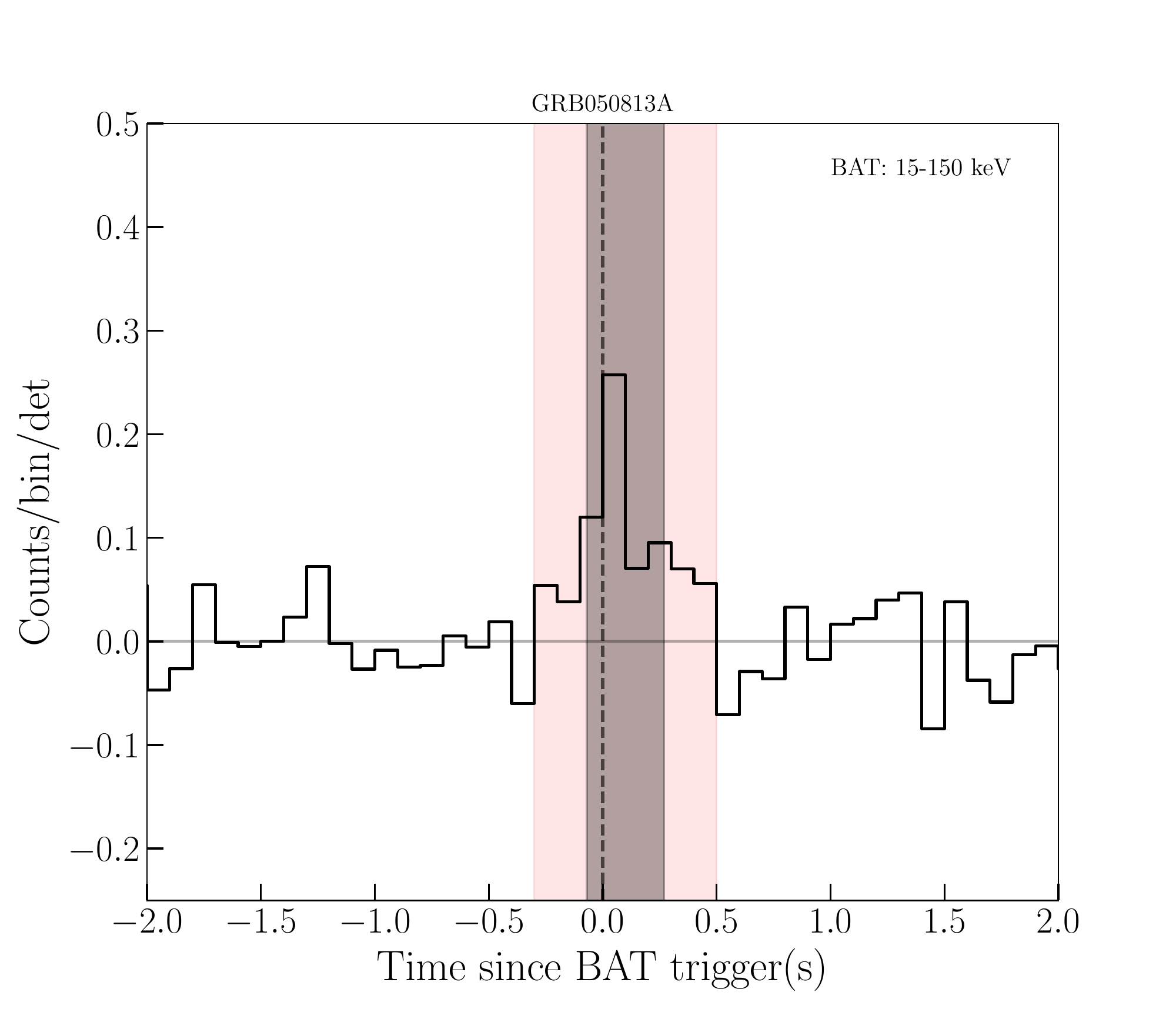}
    \includegraphics[width=.3\linewidth]{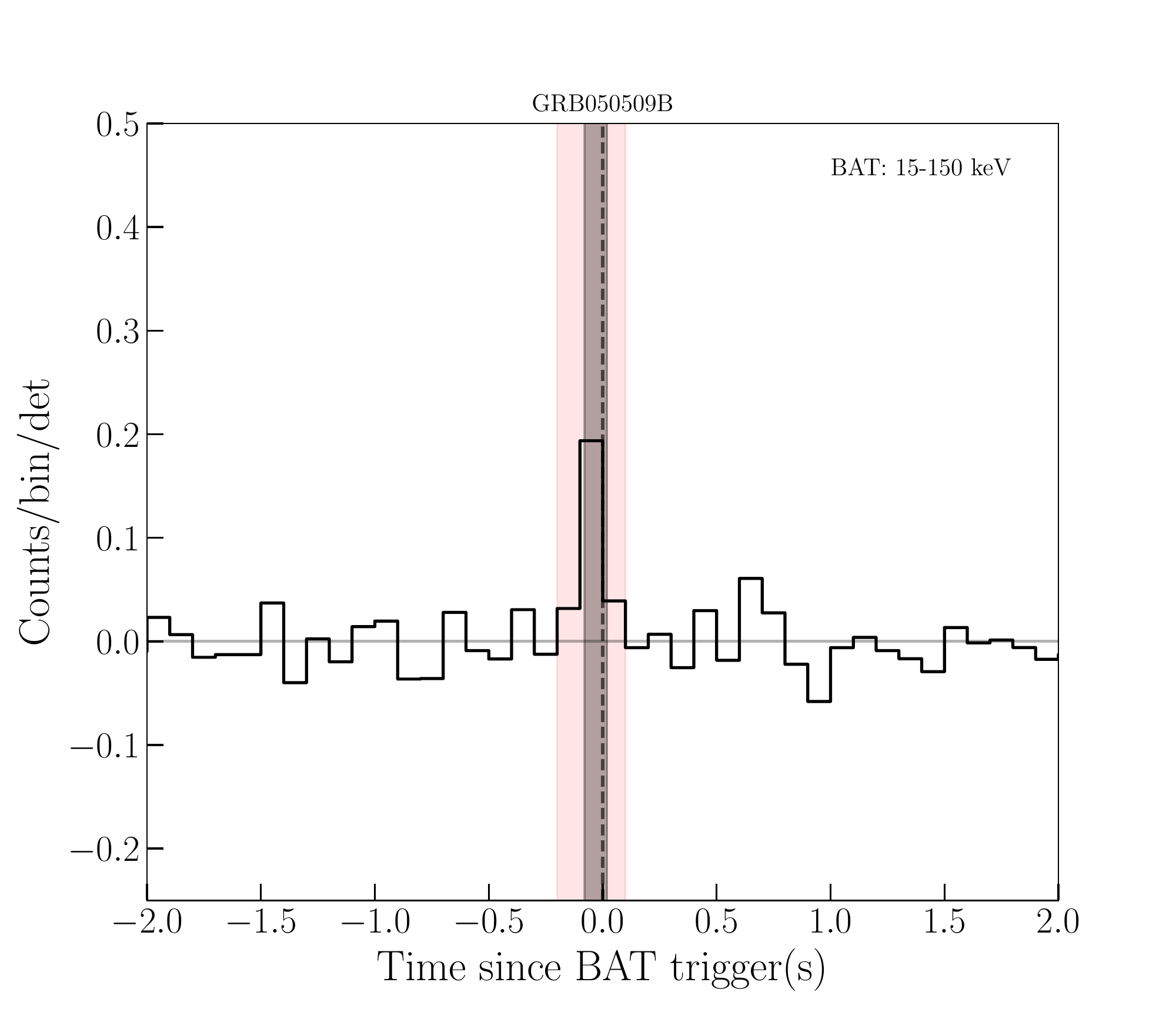}
     \caption{Figure \ref{lcs_1} continued.}
    \label{lcs_4}
\end{figure}

\newpage

\begin{table*}
	\centering
	\caption{The AIC statistics of model fits of integrated spectrum.}
	\label{AIC_integ}
	{\tiny
	\begin{tabular}{|c|c|c|c|c|c|c|c|c|c|c|c|}
		\hline
		 \multirow{1}{*}{GRB name} & \multicolumn{2}{c|}{CPL} &\multicolumn{2}{c|}{PL}& \multicolumn{2}{c|}{BB} & \multicolumn{2}{c|}{mBB} & \multicolumn{2}{c|}{Band} & \multirow{1}{*}{Chosen best model} \\
		\cline{2-11}
		 & AIC & -2log{$\cal{L}$}/DOF & AIC & -2log{$\cal{L}$}/DOF & AIC & -2log{$\cal{L}$}/DOF & AIC & -2log{$\cal{L}$}/DOF & AIC & -2log{$\cal{L}$}/DOF & \\
		\hline
		170817A &2826 & 2820/362 &2837 & 2833/363 &2826 & 2822/363 &2826 & 2820/362 & - & - & mBB\\
		170428A &59 & 53/57 &59 &55/58 &73 & 69/58 &59 & 53/57 &- & - & mBB   \\
		161104A &- & - &70 & 66/58 &80 & 76/58 &73 &67/57 &- & - & PL \\
		160821B &2654 &2648/522 &2658 &2654/523 &2687 &2683/523 &2653 &2647/522 &- & - & mBB\\
		160624A &1892 & 1886/523 &1919 &1915/524 &1901 &1897/524 &1893 &1887/523 &- &- & mBB \\
		150423A &68 &62/57 &67 &63/58 &70 &66/58 &68 & 62/57 &- & - & mBB\\
		150120A &3804 &3798/525 &3805 &3801/526 &3871 &3867/526 &3804 &3798/525 &3807 &3799/524 &mBB   \\
		150101B &175 &169/523 &177 &173/524 &192 &188/524 &175 &169/523  &177 &169/522 & mBB \\
		141212A  &67 &61/57 &67 &63/58 &72 &68/58  &67 &61/57 &- & - & mBB \\
		140903A &42 & 36/57 &40 & 36/58 &98 & 94/58 &41 & 35/57 &- & - & mBB  \\
		140622A &46 &40/57 &48 &44/58 &44 & 40/58 &46 &40/57 &- & - & BB \\
		131004A &3293 &3287/522 &3294 & 3290/523 &3399 &3395/523 &3295 & 3289/522 &3288 & 3280/521 & Band \\
		130603B &71 & 65/57 &70 & 66/58 &204 & 200/58 &71 & 65/57 &- & - & mBB\\
		120804A &63 & 57/57 &71 & 67/58 &116 & 112/58 &63 & 57/57 & -& - & mBB  \\
		111117A &2452 & 2446/525 &2468 &2464/526 &2478 &2474/526 &2452 &2446/525 &2454 &2446/524 & mBB   \\
		101219A &58 & 52/57 &56 & 52/58 &120 &116/58 &59 &53/57 & - & - &PL  \\
		100724A &41 &35/57 &39 &35/58 &48 &44/58 &41 & 35/57 &- & - &mBB  \\
		100628A &58 &52/57 &67 &63/58 &61 &57/58 &64 &57.6/57 &- &- & CPL\\
		100625A &1079 &1073/257 &1149 &1145/258 &1217 &1213/258 &1078 &1072/257 &1081 &1073/256 & mBB  \\
		100206A &834 & 828/526 &935 & 931/527 &936 &932/527 &835 &829/526 &833 &825/525 &mBB \\
		100117A &618 & 612/408 &642 & 638/409 &646 & 642/409 &618 & 612/408 &620 & 612/407 &mBB \\
		090927 &3937 &3931/525 &3936 & 3932/526 &3974 & 3970/526 &3938 &3932/525 &3938 &3930/524 &mBB  \\
		090515 &55 &49/57 &56 &52/58 &53 &49/58 &56 &49/57 & - & - &BB  \\
		090510 &2867 & 2861/532 & - & -  & - & - & 2918 & 2912/532 &2772 & 2764/531 & mBB+PL$^*$ \\
		 & & & & & &  &2767$^*$ & &2767&  &  \\
		  & & & & & &  &  & &(Band +PL)&  &  \\
		090426 &63 &57/57 &63 &59/58 &72 & 68/58 &63 & 57/57 & - & - &mBB  \\
		090417A &64 & 58/57 &66 & 62/58 &62 & 58/58 &64 & 58/57 & - &- & BB  \\
		080905A &2705 &2699/523 &2767 &2763/524 &2738 &2734/524 &2705 &2699/523 &2707& 2699/522 &mBB  \\
		071227 &50 &44/57 &48 &44/58 &63 &59/58 &50 &44/57& -&- &mBB  \\
		070923 &71 & 65/57 &69 & 65/58 &76 &72/58 &71 & 65/57 & - & - &mBB  \\
		070729 &53 &47/57  &51 & 47/58  &55 &51/58 &54 &48/57 & - & - &PL  \\
		070724A &62 & 56/57 &66 &62/58 &61 &57/58 &62 &56/57 & - & - &mBB  \\
		070429B &48 & 42/57 &50 &46/58 &59 &55/58 &49 &43/57 & -& - &mBB   \\
		061217 &56 &50/57 &54 &50/58 &58 & 54/58 &56 &50/57 & - & - &mBB  \\
		061201 &67 &61/57 &65 &61/58 &106 &102/58 &68 &62/57 & - & - &PL  \\
		060801 &76 &70/57 &74 &70/58 &78 & 74/58 &76 &70/57 & - & - &mBB  \\
		060502B &67 & 61/57 &66 &62/58 &70 &66/58 &68 &62/57& - & - &mBB  \\
		051221A &74 & 68/57 &72 &68/58 &562 & 558/58 &75 &69/57 & - & - & PL  \\
		050813 &73 &67/57 &74 &70/58 &71 &67/58 &73 &67/57 & - & - &BB  \\
		050509B &60 &54/57 &58 &54/58 &59 &55/58 &60 &54/57 & - & - &mBB  \\
		
				\hline
	\end{tabular}
	}\\
	Note: The AIC values are not reported for those models fits where the spectral parameters are not well constrained. 
\end{table*}

\begin{table*}
	\centering
	\caption{The AIC statistics of model fits of peak count spectrum.}
	\label{AIC_peak}
	{\tiny
	\begin{tabular}{|c|c|c|c|c|c|c|c|c|c|c|c|}
		\hline
		 \multirow{1}{*}{GRB name} & \multicolumn{2}{c|}{CPL} &\multicolumn{2}{c|}{PL}& \multicolumn{2}{c|}{BB} & \multicolumn{2}{c|}{mBB} & \multicolumn{2}{c|}{Band} & \multirow{1}{*}{Chosen best model} \\
		\cline{2-11}
		 & AIC & -2log{$\cal{L}$}/DOF & AIC & -2log{$\cal{L}$}/DOF & AIC & -2log{$\cal{L}$}/DOF & AIC & -2log{$\cal{L}$}/DOF & AIC & -2log{$\cal{L}$}/DOF & \\
		\hline
		170817A &1265 &1259/362 &1277 &1273/363 &1273 &1269/363 &1264 &1258/362 & - & - &mBB \\
		170428A &65 & 59/57 &63 &59/58 &65 &61/58 &65 &59/57 & - & - &mBB    \\
		161104A &57 & 50/41 &71 & 67/42 &57 &53/42 &61 &54/41 & - & - &BB \\
		160821B &-133 &-139/420 &-126 &-130/421 &-127 &-131/421 &-132 &-138/420 & - &-  &mBB \\
		160624A &426 & 420/523 &458 &454/524 &430 &426/524 &426 &420/523 &427 &419/522 &mBB \\
		150423A &73 & 67/57 &71 &67/58 &76 & 72/58  &73 &67/57 & - & - &mBB \\
		150120A &-62 & -68/525 &-60 & -64/526 &-52 & -57/526 &-62 & -68/525 &-60 & -68/524 &mBB \\
		150101B &-536 &-542/463 &-538 &-542/464 &-521 &-525/464 &-538 &-544/463 &-536 &-544/522 &mBB\\
		141212A &47 &40/37 &45 &41/38 &53 &49/38 &47 &40/37 & - & - &mBB \\
		140903A &48 &42/57 &46 &42/58 &63 &59/58 &48 &42/57 & - & - &mBB\\
		140622A &35 & 29/53 &50 &46/54 &32 & 28/54 &37 &36/53 & - & - &BB\\
		131004A &-1740 &-1746/502 &-1742 &-1746/503 &-1745 &-1749/503 &-1744 &-1750/502  &-1747 &-1755/521 &Band\\  
		130603B &45 &39/57 &50 &46/58 &51 &47/58 &45 &39/57 & - & - &mBB\\ 
		120804A &67 &61/57 &67 &63/58 &77 &73/58 &67 &61/57 & - & - &mBB\\ 
		111117A &-960 &-966/525 &-947 &-951/526 &-946 &-950/526 &-960 &-966/525 &-958 &-966/524 & mBB\\ 
		101219A &53 &47/57 &50 &46/58 &61 &57/58 &53 &47/57 & -  & - &PL\\ 
		100724A &59 &53/57 &59 &55/58 &60 &56/58 &59 & 53/57 & - & - &mBB\\ 
		100628A &59 & 53/57 &63 &59/58 &57 & 53/58 &60 &54/57 & - & - &BB\\ 
		100625A &-960 &-966/202 &-951 &-955/203 &-961 &-965/203 &-960 &-966/202 &-960 &-968/256 &BB\\ 
		100206A &-884 &-890/526 &-778 &-782/527 &-799 &-803/527 &-883 &-889/526 &-882 &-890/525 &mBB\\ 
		100117A &-512 &-518/408 &-511 &-515/409 &-472 &-476/409 &-512 &-518/408 &-510 &-518/407 &mBB\\ 
		090927 &1179 &1173/525 &1181 &1177/526 &1203 &1199/526 &1180 &1174/525 &1181 &1173/524 &mBB\\
		090515 &63 &57/57 &61 &57/58 &53 &49/58 &56 &50/57 & - & - &BB\\ 
		090510 &879 & 873/532 &919 &915/533 &922 &918/533 & 879 &873/532 &881 &873/531 &mBB\\ 
		090426 &41 &34/41 &45 &41/42 &40 &36/42 &44 &36/41  & - &-  &BB\\ 
		090417A &56 &49/41  &54 &50/42  &56 &52/42 &56 &50/41 & - &- &mBB\\
		080905A &-2255 &-2261/402 &-2257 &-2261/403 &-2269 &-2273/403 &-2267 & -2273/402 &-2264 &-2272/522 &BB\\ 
		071227 &69 &63/57 &67 &63/58 &68 &64/58 &69 &63/57 & - &- &mBB\\ 
		070923 &72 &66/57 &70 &66/58 &75 &71/58 &72 &66/57 & - &- &mBB\\  
		070729 &66 &60/57 &64 & 60/58 &67 &63/58  &66 &60/57 & - & -&mBB\\ 
		070724A &63 &57/57  &61 &57/58 &69 & 65/58  &63 &57/57 & - &- &mBB\\ 
		070429B &54 &48/57  &55 &51/58 &52 &48/58 &54 &48/57 & -& - &mBB\\ 
		061217 &62 &56/53 &61 &57/54 &59 & 55/54 &62 &56/53 & - &- &BB\\ 
		061201 &60 &54/57  &58 &54/58 &67 &63/58 &60 &54/57 & - &- &mBB\\ 
		060801 &63 &57/57 &60 &56/58 &63 &59/58 &63 &57/57 & - & - &PL\\ 
		060502B &52 &46/57 &50 &46/58  &52 &48/58 &52 &46/57 & - &- &mBB\\ 
		051221A &71 &65/57 &75 &71/58 &129 &125/58 &71 &65/57 & - & - &mBB\\  
		050813 &58 &52/57 &62 &58/58 &56 &52/58 &58 &52/57 &-& -&BB\\ 
		050509B &44 &37/41 &42 &38/42 &42 &38/42 &44 &37/41 & - &- &BB\\ 
		
		\hline
	\end{tabular}
	}\\
	Note: The AIC values are not reported for those models fits where the spectral parameters are not well constrained.  
\end{table*}

\begin{table*}
	\centering
	\caption{The signal-to-noise ratio (SNR) used to obtain the peak time interval of the sGRBs in the sample are listed below.}
	\label{sample_SNR}
	\begin{tabular}{|c|c|}
	\hline
		 \multirow{1}{*}{GRB name} & SNR (sigma) \\
	\hline
	    170817A & 3.0 \\
		170428A & 10.0 \\
		161104A & 5.0 \\
		160821B & 6.0 \\
		160624A & 4.0 \\
		150423A & 6.0 \\
		150120A & 8.0 \\
		150101B & 5.0 \\
		141212A & 5.0 \\
		140903A & 9.0 \\
		140622A & 5.0 \\
		131004A & 10.0 \\
		130603B & 15.0 \\
		120804A & 12.0 \\
		111117A & 4.5 \\
		101219A & 10.0 \\
		100724A & 6.0 \\
		100628A & 6.0 \\
		100625A & 8.0 \\
		100206A & 8.0 \\
		100117A & 6.0 \\
		090927 & 8.0 \\
		090515 & 4.0 \\
		090510 & 6.0 \\
		090426 & 6.0 \\
		090417A & 4.0 \\
		080905A & 5.0 \\
		071227 & 6.0 \\
		070923 & 6.0 \\
		070729 & 6.0 \\
		070724A & 5.0 \\
		070429B & 6.0 \\
		061217 & 5.0 \\
		061201 &10.0 \\
		060801 & 6.0 \\
		060502B & 6.0 \\
		051221A & 25.0 \\
		050813 & 4.0 \\
		050509B & 3.5 \\
	
		\hline
	\end{tabular}
\end{table*}

\begin{figure}
    \centering
    \includegraphics[scale=0.5]{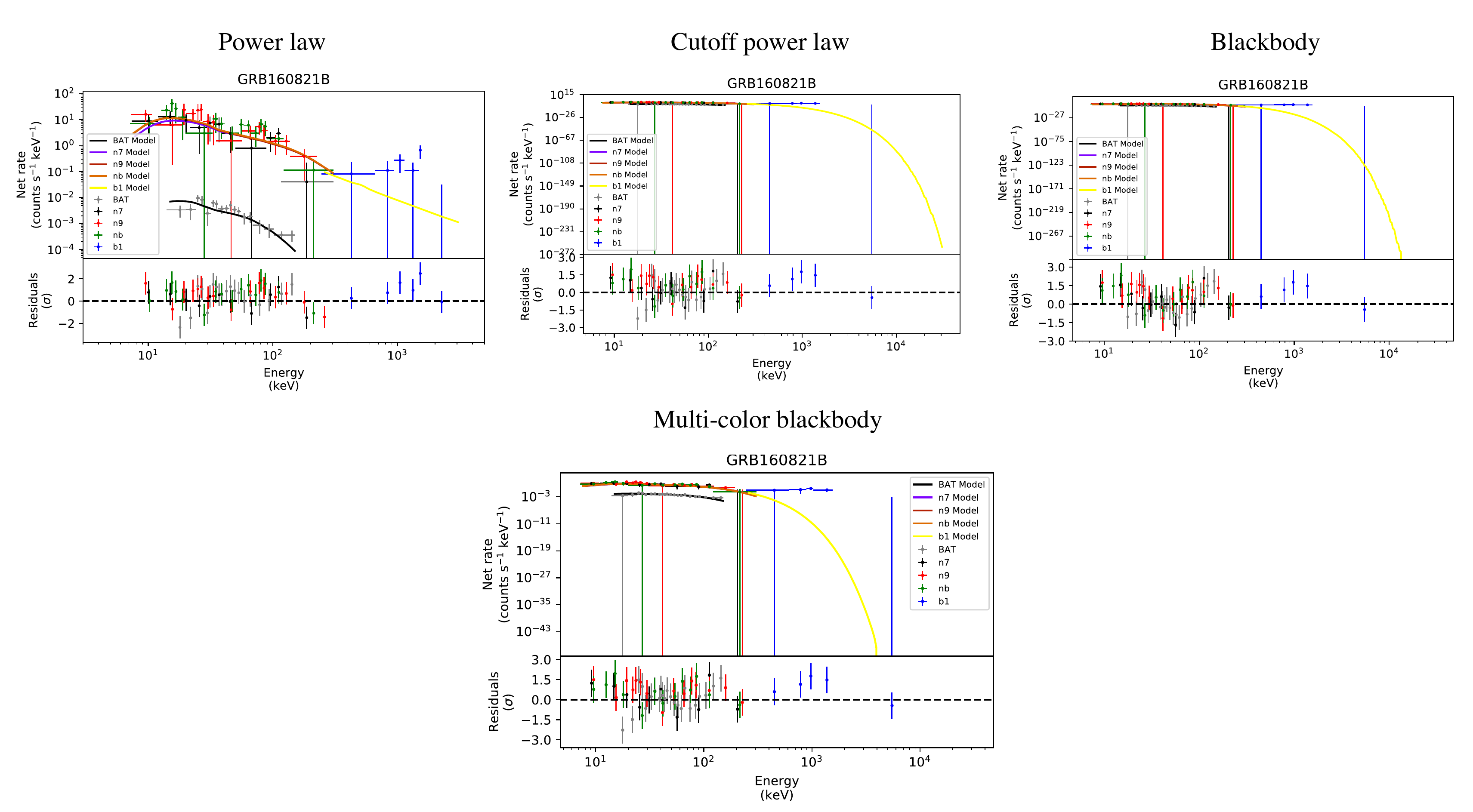}
    \includegraphics[scale=0.5]{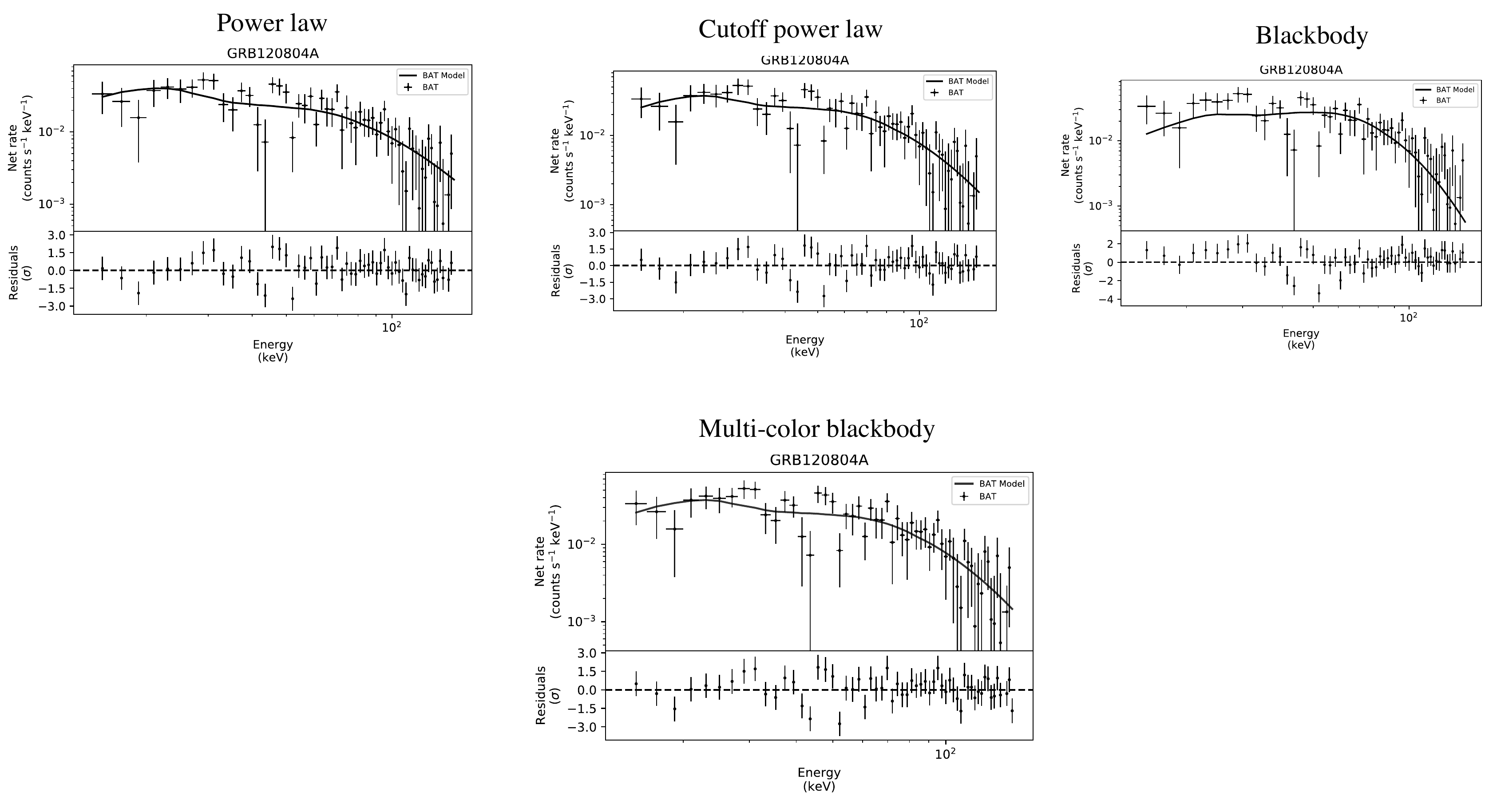}
    \caption{In order to demonstrate the aspect of degeneracy between different spectral models,  the counts spectra and the residuals obtained for the spectral fits using different models such as {\tt PL}, {\tt CPL}, {\tt BB} and {\tt mBB}, done to the peak count spectrum of (a) GRB 160821B and (b) GRB 120804A are shown. We note that all the different spectral models are consistent with the same data.}
    \label{residuals}
\end{figure}

\begin{figure}
    \centering
    \includegraphics[scale=0.4]{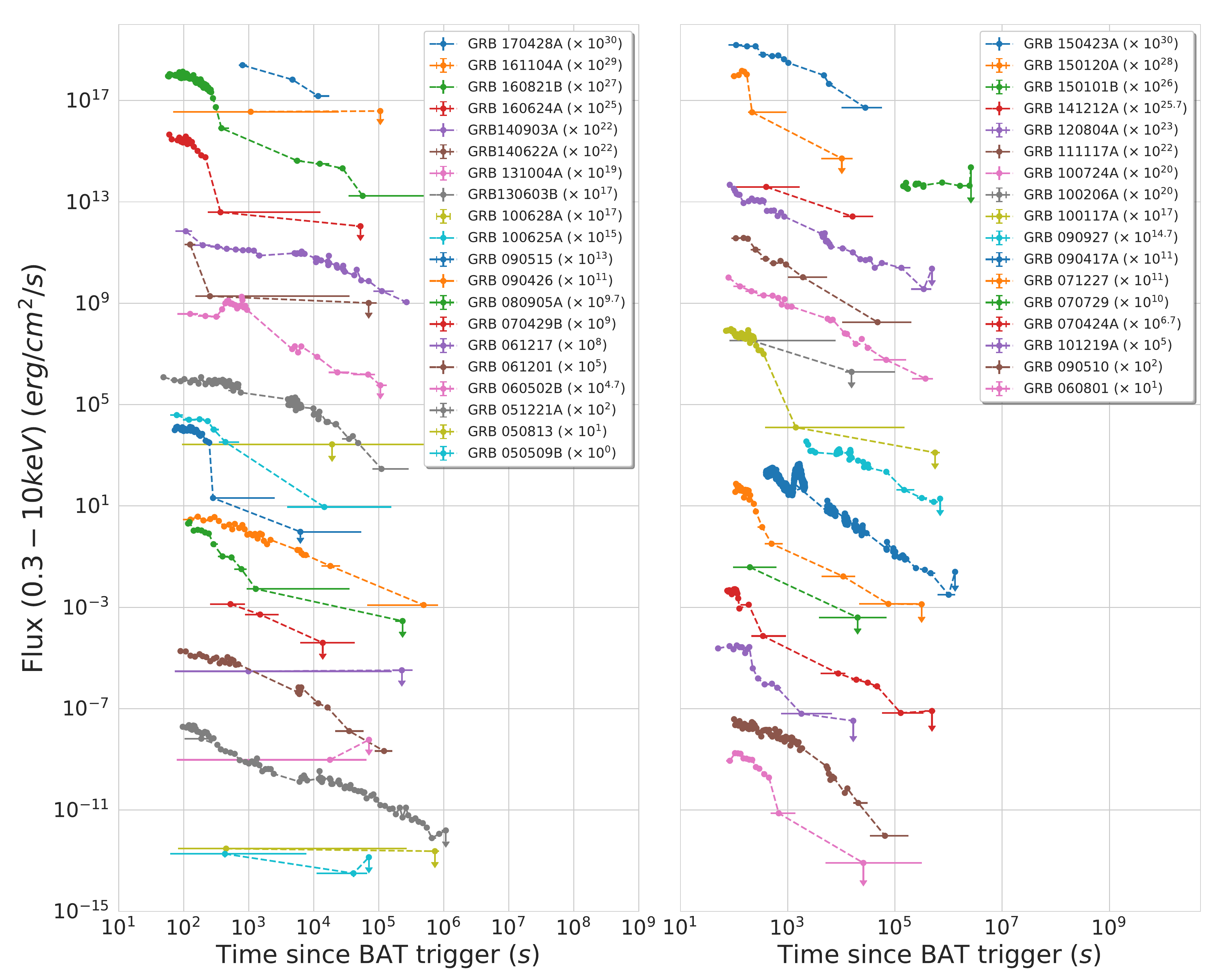}
    \caption{The {\it Niel Gehrels Swift} XRT afterglow flux light curves\textsuperscript{+} observed for the different sGRBs in the sample are shown. 
    The XRT data were not available for GRB 170817A and GRB 070923A.}
    \small\textsuperscript{+} https://www.swift.ac.uk/xrt\_curves/
    \label{xrt_afterglow}
\end{figure}

\begin{figure}
    \centering
    \includegraphics[scale=0.5]{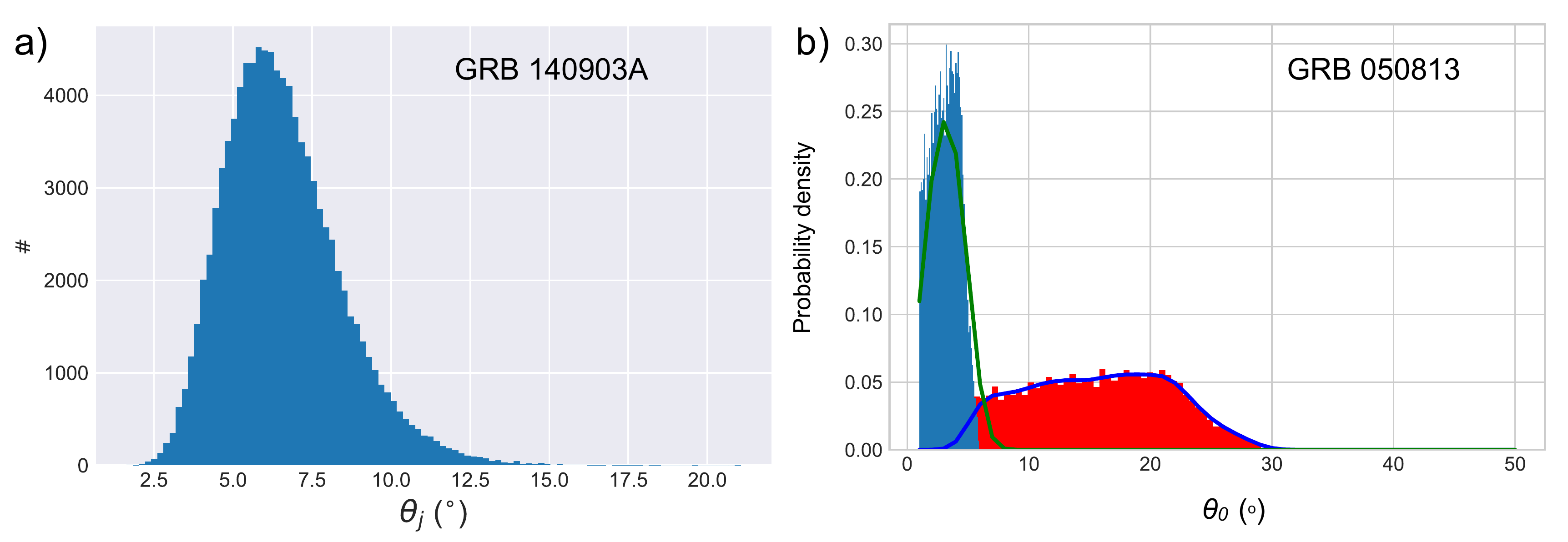}
    \caption{For an example in (a) we have plotted the distribution obtained for the $\theta_j$ estimated using the jet break observed in GRB 140903A and in (b) the distributions obtained for $\theta_0$ (red) and $\theta_0/5$ (blue) for the GRB 050813A are shown.}
    \label{theta_j_theta_0}
\end{figure}

\end{document}